\documentclass[iop,twocolumn]{aastex62}
\usepackage{rotating}
\usepackage{appendix}
\usepackage{ulem}
\usepackage{lineno}

\newcommand{\hbmage}{8.0$^{+3.2}_{-2.3}$ Gyr}

\newcommand{\GE}{\textit{Gaia}--\textit{Enceladus}--\textit{Sausage}}

\graphicspath{{./}{figures/}}

\shorttitle{Halo giant star ages}
\shortauthors{Grunblatt, Zinn, Price-Whelan et al.}

\newcommand{\acronym}[1]{{\small{#1}}}

\newcommand{\kms}{\ensuremath{\mathrm{km}~\mathrm{s}^{-1}}}

\newcommand{\pc}{\ensuremath{\mathrm{pc}}}
\newcommand{\kpc}{\ensuremath{\mathrm{kpc}}}

\def\numax{\nu_{\mathrm{max}}}
\def\dnu{\Delta \nu}

\newcommand{\gaia}{\textsl{Gaia}}
\newcommand{\tess}{\acronym{TESS}}

\begin{document}

\title{Age-Dating Red Giant Stars Associated with Galactic Disk and Halo Substructures}

\newcommand{\cca}{Center for Computational Astrophysics, Flatiron Institute, Simons Foundation, 162 Fifth Avenue, New York, NY 10010, USA}

\correspondingauthor{Samuel Kai Grunblatt}
\email{sgrunblatt@amnh.org}

\author[0000-0003-4976-9980]{Samuel K. Grunblatt}
\altaffiliation{Kalbfleisch Fellow}
\affil{Department of Astrophysics, American Museum of Natural History, 200 Central Park West, Manhattan, NY 10024, USA}
\affil{\cca}

\author[0000-0002-7550-7151]{Joel C. Zinn}
\altaffiliation{NSF Astronomy and Astrophysics Postdoctoral Fellow}
\affil{School of Physics, University of New South Wales, NSW 2052, Australia}
\affil{Department of Astrophysics, American Museum of Natural History, 200 Central Park West, Manhattan, NY 10024, USA}
\author[0000-0003-0872-7098]{Adrian~M.~Price-Whelan}
\affiliation{\cca}

\author[0000-0003-4540-5661]{Ruth Angus}
\affil{Department of Astrophysics, American Museum of Natural History, 200 Central Park West, Manhattan, NY 10024, USA}
\affiliation{\cca}

\author[0000-0003-2657-3889]{Nicholas Saunders}
\altaffiliation{NSF Graduate Research Fellow}
\affiliation{Institute for Astronomy, University of Hawai`i, 2680 Woodlawn Drive, Honolulu, HI 96822, USA}

\author[0000-0003-2400-6960]{Marc Hon}
\altaffiliation{Hubble Fellow}
\affiliation{Institute for Astronomy, University of Hawai`i, 2680 Woodlawn Drive, Honolulu, HI 96822, USA}

\author[0000-0002-5496-365X]{Amalie Stokholm}
\affiliation{Stellar Astrophysics Centre, Department of Physics and Astronomy, Aarhus University, Ny Munkegade 120, DK-8000 Aarhus C, Denmark}

\author[0000-0003-4456-4863]{Earl P. Bellinger}
\altaffiliation{SAC Postdoctoral Research Fellow}
\affil{Stellar Astrophysics Centre, Department of Physics and Astronomy, Aarhus University, Ny Munkegade 120, DK-8000 Aarhus C, Denmark}
\affil{School of Physics, University of New South Wales, NSW 2052, Australia}

\author[0000-0002-3430-4163]{Sarah L. Martell}
\affil{School of Physics, University of New South Wales, NSW 2052, Australia}

\author[0000-0002-7547-1208]{Benoit Mosser}
\affil{LESIA, Observatoire de Paris, Universit\'e PSL, CNRS, Sorbonne Universit\'e, Universit\'e de Paris, 92195 Meudon, France}

\author[0000-0002-6993-0826]{Emily Cunningham}
\affiliation{\cca}

\author[0000-0002-4818-7885]{Jamie Tayar}
\altaffiliation{Hubble Fellow}
\affiliation{Institute for Astronomy, University of Hawai`i, 2680 Woodlawn Drive, Honolulu, HI 96822, USA}

\author[0000-0001-8832-4488]{Daniel Huber}
\affiliation{Institute for Astronomy, University of Hawai`i, 2680 Woodlawn Drive, Honolulu, HI 96822, USA}

\author[0000-0001-9234-430X]{Jakob Lysgaard R{\o}rsted}
\affiliation{Stellar Astrophysics Centre, Department of Physics and Astronomy, Aarhus University, Ny Munkegade 120, DK-8000 Aarhus C, Denmark}

\author[0000-0002-6137-903X]{Victor Silva Aguirre}
\affiliation{Stellar Astrophysics Centre, Department of Physics and Astronomy, Aarhus University, Ny Munkegade 120, DK-8000 Aarhus C, Denmark}



\begin{abstract}

The vast majority of Milky Way stellar halo stars were likely accreted from a small number ($\lesssim$3) of relatively large dwarf galaxy accretion events. 
However, the timing of these events is poorly constrained, relying predominantly on indirect dynamical mixing arguments or imprecise age measurements of stars associated with debris structures.
Here, we aim to infer robust stellar ages for stars associated with galactic substructures to more directly constrain the merger history of the Galaxy.
By combining kinematic, asteroseismic, and spectroscopic data where available, we infer stellar ages for a sample of 10 red giant stars that were kinematically selected to be within the stellar halo, a subset of which are associated with the \GE\ halo substructure, and compare their ages to 3 red giant stars in the Galactic disk.
Despite systematic differences in both absolute and relative ages determined here, age rankings of stars in this sample are robust. Passing the same observable inputs to multiple stellar age determination packages, we measure a weighted average age for the \GE\ stars in our sample of 8 $\pm$ 3 (stat.) $\pm$ 1 (sys.) Gyr. We also determine hierarchical ages using \texttt{isochrones} for the populations of \GE, {\it in situ} halo and disk stars, finding a \GE\ population age of \hbmage. Although we cannot distinguish hierarchical population ages of halo or disk structures with our limited data and sample of stars, this framework should allow distinct characterization of Galactic substructures using larger stellar samples and additional data available in the near future.

\end{abstract}



\section{Introduction} \label{sec:intro}


The formation of galaxies proceeds hierarchically, as objects grow their stellar components through accretion of gas (which then forms stars) or through direct accretion of stars from lower mass satellites \citep{searle1977, bland-hawthorn2002}.
In the Milky Way, we have substantial evidence for this process through observations of the many dwarf galaxy satellites around the Galaxy \citep[e.g.,][and references therein]{mcconnachie2012}, through the discovery of phase-coherent stellar and gas streams and debris structures \citep[e.g.,][and references therein]{grillmair2016}, and through more recent suggestions about the origins of the dynamically-older, more mixed inner stellar halo, which is now thought to have formed from a small number of large merger events \citep{deason2015, helmi2018, belokurov2018, myeong2019}.
While many dynamical arguments have been made to attempt to date the mergers that led to the formation of the bulk of the Milky Way's stellar halo \citep[e.g.,][]{koppelman2020}, directly measuring stellar ages provides a powerful and complementary approach for timing these merger events, thus providing constraints on the history of galaxy formation and, in turn, cosmological models \citep{marquez1994, helmi2008}.

Kinematic properties of stellar populations are essential for inferring the history of our Galaxy's formation. The second data release from the \gaia\ mission (\gaia\ DR2) provides superb astrometric parameters, radial velocities, and photometry for more than 1 billion stars, revolutionizing our insight into the kinematics of stars in our Galaxy \citep{gaia2018}. These observations have already uncovered a number of previously unknown stellar streams and stellar halo substructures \citep[e.g.,][]{koppelman2018,malhan2018, malhan2021}. One of the most intriguing discoveries from this dataset was that a significant fraction of halo stars located near the Sun are associated with an extended kinematic structure that has negligible or retrograde mean motion around the Galaxy. This structure, now referred to as the ``\GE'' \citep[e.g.,][]{vincenzo2019}, is readily apparent in the velocity distribution of stars observed by \gaia\ DR2 (and later data releases), and its most easily observable components are generally giant stars \citep[due to Malmquist bias,][]{malmquist1922}. Previous studies of the halo velocity structure have shown that the radial and/or mildly retrograde substructure could be the result of stars originating in an external galaxy which merged with the Milky Way in the past \citep{helmi2018, belokurov2018}. In addition, using a sample of \gaia\ DR1 stars with spectroscopy from RAVE and APOGEE, \citet{bonaca2017} found a population of metal-rich stars on eccentric orbits, and argued that these stars formed in the disk of the Milky Way and were perturbed onto halo-like orbits. Several studies have suggested that the \GE\ merger was potentially responsible for perturbing this {\it in situ} halo population \citep[e.g.,][]{belokurov+2020, bonaca2020}. In addition, structures with even stronger retrograde motion have also been identified, which trace the initial formation and early growth of our Galaxy \citep{myeong2019, koppelman2019, helmi2020}. Therefore by using newly available kinematic information, we can now identify thousands of stars that fall into these newly discovered kinematic populations.

However, kinematics alone are insufficient to determine the origins of a star. By combining kinematic data with additional types of information, stellar properties can be constrained much more tightly. Large spectroscopic surveys provide new, invaluable insights into the compositions of stars, which can be combined with kinematic information to trace individual stellar populations of our Galaxy more effectively. Related to this, different kinematic groups in the galactic halo have been shown to have distinct chemical properties \citep{nissen2010,hayes2018,mackereth2019}. Identification of these chemical properties subsequently allows more precise inference about the mass distribution of the Galaxy \citep{pricewhelan2020}. This information can be furthermore combined with time-series observations of stars to obtain constraints on time-series variability, and thus constrain stellar properties independently of either spectroscopic or kinematic information. Asteroseismology, or the study of oscillations in stars, uses such time-series information to provide additional independent constraints on stellar properties \citep{miglio2013,silvaaguirre2015,pinsonneault+2018}. By combining asteroseismic, spectroscopic, and kinematic information, we can produce a uniquely well-characterized sample of stars with examples in physically and chemically distinct regions of our Galaxy. Furthermore, by constraining the ages and compositions of stars in our galactic halo which are remnants of dwarf galaxies that formed at high redshift, we can study the star formation and chemical enrichment histories of the early Universe in great detail \citep[e.g.][]{naidu2020}.  

In the past, these studies were generally limited to certain regions of the sky, which made it challenging to observe widely dispersed halo populations. However, the recently launched NASA Transiting Exoplanet Survey Satellite \citep[\tess,][]{ricker2014} opened the brightest stars across $\gtrsim$80\% of the sky to photometric studies with micro-magnitude, high-cadence precision in its two-year nominal mission. Within the \tess\ survey, there are tens to hundreds of thousands of giant stars which are bright enough and oscillate on timescales of hours or longer, sufficient for asteroseismic properties to be estimated from 30-minute (or higher) cadence \tess\ data, presenting an abundance of opportunities for detailed characterization, analysis and follow-up. By combining a sample of red giant stars observed to be oscillating in \tess\ data with astrometry and kinematics from \gaia\ DR2, we can identify evolved stars that are part of the \GE\ or other distinct kinematic structures. We can then use asteroseismic information to determine precise stellar densities and surface gravities, and thus masses and radii, and then combine this with spectroscopic, photometric, and kinematic information to apply power-law relations, isochrone and/or asteroseismic frequency models to determine stellar ages. These ages will allow us to place timing constraints on the \GE\ merger event, and may reveal age differences between kinematically and chemically distinct structures in our Galaxy.


In this study we combine multiple datasets and analysis packages to robustly determine effective temperatures, masses, radii, and ages for 13 red giant stars over a range of evolutionary states and densities that are widely distributed and associated with distinct features in kinematic space. We verify these age estimates using multiple models to estimate stellar ages to ensure that our values are robust \citep{morton2016,sharma16, huber2017,bellinger2020,dasilva2006,rodrigues2017, silvaaguirre2018}. We then use this sample to estimate the age distribution of stars within the \GE\ structure relative to other stars within the stellar halo or other components of the Galaxy. We compare our estimates of the \GE\ merger epoch to other such estimates made for this population \citep[e.g.,][]{chaplin2020, matsuno2020, montalban2020}, explore potential sources of error in various age determination methods, and discuss implications of the results of this and other complementary studies of Galactic archaeology.




\section{Target Selection}

\begin{figure*}
\centering
\includegraphics[width=0.495\textwidth]{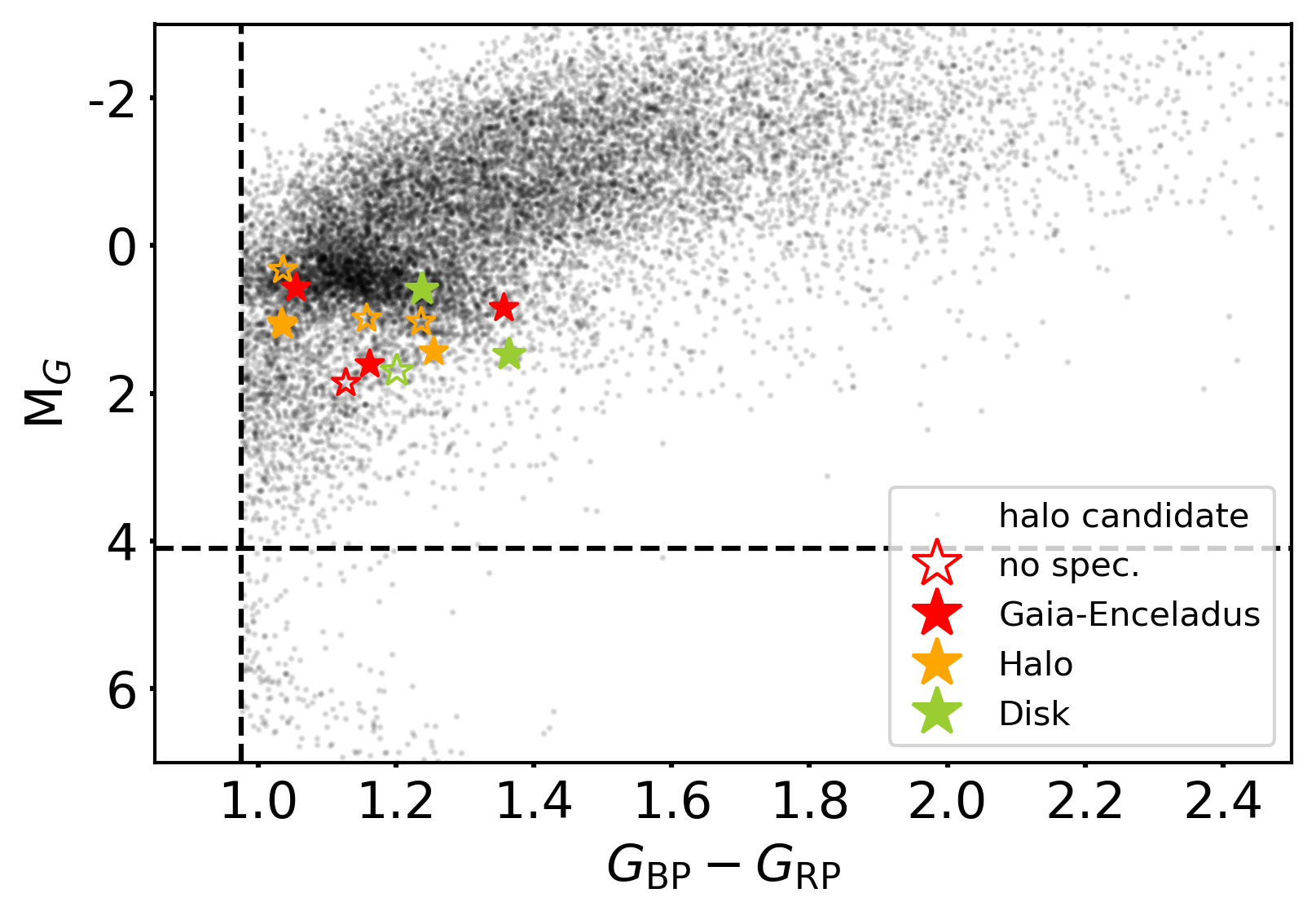}
\includegraphics[width=0.495\textwidth]{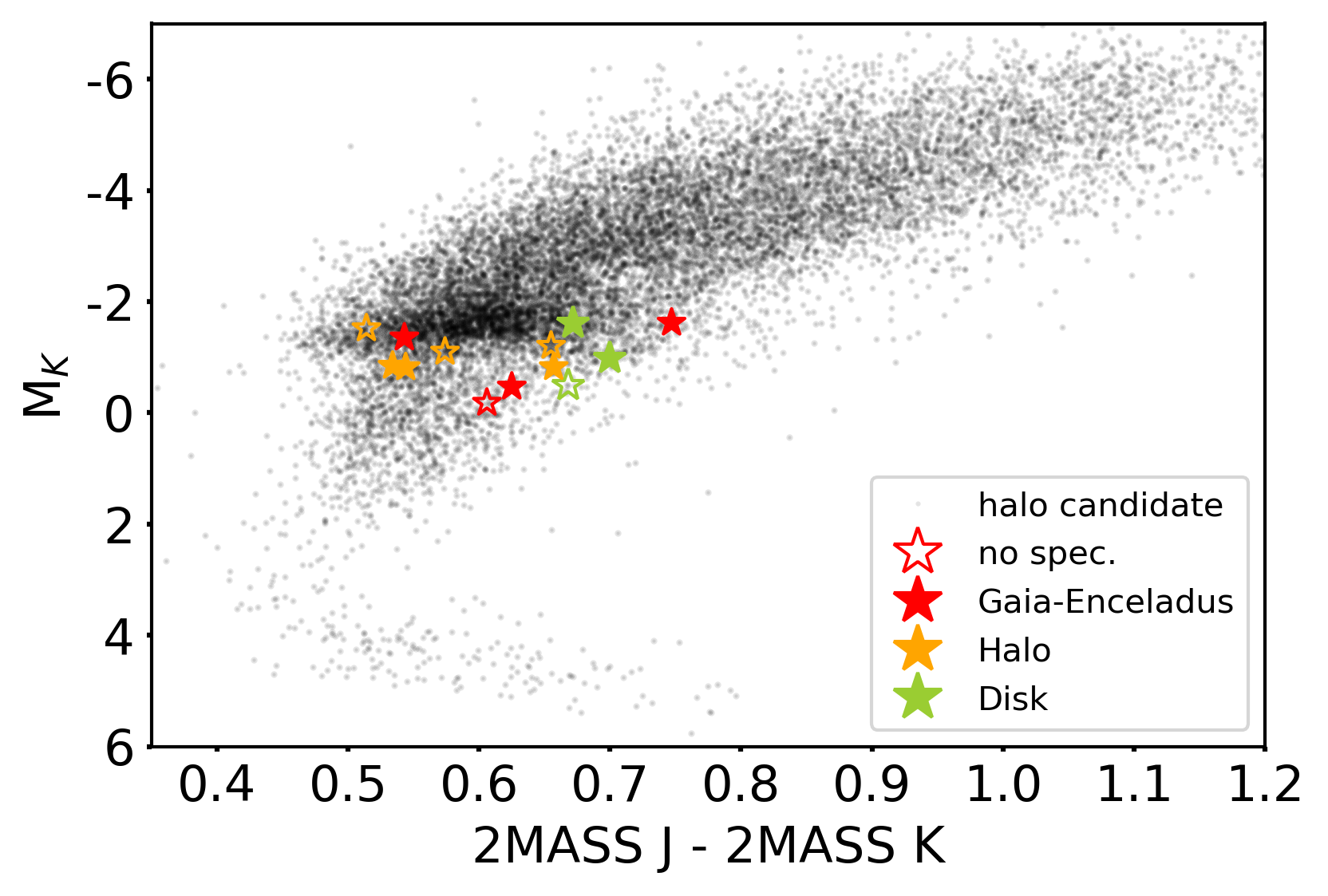}
\vspace{-0.5cm}
\caption{Absolute color-magnitude diagrams of \gaia-selected stars in our sample. Points in black were selected based on pure \gaia\ observables (colors, absolute magnitudes, radial velocities and proper motions). The colored points represent stars for which we detect unambiguous asteroseismic oscillations, and those with filled in colored symbols also have spectroscopic information. The left panel shows \gaia\ $B_{p}-R_{p}$ color vs. absolute G magnitude, and the right panel shows 2MASS J-K color vs. absolute K magnitude. Cuts were made in the \gaia\ photometric plane and are shown as black dotted lines in the left panel. Our asteroseismically constrained targets span the full range of the distribution of giant colors in this regime.}
\label{fig:gaiahrd}
\end{figure*}

To select red giant stars observed by the \tess\ mission, we crossmatch the \tess\ input catalog \citep{stassun2019} to the source catalog in \emph{Gaia} Data Release 2 \citep[DR2;][]{gaia2018,gaia2018b} and make use of \emph{Gaia} absolute magnitudes and colors. In detail, our initial target selection was done by querying the \tess\ Input Catalog v8.01 \citep{stassun2019} though the Mikulski Archive for Space Telescopes (MAST) on November 15th, 2019  using \emph{Gaia} DR2 photometry. We required stars to have apparent magnitude $G < 13$, parallax signal-to-noise $\varpi / \sigma_\varpi > 6$, absolute magnitude $M_{G}< 4.1$ (as estimated using distance moduli computed directly from the catalog parallax values), colors in the range $0.975 < (G_{\rm BP} - G_{\rm RP}) < 3.0$, and measured radial velocities (similar to the sample in \citealt{grunblatt2019}). This selection isolates 1,771,924 probable red giant branch stars in the luminosity range where asteroseismic signals are detectable in \tess, and removes main sequence stars and subgiants.  We illustrate the stars which pass this initial color-magnitude selection as well as our subsequent kinematic selection (detailed below), as well as our final detailed age analysis sample, in color-magnitude space in Figure~\ref{fig:gaiahrd}.





\label{sec:targets}

After our color-magnitude cuts, we then use \gaia\ parallax, proper motion and radial velocity information to remove stars that are clearly rotating with the disk by converting measured quantities to velocities in a Galactocentric reference frame.
In particular, we convert heliocentric \gaia\ astrometry and radial velocity measurements \citep{lindegren2018, gaia2019} into Galactocentric cylindrical velocity components, $(v_R, v_\phi, v_z)$, using the \texttt{astropy} coordinate transformation framework \citep{astropy13,astropy18}.
We use the default \texttt{v4.0} Galactocentric frame parameters,\footnote{This is a right-handed system with the Sun along the $-x$ axis and the solar velocity in the $+y$ direction.} which assume a solar Galactocentric distance of $8.122~{\rm kpc}$ \citep{gravity2018}, total solar velocity of $(-12.9, 245.6, 7.78)~\kms$ \citep{drimmel2018, gravity2018, reid2004}, and height above the midplane of $20.7~\pc$ \citep{bennett2019}.

\begin{figure*}[t]
  \centering
  \includegraphics[width=.49\linewidth]{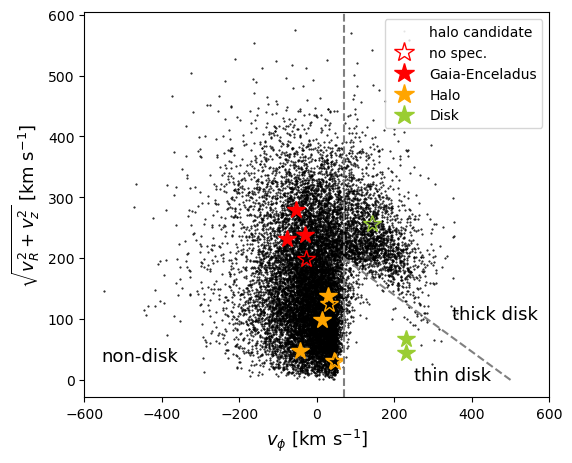}
  \centering
  \includegraphics[width=.49\linewidth]{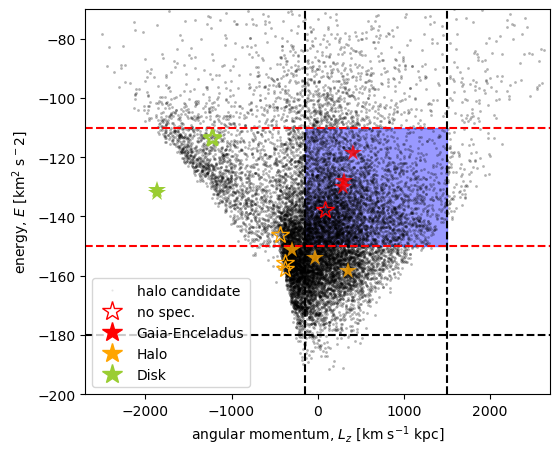}

\caption{{\it Left}: Toomre diagram illustrating the rotational and perpendicular velocity of stars in our initial sample. Our initial sample was selected based on cuts described in \S 2 on \emph{Gaia} parameters to select stars that are not part of the thin disk population (black points). Stars which we have selected for detailed asteroseismic investigation are shown by the colored points, where those stars with spectra available are filled in. {\it Right}: Vertical angular momentum versus orbital energy for all stars in our sample, using the same legend scheme as the figure to the left. We define \GE\ stars as those with kinematics located between the \citet{helmi2018} (dashed black lines) and \citet{koppelman2019} (dashed red lines) definitions of the \GE\ on this plot (shaded region). Stars which do not meet these criteria but have particularly low angular momenta we designate as `halo' stars. Two asteroseismic stars which did not make initial kinematics cuts are added to represent a disk population, as well as one star with low binding energy but high rotational velocity to represent the thick disk. These populations of stars all occupy distinct clusters of kinematic parameter space, as shown in both of these panels.  \label{fig:kinem}}

\end{figure*}

We perform a set of loose cuts to remove stars that are associated with the kinematic thin and thick disk populations by only keeping stars with $v_\phi < 50~\kms$ or $\sqrt{v_R^2 + v_z^2} > 200~\kms$ (see left panel of Figure~\ref{fig:kinem}). For these stars, we then also compute their orbital angular momenta ($z$-components, $L_z$) and energies ($E$, assuming the three-component Milky Way model implemented in \texttt{gala}, \citealt{gala, bovy2015}).
To remove some kinematic thick disk stars that pass our initial velocity selection, we additionally require $L_z > -600~\kms~\kpc$ for our final sample of probable stellar halo giant stars. 
Figure~\ref{fig:kinem} (left panel) shows Galactocentric velocity components for the full sample of TIC giant stars (background density), where colored symbols correspond to our detailed age analysis sample. The wedge-shaped area seemingly missing from the background density corresponds to thin disk velocities excluded form our analysis. The right panel of Figure~\ref{fig:kinem} shows the same stars in the space of energy and $z$ angular momentum (in units of solar angular momentum, $L_{z, \odot} \approx 598.4~\kpc~\kms$, and solar kinetic energy, $E_{k, \odot} \approx 60546~{\rm km}^2~{\rm s}^{-2}$). Using these conventions, prograde orbits are on the left and retrograde orbits are on the right. We highlight our \GE\ kinematic selection as the shaded region in the figure. 

While the majority of giant stars observed by both \tess\ and \gaia\ are associated with the Galactic disk, 13,205 giants remain in our sample that are likely associated with the Galactic stellar halo and its substructures (all background points visible in Figures~\ref{fig:gaiahrd} and \ref{fig:kinem}).
The distribution of these \tess\--\gaia\ halo giants appears to be reasonably sampling the local stellar halo, with a mean rotation near zero and a wide dispersion in rotational velocity. The \GE, initially identified as a ``blob'' in \citet{koppelman2018}, can be seen as an overdensity near the center of the right panel of Figure \ref{fig:kinem}. We also indicate the \GE\ parameter space as defined by \citet{helmi2018} in angular momentum and \citep{koppelman2019} in binding energy with dotted lines, choosing a more conservative overlap region of the parameter spaces defined by the above studies to define the \GE\ in our sample. Stars on disk orbits would occupy the lower envelope of the prograde distribution, but have been effectively removed by our earlier velocity selection (aside from the two disk stars we have intentionally re-included and a sub-population of thick disk stars with particularly low binding energy).
In both panels, the 13 large star markers indicate the sub-sample of stars for which we have reliable asteroseismology, and spectroscopy in some cases (as indicated in the legend).
Included in this sub-sample are two thin-disk, red giant stars that we have added back into our sample which also have spectral information (i.e., that fail our halo selections) as comparison stars. These stars, in conjunction with one additional star in our sample that is kinematically consistent with the thick disk, will be used as comparative benchmarks for the rest of the stars in our sample.




\section{Determination of Stellar Properties}

\subsection{Kinematic Structure Associations}

Kinematic information is essential to understanding the structure and formation of our Galaxy. Kinematics can be used to distinguish membership in the major components of the Milky Way, such as the thin/thick disk, bar/bulge, and halo, and to identify smaller substructures within the Galaxy, which often provide crucial information about its formation and interaction history. Here, we investigate whether our halo star sample is consistent into any of those substructures, and how that association may affect the interpretation of the ages of the stars studied here.

We compare the kinematics of our stellar sample to kinematics of stellar substructures identified in \gaia\ DR2 in Figure~\ref{fig:kinem}. In the left panel, we illustrate the rotational and perpendicular velocities of the stars in our sample relative to the Galactic disk. Stars selected on only \emph{Gaia} observables are shown in black. Stars selected for detailed asteroseismic age analysis are shown by colored markers. \citet{helmi2018} define the \GE\ in kinematics space as $-$150 km s$^{-1}$ kpc $<$ $L_z$ $<$ 1500 km s$^{-1}$ kpc and E $>$ $-$1.8$\times$10$^5$ km$^2$/s$^2$, corresponding to the black dashed lines in the right panel of Figure~\ref{fig:kinem}. \citet{koppelman2019} take a stricter definition of the \GE\, requiring a lower binding energy of $-$1.1$\times$10$^5$ km$^2$/s$^2$ $>$ E $>$ $-$1.5$\times$10$^5$ km$^2$/s$^2$ (red lines in right panel). We adopt the most stringent combination of these definitions to distinguish the different regimes in our sample. 

After using the asteroseismic constraints (determined in the following section) to vet our sample further, we returned to kinematic information to classify this detailed age analysis sample by distinct galactic substructure membership. Using the angular momentum definition of \citet{helmi2018} and the binding energy definition of \citet{koppelman2019}, we identify the 4 stars in our age analysis sample which are most likely to be members of the \GE\ substructure in red, whereas stars with similarly low angular momenta but higher binding energies are classified as `halo' stars in orange. We note that two of the six orange `halo' stars we have chosen for our analysis overlap with the \citet{helmi2018} definition of the \GE\, but not the \citet{koppelman2019} definition, highlighting the need for additional information beyond kinematics to distinguish between stellar origins. Finally, we highlight two thin disk stars and one thick disk star with angular momenta and rotational velocities consistent with the galactic disk in green. We do not identify significant overlap between our designated \GE\ or `halo' stars and the Thamnos, Sequoia, or Helmi streams identified in the inner halo and/or thick disk \citep{helmi1999, myeong2019, mackereth2019, koppelman2019}.

\subsection{Chemical Analysis}

The six stars in our sample that are also found in the GALAH and APOGEE catalogues are listed in the \tess\ Input Catalog (TIC) as TIC20897763, TIC453888381, TIC341816936, TIC393961551, TIC279510617, and TIC300938910. Effective temperatures, metallicities, and $\alpha$-element abundances derived by these survey projects are used to determine stellar masses and radii in \S\ref{sec:seis}, and stellar ages in \S\ref{sec:ages}. 

These survey projects use different target selection and instrumentation to study complementary samples of stars in the Milky Way. GALAH \citep{desilva2015} is acquiring R~$\approx$~28,000 optical spectra for stars with apparent magnitudes in the range $12 < V < 14$ and Galactic latitude of at least 10$^\circ$. The GALAH DR2 catalog \citep{Buder2018} contains radial velocities, stellar parameters, and abundances of up to 23 elements for 342,682 stars, and the DR3 catalog with 30 elements in 567,115 stars has recently been made publicly available as well \citep{Buder2020}. APOGEE and APOGEE-2 \citep{Majewski2017} collected R~$\approx$~22,500 infrared spectra (1.51-1.70 $\mu$m) for stars with apparent magnitudes $H<12.2$ and absolute colors $(J-K_{s})_{0}\ge0.5$ across a grid of sightlines through the Galaxy. Working in the infrared allows APOGEE to avoid much of the extinction that limits optical surveys, giving it a unique grasp on stars in the Galactic disk and bulge. The SDSS-IV DR16 catalog \citep{Ahumada2019} provides radial velocities, stellar parameters, and abundances of up to 24 elements for 430,000 APOGEE stars \citep{jonsson2020}. 

\begin{figure}
    \centering
    \includegraphics[width=\linewidth]{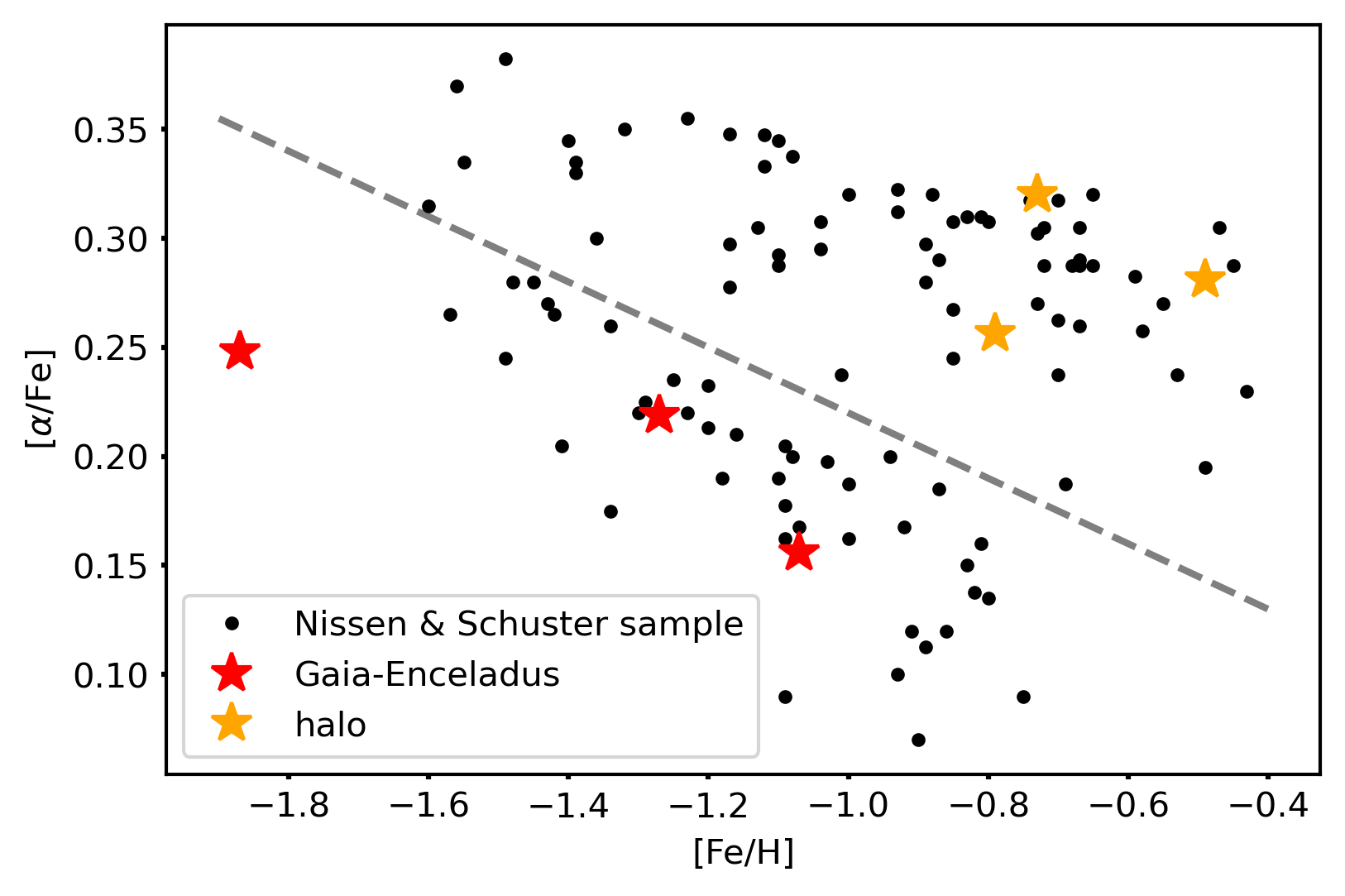}
    \caption{Metallicity vs. $\alpha$-element abundance for all stars with spectra in our sample {\bf (red and orange symbols)}, overplotted with literature data. Of the six stars for which we have spectra, three are clearly part of the high-$\alpha$ population, (above gray line) two are clearly part of the low-$\alpha$ population (below line), and the most metal-poor star (TIC341816936) is consistent with the low-alpha population but outside the metallicity range of \citet{nissen2010}. \label{fig:alphamet}}
\end{figure}

Elemental abundances are important as a criterion for identifying stars accreted from dwarf galaxies. The star formation and feedback that drive galactic chemical evolution proceed differently in dwarf galaxies than in more massive galaxies. Lower mass galaxies cannot sustain steady star formation, and so it proceeds in bursts. This shortens the early enrichment era that is dominated by Type~II supernovae and allows Type~Ia supernovae to become important contributors earlier. SN~II produce $\alpha$-elements (e.g., Mg, Si, S, Ca, Ti) as well as iron peak elements, while SN~Ia produce iron peak elements without $\alpha$-elements. As a result, dwarf galaxy stars tend to have lower $\alpha$-element abundances at a given metallicity than their massive-galaxy counterparts. 

A bifurcation in the [$\alpha$/Fe] vs [Fe/H] distribution has been observed in Galactic halo stars, correlated with orbital kinematics \citep[e.g.,][]{nissen2010, matsuno2020}, and is interpreted as an artifact of the two possible origins of halo stars---those formed \textit{in situ} and those accreted from dwarf galaxies in the past. In Figure~\ref{fig:alphamet} we illustrate the metallicities and alpha abundances for the stars in our sample, relative to the high-precision study of \citet{nissen2010}. Three of the stars for which we have spectroscopic data have relatively high $\alpha$ abundances, chemically consistent with the population of stars formed in our own Galaxy, while two stars---TIC20897763 and TIC393961551---have lower $\alpha$-element abundances, more consistent with the accreted population chemistry. The case of the star with the lowest metallicity (TIC341816936) is more ambiguous, since it falls below the line but is outside the metallicity range of the \citet{nissen2010} data. More recent studies have indicated that the low-$\alpha$ element halo population extends to the metallicity of TIC341816936, and provide additional evidence the star is chemically distinct from other known stellar streams \citep[e.g., the Sagittarius stream,][]{hayes2018}, but studies of the metallicity distribution function indicate that TIC341816936 may be atypically metal-poor for the \GE, while TIC20897763 and TIC393961551 are in strong agreement with the mean metallicity of the \GE\ \citep{feuillet2020}. As a result, TIC20897763 and TIC393961551 are the most likely \GE\ dwarf galaxy remnants in our spectroscopic sample, and they should provide the clearest constraints on the age of the accreted \GE\ stars in the Galactic halo.
 
 
For those stars in our sample which have spectra, we adopt the spectroscopic catalogue values of effective temperature for those stars for mass and radius determination. For the stars without spectroscopic effective temperatures, we used effective temperatures calculated with \texttt{isoclassify} based on 2MASS $J$ and $K$ photometry \citep{Skrutskie2006}, \emph{Gaia} DR2 parallax, and the reddening map of \citet{green2018}, using the $J-K$ color-temperature relation of \citet{gonzalez2009} (see \citealt{huber2017} for additional details). We also use this method to measure an effective temperature independently of any spectroscopically-defined temperature. We find that for the subset of stars with spectra, all spectroscopically-determined temperatures agree with temperatures determined with \texttt{isoclassify} within stated uncertainties. For those stars with APOGEE spectroscopic temperatures, uncertainties were adopted to be $30$~K (M.~H. Pinsonneault, pers. comm.; Pinsonneault et al., in prep.). Systematic uncertainties in the temperature scale may be possible at the level of $20$ K \citep{zinnrad}, which reflects the fundamental uncertainty in the infrared flux method temperature scale to which APOGEE temperatures are tied \citep{holtzman+2015}.

\subsection{Asteroseismology}
\label{sec:seis}


Asteroseismology is the study of relating observed oscillations to the physical properties of a star \citep{christensen1983}. These oscillations can be most easily identified in the power spectra of stellar light curves. Numerous analysis packages have been developed to derive precise stellar properties from asteroseismic oscillation signals, by analysis of power spectra of oscillating stars \citep{huber09,mosser2011,garcia2014,hon2018,zinn2019}. In order to precisely determine the stellar radii and masses of all the stars in our sample, we produce power density spectra of all of our targets from light curves which we generated using our \texttt{giants} pipeline (Saunders et al., {\it in prep.}). 

\subsubsection{Data Preparation}

\begin{figure*}[t]
\centering
  \centering
  \includegraphics[width=.49\linewidth]{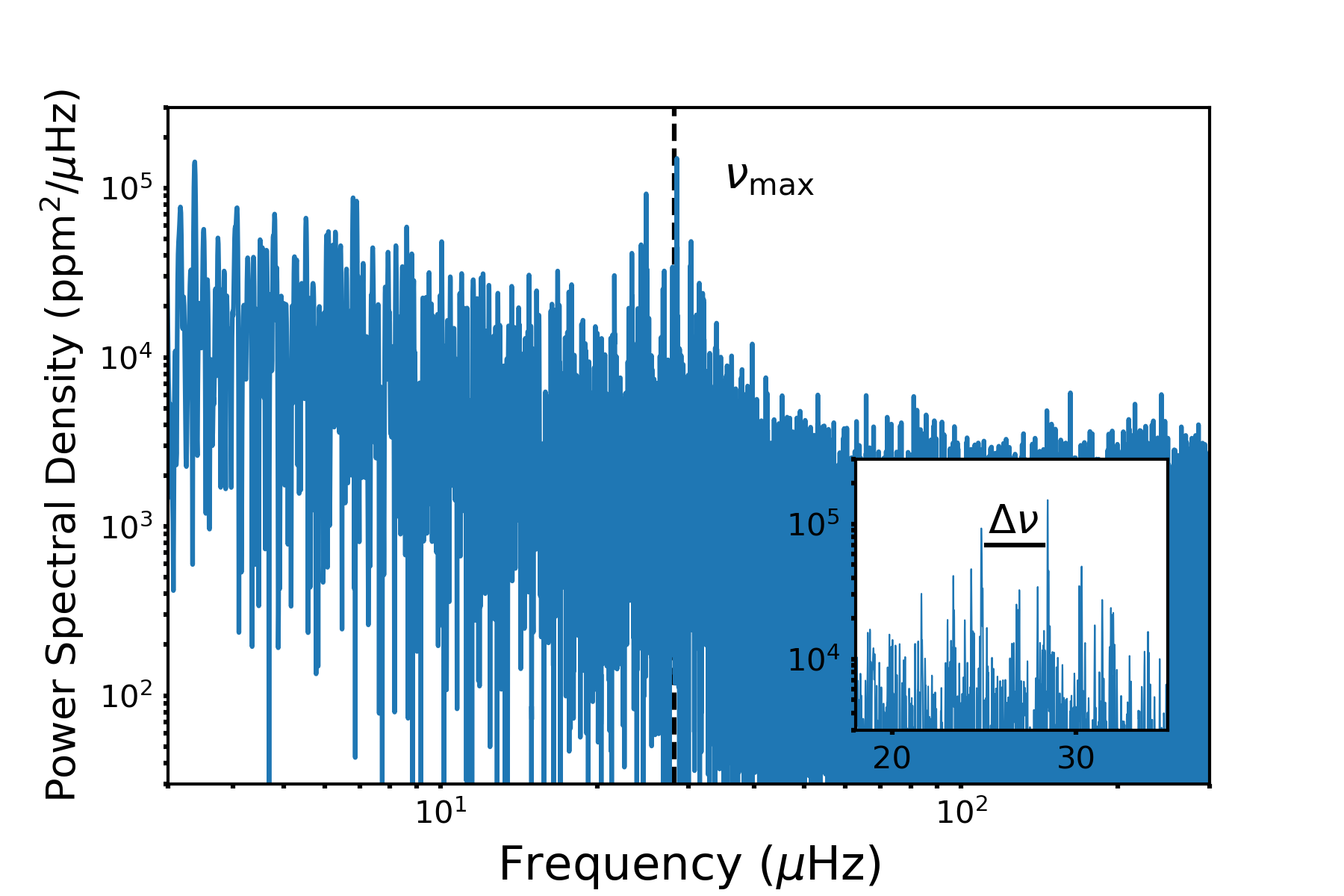}
  \centering
  \includegraphics[width=.49\linewidth]{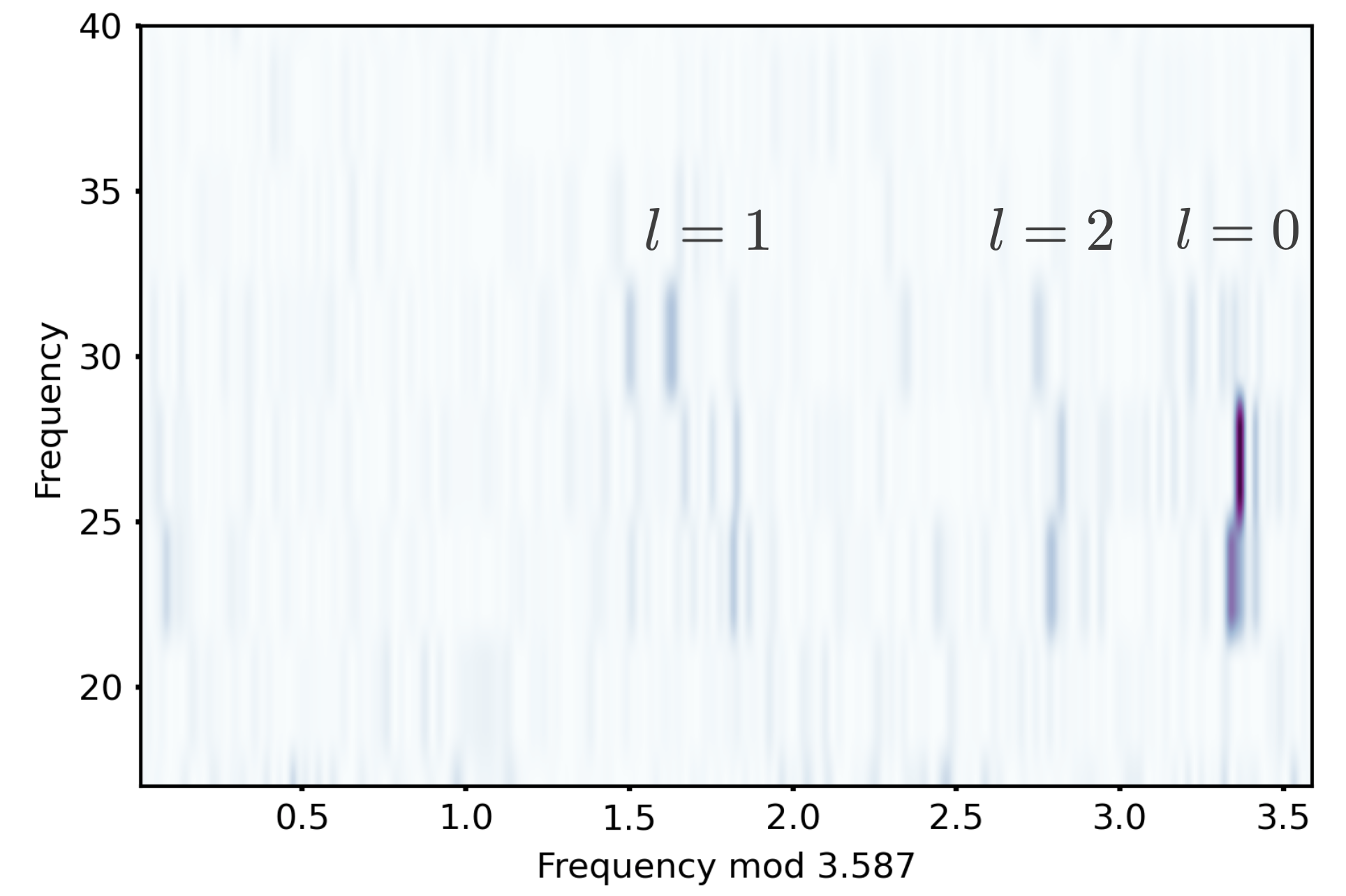}
\caption{{\it Left:} Power spectral density of TIC279510617, one of the stars in our sample. Oscillations centered at 28 $\mu$Hz can clearly be seen and are highlighted in the inset plot. $\nu_\mathrm{max}$ and $\Delta\nu$ have been labeled. {\it Right:} Echelle diagram of TIC279510617. Vertical ridges corresponding to the $l$ = 0, 1, and 2 degrees of oscillation are visible vertically and have been labeled accordingly.}
\label{fig:psd_echelle}
\end{figure*}


As the \tess\ Mission only recently produced light curves for full-frame image targets for the community \citep{huang2020}, our analysis optimizes and combines alternative preexisting tools to extract and detrend light curves from full frame image cut-outs.

Once \GE\ members had been selected, we used the \texttt{TESScut} software package to produce light curves \citep{lightkurve,brasseur2019}. To clean the light curves more thoroughly, we identified and removed data points that were greater than or less than the median flux by at least 6-$\sigma$, where $\sigma$ is the standard deviation of the flux in the light curve. Additionally, we applied a Gaussian filter to smooth trends on timescales greater than 3 days. Light curves were then normalized, further smoothed and processed to remove outliers using Pixel Level Decorrelation (PLD) techniques built into the \texttt{giants} package developed by our team, now embedded in \texttt{lightkurve} \citep[][Saunders, et al {\it in prep.}]{lightkurve}. This method expands on previous PLD detrending methods used for other telescopes \citep{deming2015,luger2016} and is described in more detail in Saunders et al. ({\it in prep.}). We produced light curves for over 9,000 stars in this sample using this technique. 

Known \tess\ lightcurve features, such as those produced by momentum dumps every 2 days that keep the spacecraft pointing accurate, or gaps in coverage every $\sim$2 weeks to downlink data to Earth, can mimic or modify asteroseismic signals. In addition, astrophysical false positive signals can also be produced by eclipsing binary systems or classical pulsators such as RR Lyrae variables. To exclude these unwanted signals from our analysis, in addition to the initial detrending done by the \texttt{giants} algorithm, we also exclude data within 1 day of any gap in data acquisition within a campaign, as well as within 1 day of the start and end of each campaign to remove spurious signals near stellar oscillation or transit timescales. 

Figure~\ref{fig:psd_echelle} illustrates an example star showing oscillations in our study. TIC279510617 is a star in our sample in the southern Continuous Viewing Zone (CVZ) of \tess\, and has been observed as part of the GALAH survey \citep{Buder2018}. The left panel of Figure~\ref{fig:psd_echelle} shows the power spectral density of the \tess\ lightcurve of this star. The stellar oscillations appear as a cluster of peaks around 28 $\mu$Hz, highlighted in the inset plot. $\nu_\mathrm{max}$ has been labeled in the main plot, and $\Delta\nu$ has been highlighted in the inset. On the right, we have folded the power spectral density at the measured $\Delta\nu$ value to reveal ridges corresponding to the spherical degrees of oscillation at multiple orders as labeled, resulting in the pattern seen here.

\subsubsection{Asteroseismic Measurement}
After producing light curves with \texttt{TESScut} and \texttt{lightkurve} tools, we then perform an asteroseismic analysis on all power spectra that pass the filters above, calculating the best-fit frequency of maximum power ($\nu_\mathrm{max}$) and regular frequency spacing ($\Delta\nu$) between sequential radial oscillation modes using the \citet{huber2009} SYD pipeline, which has been well established for the asteroseismic analysis of \emph{Kepler} and \emph{K2} photometry \citep{huber2011, huber2013,stello2017, grunblatt2019}. We calculate uncertainties for our asteroseismic quantities using a Monte Carlo method, producing 500 realizations of each asteroseismic fit and using the standard deviation of the sample of asteroseismic fits for each star to determine parameter uncertainties as described in \citet{huber2011}. These $\nu_\mathrm{max}$ and $\Delta\nu$ uncertainties are propagated into the uncertainties on stellar mass and radius. We report our measured $\nu_\mathrm{max}$ and $\Delta\nu$ best-fit values and uncertainties in Table \ref{table1}.


\begin{figure*}[t]
\centering
  \centering
  \includegraphics[width=.49\linewidth]{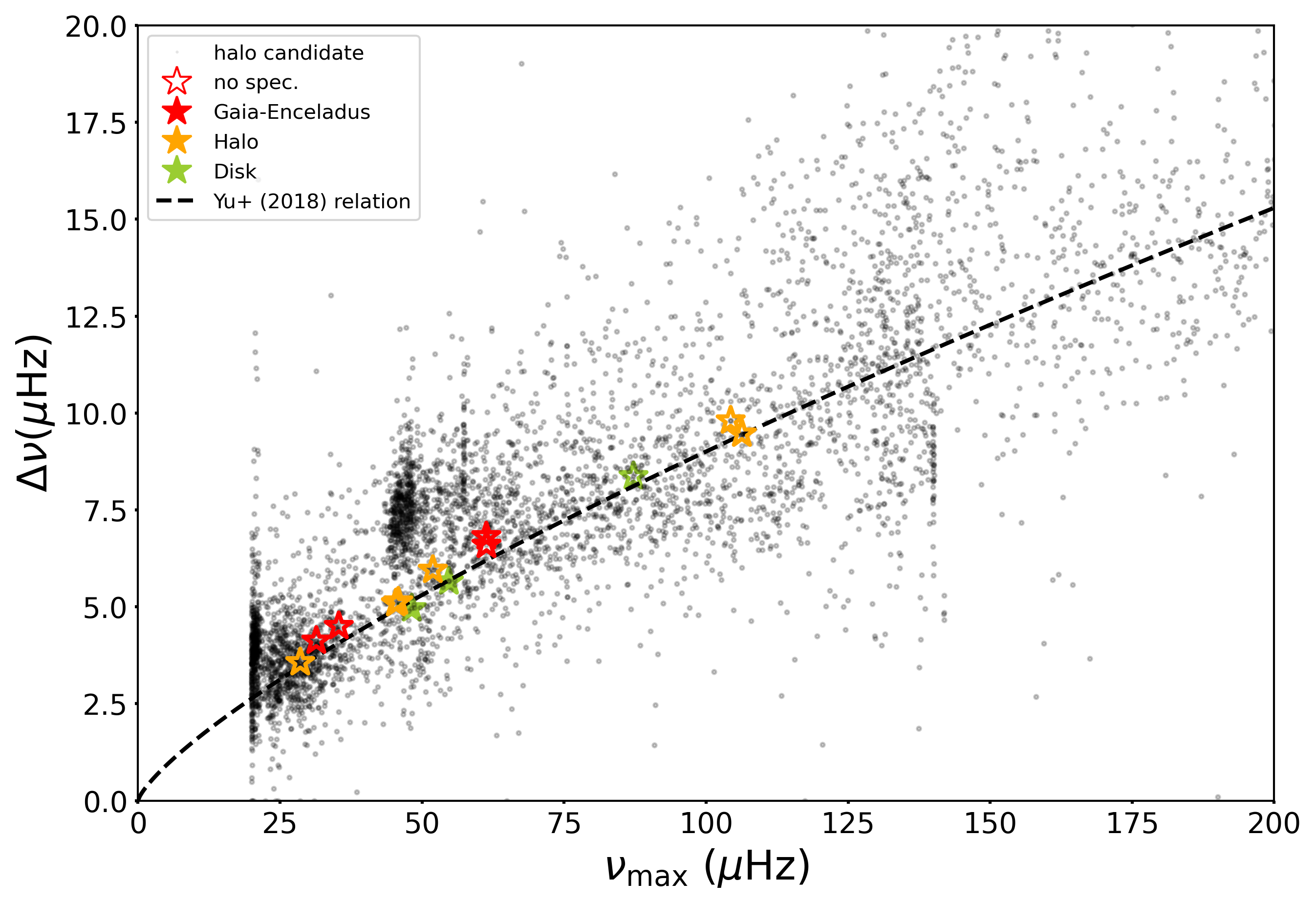}
  \centering
  \includegraphics[width=.49\linewidth]{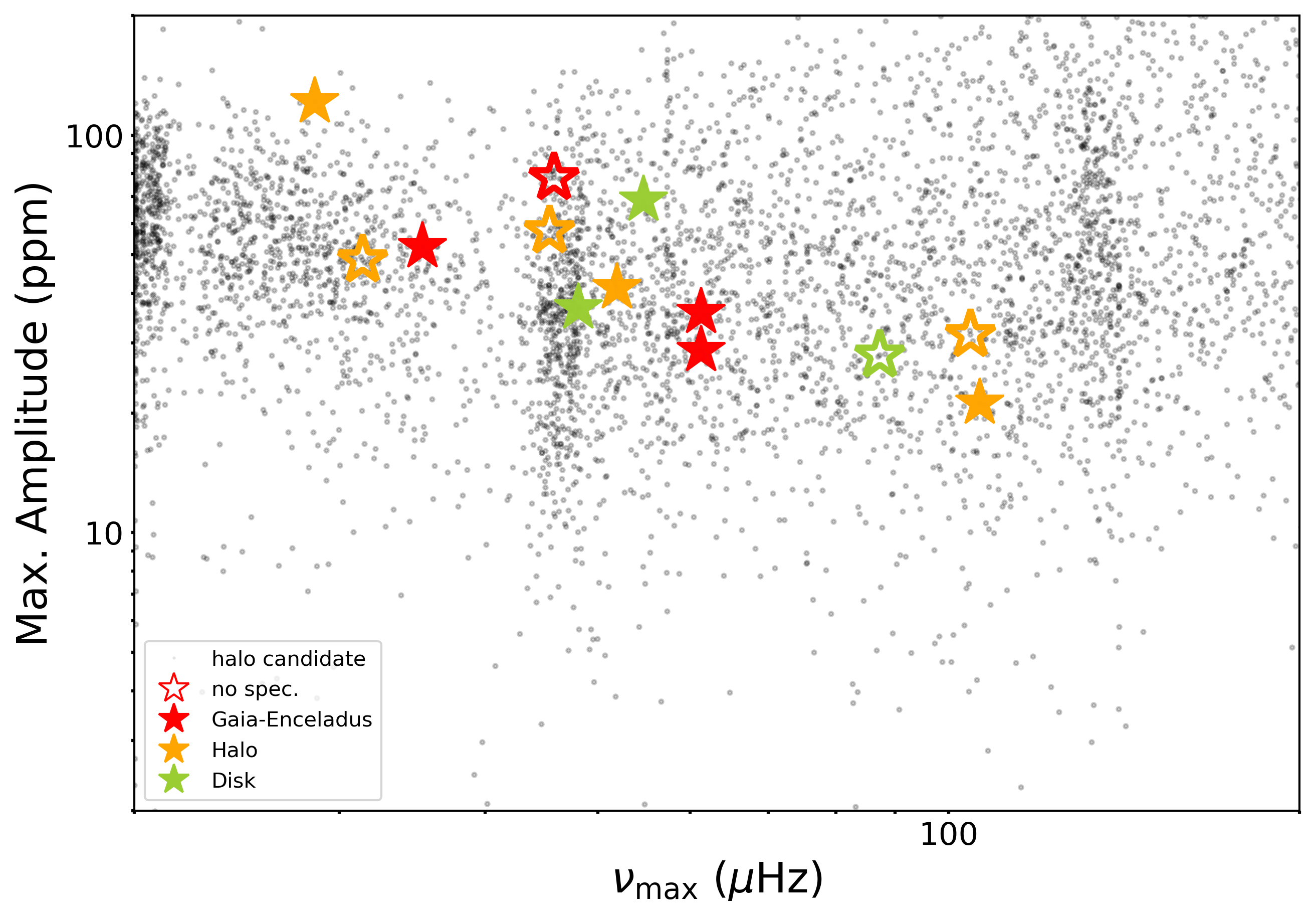}
\caption{{\it Left:} $\nu_{\max}$-$\Delta\nu$ relation for all stars in our sample. All stars selected in \emph{Gaia} observables are shown in black. Stars which passed both our initial automatic asteroseismic vetting with \texttt{SLOSH} as well as our more stringent visual inspection are shown as red, orange and green stars, where filled in symbols indicate spectral information is also available. {\it Right:} $\nu_{\mathrm{max}}$ vs. maximum oscillation amplitude for the stars in our sample, with the same color coding as the left panel. Though the stars in our sample follow the expected relative trend (amplitudes appear larger in stars with lower $\nu_{\mathrm{max}}$), the large underlying scatter of amplitude measurements highlights the difficulty of precision asteroseismology with current \tess\ datasets. }
\label{fig:numaxdnu}
\end{figure*}


As the noise properties of \tess\ are currently poorly understood, we attempt to use only the highest quality asteroseismic detections for our analysis. To do so, we use the automated asteroseismic detection pipeline \texttt{SLOSH}, developed for the analysis of \emph{Kepler} and \tess\ datasets \citep{hon2018, hon2018a}. This pipeline returns a score between 0 and 1 indicating the likelihood that oscillations are visible within a power spectrum, with 1 being extremely likely and 0 being a non-detection. To choose only stars with clearly visible oscillations, we select those with a score exceeding 0.975.




However, it is possible that additional noise may masquerade as an asteroseismic signal and thus pass our automatic vetting process. Thus, we perform a detailed visual inspection of all stars which pass our automatic vetting, and select only 11 stars with the clearest and least ambiguous asteroseismic signals in our final analysis. We also identify two additional stars with disk-like kinematics and similarly high-quality asteroseismic signals in light curves produced using \texttt{TESScut} and \texttt{giants}, which also pass our automatic vetting and lie in a similar region of asteroseismic parameter space as our kinematically selected halo stars. We confirm that these stars have true oscillations by checking them against the empirical $\nu_\mathrm{max}$--$\Delta\nu$ relation determined by \citet{yu2018} for all \emph{Kepler} stars. We find that none of the stars in our final sample deviate from this relation by more than 20\%. 

We illustrate our automatically and visually vetted samples relative to the entire target population in Figure~\ref{fig:numaxdnu}. The right panel illustrates that our sample of stars appears to form the upper envelope of the amplitude distribution of stellar oscillations detected. This is likely due to biases of our visual inspection technique. We note that future studies of much larger samples of \tess\ stars should enlarge and improve training sets for asteroseismic signal identification in similar studies, and hope that future characterization of systematic noise in \tess\ will make extension of this analysis through automated asteroseismology with \tess\ light curves more reliable \citep[][Stello, {\it in prep.}]{mackereth2021}.

We use three independent asteroseismic pipelines to calculate asteroseismic parameters. Our primary analysis uses the SYD pipeline \citep{huber09}, which searches for $\numax$ by smoothing the power spectrum and detecting a region of power excess. The power excess is modelled using a Gaussian, the center of which is taken to be $\numax$. The large frequency separation $\dnu$ is then fitted using an autocorrelation in the vicinity of $\numax$. We check the SYD results against two other pipelines: BAM and COR. A full description of the BAM methodology is found in \cite{zinn+2019bam}. In short, BAM takes advantage of empirical, intrinsic correlations between the amplitude/time-scale of the red noise in a solar-like oscillator power spectrum and $\numax$. Upon fitting $\numax$, $\dnu$ is computed via an autocorrelation of the power spectrum in the vicinity of $\numax$. BAM then uses Bayesian evidence to provide an indication of whether or not the fitted oscillations are significant. 
COR \citep{mosser_appourchaux2009} is a complementary pipeline in the sense that, unlike BAM, it operates by first searching for $\dnu$ instead of $\numax$. This is done using a Fourier spectrum of the filtered Fourier spectrum of the light curve. A null hypothesis test is performed to evaluate the significance of the $\dnu$ detection above the noise level. In the case of a $\dnu$ detection, the power excess is characterized by a Gaussian centered around $\numax$, the significance of which is assessed using an amplitude-to-background ratio. We find that asteroseismic determinations of $\Delta\nu$ for all of the above pipelines agree within estimated uncertainties. 


Furthermore, we use COR to determine whether the stars in our sample are ascending the red giant branch for the first time, or have begun burning helium in their cores \citep{white11, mosser2011}. To determine evolutionary states, COR fits a theoretically-motivated template of modes to the power spectrum to yield a best-fitting period spacing. The period spacing describes the separation between so-called `mixed modes', which are due to coupling between p-modes and g-modes, and has been demonstrated to be an effective discriminator of evolutionary state, given its sensitivity to the conditions in the core \citep{bedding+2011}. We list the asteroseismically-determined evolutionary states in Table~\ref{table2}. For those stars which have APOGEE data available, we can also appeal to spectroscopically determined evolutionary states based on the method described in \cite{elsworth+2019} and modified in \cite{warfield+2021}. The spectroscopic evolutionary states agree with those from asteroseismology where there is overlap.

\subsubsection{Stellar Parameter Estimation}

To estimate stellar masses and radii from the measured $\nu_\mathrm{max}$ and $\Delta\nu$ values that passed our asteroseismic vetting (Figure~\ref{fig:numaxdnu}, red points), we use the asteroseismic scaling relations of  \citet{brown1991} and \citet{kjeldsen1995}:

\begin{equation}
\frac{\Delta \nu}{\Delta \nu_{\odot}} \approx f_{\Delta \nu} \left(\frac{\rho}{\rho_{\odot}}\right)^{0.5} \: ,
\end{equation}

\begin{equation}
\frac{\nu_{\rm max}}{\nu_{\rm max, \odot}} \approx \frac{g}{{\rm g}_{\odot}} \left(\frac{T_{\rm eff}}{T_{\rm eff, \odot}}\right)^{-0.5} \: .
\end{equation}

\noindent
where $f_{\Delta\nu}$ is the correction factor suggested by \citet{sharma16} to account for known deviations from the previously established asteroseismic scaling relation, and which is calculated using \texttt{asfgrid} \citep{asfgrid}.\footnote{\url{http://www.physics.usyd.edu.au/k2gap/Asfgrid/}} Equations (2) and (3) can be rearranged to solve for mass and radius \citep{2008ApJ...674L..53S, 2010A&A...509A..77K}:

\begin{equation}
\frac{M}{\rm {\rm M}_\odot}   \approx   \left(\frac{\nu_{\rm
max}}{\nu_{\rm max,
\odot}}\right)^{3}\left(\frac{\Delta \nu}{f_{\Delta \nu}
\Delta \nu_{\odot}}\right)^{-4}\left(\frac{T_{\rm eff}}{T_{\rm
eff, \odot}}\right)^{1.5}   
\end{equation}

\begin{equation}
\frac{R}{\rm R_\odot}   \approx  \left(\frac{\nu_{\rm
max}}{\nu_{\rm max, \odot}}\right)\left(\frac{\Delta
\nu}{f_{\Delta \nu} \Delta \nu_{\odot}}\right)^{-2}\left(\frac{T_{\rm
eff}}{T_{\rm eff, \odot}}\right)^{0.5}.
\end{equation}

Our adopted solar reference values are $\nu_{\rm max, \odot}=3090\,\mu$Hz, $\Delta \nu_{\odot}=135.1\,\mu$Hz, and $T_{\rm eff, \odot}=5777\,$K \citep{huber2011}. As our stars have effective temperatures between 4500 and 5500 K, typical asteroseismic correction factor $f_{\Delta\nu}$ values for all of the stars in our analysis are between 0.98 and 1.02 \citep{sharma16}. An intrinsic scatter of $\sim$1.7\% in mass and $\sim$0.4\% in radius has been found for this relation in \emph{Kepler} red giants \citep{li2021}, placing a fundamental limit on our parameter determination.

We also note that the \citet{sharma16} formulation of $f_{\Delta\nu}$ is metallicity-dependent. Because this scaling relation approach to determining stellar properties assumes Solar abundances, we confirmed that the variation in bulk metallicity due to non-Solar alpha abundances (calculated following \citealt{salaris+1993a}) does not significantly affect our mass estimates. Furthermore, we note that only one of the age estimation techniques discussed in the following section, \texttt{isoclassify}  \citep{huber2017}, explicitly allows for the inclusion of $f_{\Delta\nu}$ when determining stellar ages. In addition, the determination of $f_{\Delta\nu}$ is model dependent, which can affect our final age determination. However, \citet{pinsonneault+2018} demonstrate that the effects of using different models results in $<$1\% deviations in $\Delta\nu$, and thus is not the dominant source of uncertainty in our age determination.





Furthermore, though there is a consensus that theoretically motivated $\Delta \nu$ corrections, $f_{\Delta \nu}$, improve the accuracy of asteroseismic scaling relations, such corrections are not currently possible to compute from theory for $\nu_{\rm{ max}}$ \citep[][]{belkacem2011,sharma16,huber2017}. While such corrections cannot yet be computed in detail using frequency modeling, \cite{viani2017} uses the logic implied in the scaling relations that the $\nu_{\rm{max}}$ scaling relation requires percent-level, metallicity-dependent corrections. Therefore, it is possible that biases exist in the scaling relations due to mismatches between the empirically predicted and measured $\nu_{\rm {max}}$. \cite{pinsonneault+2018} and \cite{zinnrad} established that SYD $\nu_{\rm {max}}$ measurements with \textit{Kepler} data agree with absolute scales calibrated to open cluster dynamical masses and \emph{Gaia} radii. The latter study quantified the accuracy of $\nu_{\rm{max}}$ for red giant branch (RGB) stars to be within $2\%$, which naively implies a scaling relation mass accuracy of $6\%$.\footnote{\cite{zinnrad} found larger biases for stars more evolved than those we consider here ($R \gtrsim 30\,\text{R}_{\odot}$).} We take this as a systematic uncertainty in our ages of $\sim 20\%$, which we include in our inferred \GE\ age in \S\ref{sec:ages}. However, indications from \textit{K2} suggest that there could be $\sim 1\%$ systematic biases in $\nu_{\rm{max}}$ when working with short ($\sim 90$d) time series compared to years-long time series (e.g., from \textit{Kepler}; Zinn et al., {\it in prep.}). Presumably, this bias exists in $\sim 30$d \tess\ light curves, and we therefore take extra precautions to minimize $\nu_{\rm{max}}$ biases on final age estimates.

As mentioned earlier, the above asteroseismic scaling relations have been shown to have systematic issues for stars with chemical abundance patterns significantly different from the Sun. Specifically, these relations over-estimate masses for metal-poor stars \citep{epstein2014}. Given that our sample of stars is measured and/or expected to be metal-poor, this bias would tend to result in over-estimated masses and under-estimated ages. The magnitude of this effect is uncertain, although mass over-estimation by 10\% is consistent with indications from \cite{epstein2014} and \cite{zinnrad}. This effect may introduce a $\lesssim +30\%$ age systematic uncertainty, consistent with our systematic uncertainties we estimate in \S\ref{sec:ages}. To minimize the effects of such systematic uncertainties, we validate our mass determinations as detailed in the following Section.

\subsubsection{Validation}

In order to ensure the robustness of asteroseismic mass estimates, we therefore also determine stellar masses according to 
\begin{equation}
M_{Gaia} = \left(\frac{\Delta \nu}{f_{\Delta \nu} \Delta \nu_{\odot}}\right)^2 \left(\frac{R_{Gaia}}{R_{\odot}}\right)^3,
\end{equation}
where the \textit{Gaia} radius is computed according to the method from \cite{zinn+2017}. In short, the Stefan-Boltzmann equation is used to calculate the radius given a spectroscopic temperature and a luminosity from the \emph{Gaia} parallax. The luminosity is computed using a $K$-band bolometric correction, taking into account dust extinction using the \cite{green2015} dust map. We assume a \emph{Gaia} parallax zero-point of $-50\mu as$, consistent with that from \cite{zinnzp}, although the zero-point certainly varies as a function of position on the sky \citep[e.g.,][]{lindegren2018,khan+2019}. We require spectroscopic information, as well as SDSS photometry for this exercise, which limits the check to TIC341816936, TIC453888381, TIC393961551, and TIC20897763.

The advantage of this check is that the \emph{Gaia} mass is independent of $\nu_\mathrm{max}$ and thus unaffected by any percent-level potential corrections to the $\nu_{\rm max}$ scaling relation suggested in \cite{zinnrad}, or metallicity-dependent corrections proposed by \cite{viani2017}. We find a variance-weighted mean offset between these masses and our asteroseismically determined masses of $<M_{Gaia}/M_{\rm{seis}}> = 0.93 \pm 0.09$\%. This agreement suggests that our asteroseismic masses are not significantly biased by more than $10\%$, even when using $\nu_{\rm{max}}$. Considering alpha-corrected $f_{\Delta \nu}$ results in a statistically equivalent agreement. That the asteroseismic masses seem over-estimated compared to \textit{Gaia} masses is consistent with expectations that our low-metallicity stars may have biased asteroseismic masses (all but one in this test has [Fe/H] $< -1$). We take this potential bias into account explicitly in our systematic age uncertainty in \S\ref{sec:ages}. We list our determined $\nu_\mathrm{max}$ and $\Delta\nu$ values and uncertainties as well as observable properties and effective temperatures calculated with \texttt{isoclassify} in Table~\ref{table1}. We list our determined masses and radii with uncertainties in Table \ref{table2}. 

We note that while we use both $\nu_\mathrm{max}$ and $\Delta\nu$ determined here to calculate stellar masses and radii, we only consider $\Delta\nu$ when calculating stellar ages to avoid unknown systematic uncertainties in determining $\nu_\mathrm{max}$ from \tess\ observations. We elaborate on this choice in the following Section. When calculating stellar ages, we use only observable quantities such as $\Delta\nu$, and do not use the masses and radii derived from asteroseismology directly in order to avoid additional model-dependent age biases.

\begin{deluxetable*}{ccccc}

\tablecolumns{5} 
\tabletypesize{\scriptsize}
\tablecaption{Directly Determined Stellar Parameters \label{table1}}
\tablewidth{0pt}
\tablehead{
\colhead{TIC ID} & \colhead{\emph{Gaia} ID} & \colhead{\emph{Gaia} mag} &  \colhead{Distance (pc)} &   \colhead{$\nu_\mathrm{max}$ ($\mu$Hz)} \\ \colhead{$\Delta\nu$ ($\mu$Hz)} & \colhead{T$_\mathrm{eff}$ (K)} & \colhead{[Fe/H]} & \colhead{[$\alpha$/Fe] \tablenotemark{*}} & \colhead{Galactic Substructure} \\ \colhead{Spectral Source}}
\startdata
TIC20897763 & Gaia 2365649471033828096 & 9.41484  & 457.879 $\pm$ 9.435 &   61.31383   $\pm$  1.21768 \\ 6.81739   $\pm$  0.24990  &4988$\pm$127 & -1.274 $\pm$ 0.019 & 0.219 $\pm$ 0.021 & \GE\ \\ APOGEE  \\
TIC341816936 & Gaia 1421776046335723008 & 11.623  & 1547.85 $\pm$ 45.63 &  36.34  $\pm$ 0.76 \\ 4.2969 $\pm$ 0.0715 & 5068 $\pm$100 & -1.873 $\pm$ 0.107 & 0.248 $\pm$ 0.023 & \GE\ \\ APOGEE  \\ 
TIC393961551 & Gaia 1506387627917936896 & 9.57166 & 500.533 $\pm$ 6.312 & 61.34 $\pm$ 1.75 \\ 6.68 $\pm$ 0.41 & 5121 $\pm$ 105 & -1.0751 $\pm$ 0.0123 & 0.156 $\pm$ 0.014 & \GE\ \\ APOGEE\\
TIC453888381 & Gaia 5230256730347457152 & 10.9714  & 788.614 $\pm$ 15.999 &  50.37  $\pm$   1.59 \\ 5.953 $\pm$ 0.049 & 4741 $\pm$ 100 & -0.728 $\pm$ 0.07 & 0.32 $\pm$ 0.021 & Halo \\ GALAH \\  
TIC279510617 & Gaia 5480550450643017216 &	10.7551 & 933.263 $\pm$ 22.520 &  28.57  $\pm$   0.16    \\   3.566  $\pm$   0.015 & 4450 $\pm$ 100 & -0.49 $\pm$ 0.05 & 0.281 $\pm$ 0.017 & Halo \\ GALAH  \\ 
TIC300938910 & Gaia 5270675018297844224 & 10.5629 & 607.156 $\pm$ 7.7075 &  106.30$\pm$0.92 \\ 9.464$\pm$0.131 & 4908 $\pm$ 100 & -0.792 $\pm$ 0.05 & 0.2566 $\pm$ 0.0165 & Halo \\ GALAH \\ %
TIC198204598 & Gaia 1629898685347273856 & 10.9455 & 952.885 $\pm$ 37.512 &  45.86$\pm$0.31 \\ 5.132$\pm$0.032 & 4979 $\pm$ 100 & -- & -- & \GE\ \\ -- \\ %
TIC1008989 & Gaia 3789639280952610304 & 9.72882 & 370.56 $\pm$ 5.85 &  104.33059  $\pm$   1.46618  \\   9.80317  $\pm$   0.15336 & 4893$\pm$100 & -- & -- & Halo \\ -- \\ 
TIC91556382 & Gaia 5065009650333147392 & 10.0855 & 870.289 $\pm$ 34.565 &  31.38665   $\pm$  1.03097  \\    4.189    $\pm$ 0.186 & 5192 $\pm$ 100 & -- & -- & Halo \\ --  \\
TIC159509702 & Gaia 1709195090281718272 & 12.1542 & 1595.67 $\pm$ 55.605 & 45.37$\pm$0.53 \\ 5.090$\pm$0.027 & 4724 $\pm$ 100 & -- & -- & Halo \\ -- \\ %
\hline
TIC25079002 & Gaia 4669316065700222976 & 9.91465 & 716.804 $\pm$ 15.5995 &   45.238 $\pm$ 0.62 \\ 4.967 $\pm$0.121 & 4797 $\pm$ 83 & 0.1636 $\pm$ 0.006 & -0.010 $\pm$ 0.006 & Disk \\ APOGEE \\
TIC177242602 & Gaia 5262295395367212288 & 10.1451 & 532.831 $\pm$ 14.294 &  54.66 $\pm$ 0.33 \\ 5.663 $\pm$0.031 & 4603 $\pm$ 100 & -0.1176 $\pm$ 0.006 & 0.125 $\pm$ 0.007 & Disk \\ APOGEE  \\
TIC9113677 & Gaia 3245485650607651584 & 10.1791 & 491.438 $\pm$ 10.040 &   87.22437  $\pm$   0.87042   \\    8.365   $\pm$  0.198 & 4764 $\pm$ 100 & -- & -- & Thick Disk \\-- \\
\enddata 
\tablenotetext{}{Directly measurable stellar parameters. These parameters were either measured directly or calculated directly from measurements of stellar parameters. These parameters were used to select these stars for this study and place them into their respective galactic substructure classes.}

\tablenotetext{*}{We note that the definition of $\alpha$-elements is a combination of a number of different elements that appear in differing amounts in different stellar populations. Thus, the definition of $\alpha$-elements for the disk stars should be interpreted differently than the definition for halo stars.}

\end{deluxetable*}

\subsection{Age Analysis}
\label{sec:ages}

We determined stellar ages using the \texttt{BASTA} \citep{silvaaguirre2015}, \texttt{isochrones} \citep{morton2015}, \texttt{isoclassify} \citep{huber2017}, \texttt{PARAM} \citep{dasilva2006,rodrigues2017} and \texttt{scaling-giants} \citep{bellinger2020} packages.  The \texttt{BASTA}, \texttt{isochrones}, \texttt{isoclassify}, and \texttt{PARAM} packages take asteroseismic, photometric, and spectroscopic parameters as inputs, while the \texttt{scaling-giants} package accepts asteroseismic parameters, metallicity, and temperature as inputs.

To determine ages with \texttt{scaling-giants}, we used seismic $\nu_\mathrm{max}$ and $\Delta\nu$ values measured by the SYD pipeline along with effective temperature determined through the direct method of \texttt{isoclassify} and a metallicity determined by either the APOGEE or GALAH surveys.

To determine ages with \texttt{BASTA}, \texttt{isochrones}, \texttt{isoclassify}, and \texttt{PARAM}, we used 2MASS K magnitudes, asteroseismic $\Delta\nu$, \emph{Gaia} parallaxes, and temperatures and metallicities from spectroscopy where available. For those stars without spectra, we relied on temperatures determined using the direct method of \texttt{isoclassify} as described earlier. \texttt{BASTA} compares the observed properties with predictions from theoretical models of stellar evolution from the recently updated BaSTI (a Bag of Stellar Tracks and Isochrones) stellar models and isochrones library \citep{hidalgo2018}, considering convective overshooting and no mass-loss. Both \texttt{isochrones} and \texttt{isoclassify} use MIST stellar isochrones in order to determine stellar parameters from observables \citep{choi2016}, while \texttt{PARAM} uses MESA isochrones constrained by individual asteroseismic radial-mode frequency models. On the other hand, \texttt{scaling-giants} establishes power-law relations for any subset of effective temperature, metallicity, $\nu_\mathrm{max}$ and $\Delta\nu$ which were calibrated on BaSTI models for red giant branch stars observed by \emph{Kepler}. Uniform priors were placed on metallicity for the \texttt{isochrones} package. Uncertainties on ages determined by the \texttt{isochrones} package were determined using \texttt{PyMultiNest} \citep{multinest}. For those stars with spectra, effective temperature and log($g$) derived from converged \texttt{isochrones} models were compared to the spectroscopic constraints from observation, and were found to agree within 2-$\sigma$ uncertainties in all cases.

\begin{deluxetable*}{ccccc}

\tablecolumns{5} 
\tabletypesize{\scriptsize}
\tablecaption{Derived Stellar Parameters \label{table2}}
\tablewidth{0pt}
\tablehead{
\colhead{TIC ID} & \colhead{\emph{Gaia} ID} &   \colhead{Radius (R$_\odot$)} & \colhead{Mass (M$_\odot$)} & \colhead{Evolutionary Type} \\ \colhead{\texttt{scaling-giants} Age (Gyr)} & \colhead{\texttt{isochrones} Age (Gyr)} & \colhead{\texttt{isoclassify} Age (Gyr)} &
\colhead{\texttt{PARAM} Age (Gyr)}  & \colhead{\texttt{BASTA} Age (Gyr)} \\ \colhead{Galactic Substructure}}
\startdata
TIC20897763 & Gaia 2365649471033828096 & 7.10 $\pm$ 0.61 &  0.94 $\pm$ 0.17 & RGB \\ 5.8 $\pm$ 3.0\tablenotemark{*} &  8.77 $\pm$ 2.9 & 5.68$^{+0.79}_{-0.83}$ & 9.29$^{+2.82}_{-3.23}$  & 9.0$^{+2.7}_{-2.5}$ \\ \GE\ \\ 
TIC341816936 & Gaia 1421776046335723008 & 9.75 $\pm$ 1.12  & 1.11 $\pm$ 0.172 & Red Clump\tablenotemark{+} \\ 2.9 $\pm$ 1.8 &  9.16 $\pm$  2.75 & 7.44$^{+1.36}_{-1.34}$ & 6.52$^{+4.46}_{-3.80}$ & 7.9$^{+2.8}_{-1.9}$ \\ \GE\ \\ 
TIC393961551 & Gaia 1506387627917936896 & 7.44 $\pm$ 0.80 &  1.03 $\pm$ 0.15 & RGB \\ 4.7 $\pm$ 6.0 &  5.93 $\pm$ 3.05 & 5.63$^{+0.84}_{-0.69}$ & 5.72$^{+4.61}_{-3.14}$ & 7.5$^{+3.3}_{-2.2}$ \\ \GE\ \\
TIC453888381 & Gaia 5230256730347457152 & 7.38 $\pm$ 0.31  & 0.828 $\pm$ 0.088 & RGB \\ 10.5 $\pm$ 3.5\tablenotemark{*} &  9.72 $\pm$  2.50 & 12.78$^{+0.78}_{-1.59}$ & 11.66$^{+1.50}_{-2.42}$ &  14.8$\pm$2.8 \\ Halo\\  
TIC279510617 & Gaia 5480550450643017216 & 10.27$\pm$0.43  & 0.866$\pm$0.044  & Red Clump\tablenotemark{+} \\ 6.4 $\pm$1.1\tablenotemark{*} &  7.99 $\pm$ 3.37 & 10.23$^{+0.95}_{-2.34}$ & 7.68$^{+4.55}_{-4.45}$ & 7.6$^{+0.9}_{-0.8}$ \\ Halo\\ 
TIC300938910 & Gaia 5270675018297844224 & 6.31 $\pm$ 0.19 & 1.27 $\pm$ 0.09 & RGB \\ 2.7 $\pm$ 0.8\tablenotemark{*} & 3.693 $\pm$ 0.702 & 3.34$^{+0.40}_{-0.33}$ & 2.11$^{+0.61}_{-0.49}$ & 2.9$^{+0.6}_{-0.4}$ \\ Halo \\ %
TIC198204598 & Gaia 1629898685347273856 & 9.24 $\pm$ 0.15 & 1.18 $\pm$ 0.05 & RGB \\ 3.3 $\pm$ 0.8\tablenotemark{*} & 8.749 $\pm$ 2.16 & 11.54$^{+0.66}_{-2.83}$ & 1.98$^{+7.13}_{-0.77}$ &  5.8$^{+1.6}_{-1.2}$ \\ \GE\ \\ %
TIC1008989 & Gaia 3789639280952610304 & 5.64$\pm$0.44 & 1.02 $\pm$ 0.083  & RGB \\ 5.8 $\pm$ 2.0 &   6.843 $\pm$ 3.33 & 6.84$^{+1.50}_{-0.90}$ & 7.27$^{+2.95}_{-2.09}$ & 10.6$^{+3.5}_{-2.9}$ \\ Halo \\ 
TIC91556382 & Gaia 5065009650333147392 & 10.26$\pm$ 1.24 & 1.013$\pm$0.28  & -- \\7.1 $\pm$ 5.0\tablenotemark{*} & 6.841 $\pm$ 3.59 & 6.43$^{+1.89}_{-0.96}$ & 5.27$^{+2.84}_{-1.77}$ & 8.9$^{+3.8}_{-2.9}$ \\ Halo\\
TIC159509702 & Gaia 1709195090281718272 & 8.52 $\pm$ 0.13 & 0.96 $\pm$ 0.04  & RGB \\ 4.8 $\pm$ 1.2\tablenotemark{*} & 9.344 $\pm$ 2.331 & 9.98$^{+1.47}_{-1.18}$ & 7.92$^{+3.47}_{-2.49}$ & 10.0$^{+3.2}_{-2.5}$ \\ Halo\\ %
\hline
TIC25079002 & Gaia 4669316065700222976 & 9.43 $\pm$ 0.79 &  1.19 $\pm$ 0.20 & Red Clump \\ 2.1 $\pm$ 3.2 &  7.07 $\pm$ 3.65 & 1.92$^{+0.17}_{-0.11}$ & 1.84$^{+0.18}_{-0.14}$ & 3.6$^{+0.3}_{-0.4}$ \\ Disk \\
TIC177242602 & Gaia 5262295395367212288 & 7.97 $\pm$ 0.38 &  1.01 $\pm$ 0.11 & RGB \\ 4.4 $\pm$ 2.7 &  7.49 $\pm$ 3.45 & 6.54$^{+0.44}_{-0.64}$ & 7.60$^{+4.02}_{-3.74}$ & 6.4$^{+0.6}_{-1.0}$  \\ Disk\\
TIC9113677 & Gaia 3245485650607651584 & 6.52 $\pm$ 0.43  & 1.08$\pm$ 0.17 & RGB \\ 5.1$\pm$2.0 &  7.07 $\pm$ 3.65 & 6.77$^{+1.9}_{-1.35}$ & 8.24$^{+3.24}_{-2.87}$ & 9.7$^{+3.3}_{-2.3}$ \\ Thick Disk\\ 
\enddata 
\tablenotetext{}{Stellar parameters derived indirectly from the measured stellar parameters. Ages were inferred using a combination of \emph{Gaia} parallax, 2MASS K magnitude, asteroseismic $\Delta\nu$, stellar effective temperature and metallicity where available. }
\tablenotetext{+}{We note that these evolutionary state designations did not agree between the \texttt{BASTA} and \texttt{COR} methods due to ambiguities in oscillation mode identification. We present the \texttt{COR} designations here.}

\end{deluxetable*}

\begin{figure*}
    \centering
    \includegraphics[width=\linewidth]{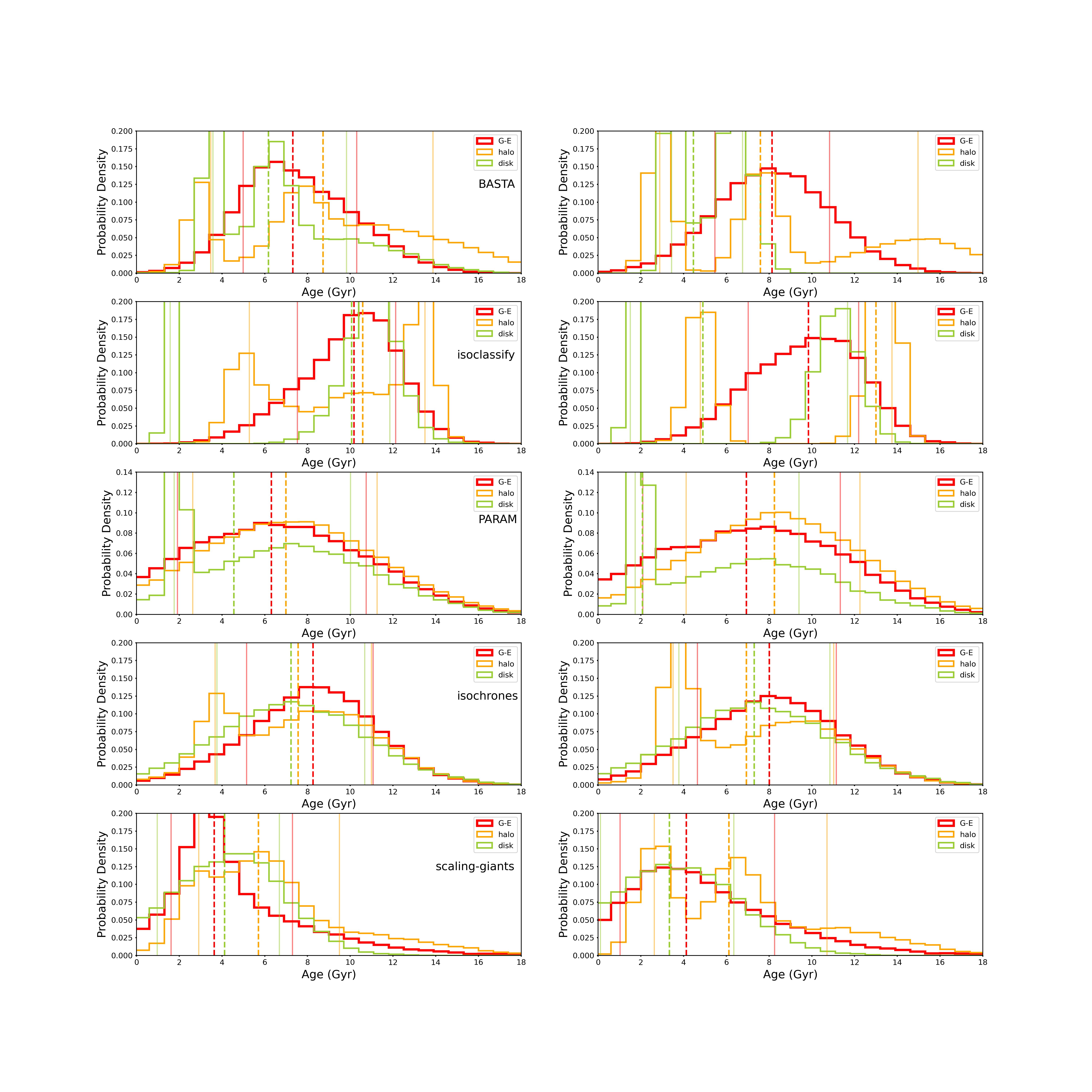}
    \caption{Stacked age distributions for all stars from all age determination methods tested here from top to bottom, where stars have been grouped based on their expected origins. \texttt{BASTA} age distributions are given on top, followed by \texttt{isoclassify}, \texttt{PARAM}, \texttt{isochrones}, and \texttt{scaling-giants}. The left column illustrates distributions for all stars in our sample, while the right column considers only stars in our sample which have spectra. Median, 16th and 84th percentiles of each set of combined age distributions have been shown by vertical lines. The \GE\ distribution is not significantly distinct from either the disk or halo populations, but it generally displays a smaller spread in age than either {\it in situ} population using the majority of age determination methods. However, proper statistical comparison of the ages of these populations requires the hierarchical analysis of the next Section.}
    \label{fig:ages}
\end{figure*}

We chose not to use the asteroseismic $\nu_\mathrm{max}$ value as an input for \texttt{BASTA}, \texttt{isochrones},  \texttt{isoclassify}, and \texttt{PARAM}. As the amplitudes of stellar oscillation are known to have a color dependence, and both oscillation and granulation amplitudes are correlated with metallicity, this may distort the observed shape of the oscillation spectrum in our sample \citep{yu2018}. Furthermore, the stochastic nature of oscillation excitation suggests that shorter time series observations will feature less Gaussian oscillation profiles. Given the different bandpass and relatively unexplored data of \tess\ relative to \emph{Kepler}, systematic differences between $\nu_{\mathrm{max},\odot}$ values measured from \tess\ and \emph{Kepler} data could affect our age estimates. This systematic difference is expected to be relatively small, given the large overlap between the \tess\ and \emph{Kepler} bandpasses, and, indeed, initial asteroseismic studies with \tess\ have not observed any systematic offset in $\nu_\mathrm{max}$ \citep[][Stello, {\it in prep.}]{silvaaguirre2020, mackereth2021}. Though a 1\% discrepancy in $\nu_\mathrm{max}$ will not influence determined stellar mass and radius by more than 3\%, it can result in a $\sim$ 10\% discrepancy in age determination. Thus, while we still present masses and radii determined with the measured $\nu_\mathrm{max}$ values in Table~\ref{table1}, we determined that these $\nu_\mathrm{max}$ values could systematically bias age estimates, and only include them for the age determination method with the largest errors and fewest input parameters, \texttt{scaling-giants}. 






Using the median, 16th and 84th percentile posteriors from each age determination method, we reproduce a posterior age distribution constructed of 10,000 samples drawn from normal distributions for each star with each age determination method. We report ages and uncertainties for each star using each method in Table \ref{table2}. We then combine these samples for each age determination method based on the expected origins of the stars based on the previous kinematic and spectroscopic analysis. We plot these combined posterior distributions in Figure~\ref{fig:ages}. We also illustrate the median, and 16th and 84th percentiles for each distribution. We find that all age distributions are significantly overlapping. However, the range between the 16th and 84th percentiles of the \GE\ stars is smaller than that of the halo and disk stars in all but one determination method (\texttt{PARAM} finds a larger age distribution in the \GE\ than the disk), where the halo star age distribution extends further to older and younger ages in all but one age determination method (\texttt{isochrones} finds effectively the same age upper bound for all three distributions, \texttt{scaling-giants} finds a younger lower bound for the \GE\ than the halo), and the disk star distribution extends to younger ages for all age determination methods (\texttt{isochrones} lower age bounds are comparable for the disk and halo stars). In addition, the median ages of both the halo and \GE\ stars are very similar, but the median age of disk stars appears slightly younger. We also compare each age determination method to one another, and have shown these comparisons in Figure~\ref{fig:agecomp} in the Appendix. We also determine the relative offsets between the various age determination methods tested. We find that ages determined using the \texttt{isochrones} package are the most consistent among methods, with median relative offsets from the \texttt{isoclassify}- and \texttt{PARAM}- determined ages of 6 and 8\%, respectively. The median relative age offset between \texttt{isoclassify} and \texttt{PARAM} is 13\%. The median relative offset in ages between \texttt{scaling-giants} and all other methods is $>$25\%. We note that this is likely related to the fact that the majority of the stars in our sample fall outside the range of the training set used to define the \texttt{scaling-giants} relations, to highlight systematic uncertainties that can be introduced when extrapolating from thin disk stars to determine absolute ages for stars with significantly non-solar masses and metallicities. However, we note that the age distribution rankings determined by \texttt{scaling-giants} appear generally consistent with those determined by the other four methods.

To test whether age rankings are robust, and our stellar populations are distinct, we determine the Spearman  correlation coefficient for all of the combinations of age ranking for the stars considered here. We find that the $p$-values recovered for this test lie between 0.21 and 0.95, implying that the null hypothesis, that the age rankings of these age determination methods are drawn from distinct distributions, is not statistically supported, implying that our age rankings are robust regardless of age determination package. 

As this Spearman coefficient age ranking analysis does not measure statistical robustness for the distinction between underlying stellar populations, we also then perform a hierarchical age analysis using the age determination package which provides access to model likelihoods and priors, \texttt{isochrones}, to establish more robust statistical constraints on ages of the underlying population from which these stars have been drawn in the following Section.



Overall, we find that our age distributions for all stars are not clearly distinct. However, there are some noteworthy features which warrant further investigation of the ages of halo stars. First, we note that the stars determined most likely to be accreted from other galaxies as part of the \GE\ merger event are not the youngest nor oldest stars investigated here, according to all of our tested methods. Secondly, those stars shown to be kinematically belonging to the disk are not the oldest stars in the sample, and according to a majority of methods, include the youngest star in our sample. Finally, we highlight the wide age distribution of both the disk and halo stars seen by all methods--it appears that the potential age range when considering all disk stars and halo stars in our sample is wider than the age range of all of the \GE\ stars. 

We report a weighted average for the \GE\ stars in our sample by weighting age distributions from all packages considered in this analysis equally except for \texttt{scaling-giants}, which we do not include due to its large age offsets relative to all other packages considered. We find a mean age of the \GE\ stars of 8 Gyr, with an average statistical uncertainty for each age determination method of 3~Gyr and average systematic uncertainty between age determination methods of 1~Gyr. We thus report an age for these stars of 8 $\pm$ 3 (stat.) $\pm$ 1 (sys.) Gyr.


\begin{figure*}
    \centering
    \includegraphics[width=.49\linewidth]{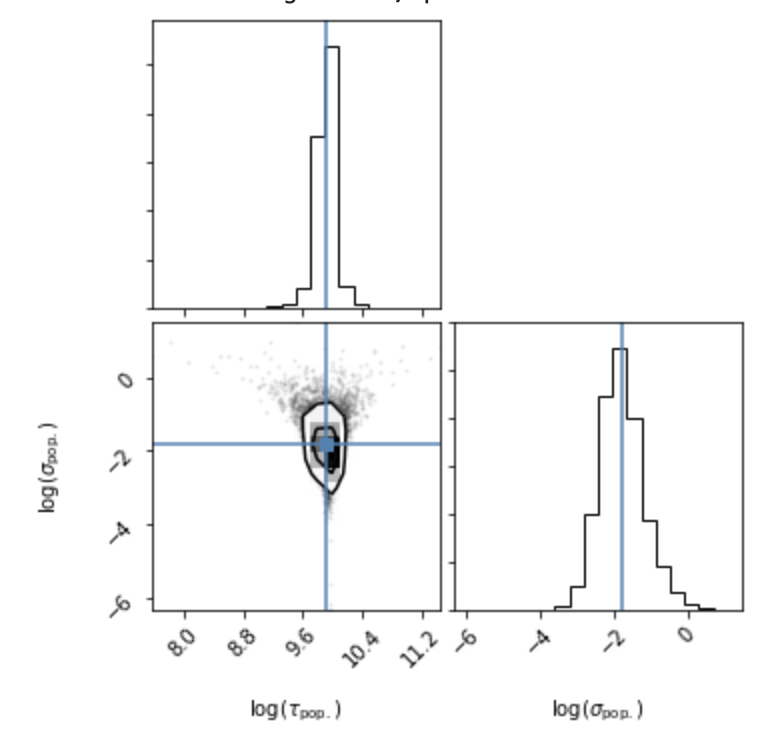}
    \includegraphics[width=.49\linewidth]{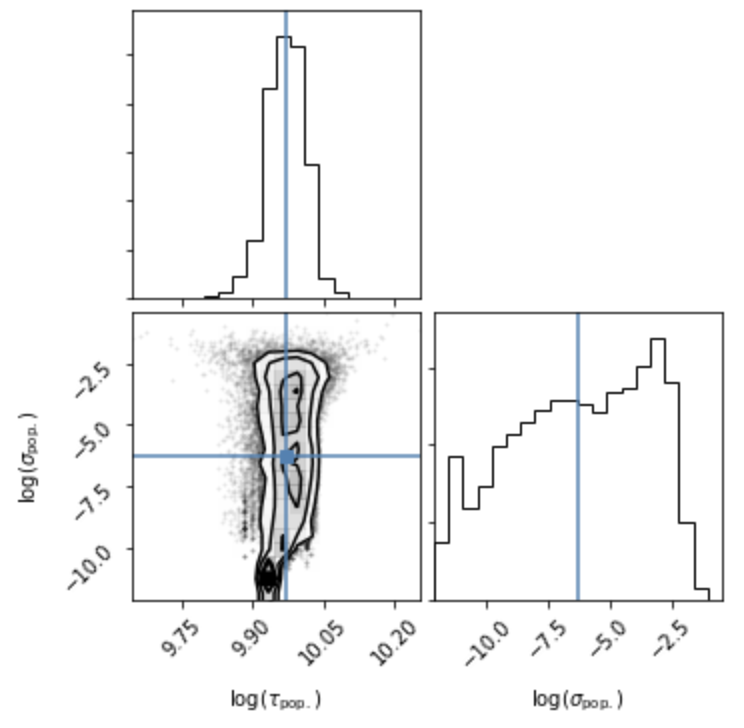}
    \caption{Left: Posterior distribution over age from a hierarchical Bayesian model of \GE\ determined from the ages of TIC20897763, TIC393961551, TIC341816936 and TIC198204598, the 4 stars most likely to be \GE\ members based on kinematics and metallicity where available. Here, all four stars have been modeled and fit for simultaneously using \texttt{emcee} to explore posterior probability distribution functions of \texttt{isochrones} models. We illustrate the posteriors for $\alpha$ = ($\tau_\mathrm{pop.}$, $\sigma_\mathrm{pop.}$), the age and variance posteriors for the larger population these stars are drawn from, in log(age) units. Simultaneously, a hierachically-determined age mean and variance for the population has been determined using a uniform age hyperprior in addition to incorporating the prior age information for each individual star. Blue lines designate 50th percentile values. We find a hierarchical age of \GE\ population of \hbmage. Right: Same as left, except using the population of {\it in situ} halo stars. We find an average age of the population of 9.5 $\pm$ 0.9 Gyr, in agreement with the average age determined for the population of \GE\ stars. \label{fig:hbm_age}}
\end{figure*}

\begin{figure}
    \centering
    \includegraphics[width=\linewidth]{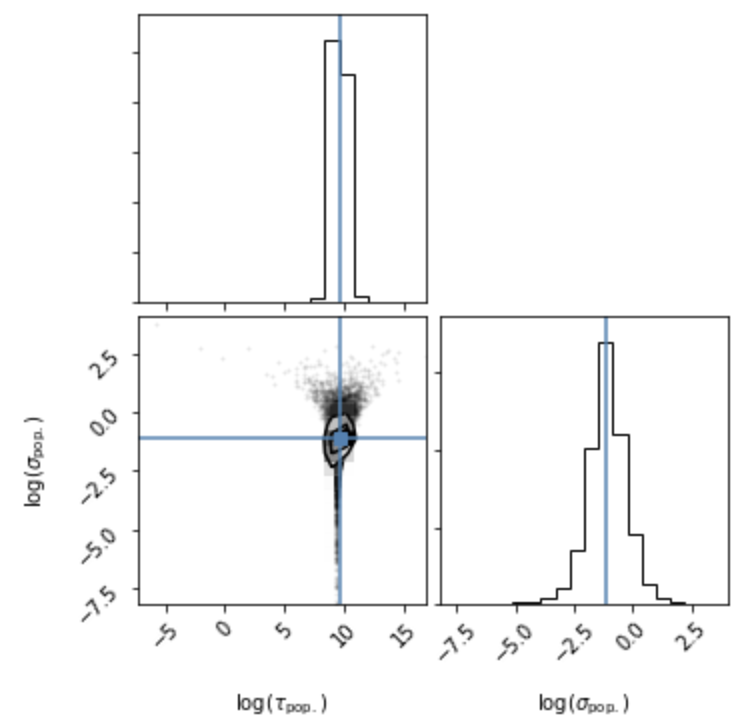}
    \caption{Same as above figure, except using the population of {\it in situ} disk stars. We find an average age of the population of 4$^{+7}_{-3}$ Gyr, in agreement with the average age determined for the population of \GE\ and {\it in situ} halo stars.  \label{fig:hbm_age2}}
\end{figure}

\subsubsection{Hierarchical Age Analysis}

We also use the subset of our stars which appear most likely to be \GE\ remnants based on their metallicities, colors and kinematics, using our more conservative kinematics cut shown in Figure~\ref{fig:kinem} to determine a hierarchical Bayesian median age and scatter for the \GE.

Using the \texttt{isochrones} package, we generate posteriors of stellar model parameters of TIC20897763, TIC393961551, TIC341816936 and TIC198204598 in an MCMC procedure using \texttt{emcee} \citep{foremanmackey2013}. We then employ the `importance sampling trick' to generate a hierarchical age and age variance estimate for an underlying \GE\ population that these stars are drawn from \citep{hogg2010,foremanmackey2014, pricewhelan2018}. We follow a simplified implementation of the procedure of \citet{pricewhelan2019} to determine the hierarchical age and variance of this population, as detailed below.

The automatic likelihood fitting routine of \texttt{isochrones} performs interpolation between the provided grid of stellar models \citep[here we utilize the MIST model isochrone grid,][]{dotter2016, choi2016}. Given our set of measured parameters and uncertainties {\bf$m$} = (T$_\mathrm{eff}$, $\pi$, $\Delta\nu$, $m_K$, [Fe/H] {\it where available}), we can generate posterior samplings of the interpolated MIST model parameters {\bf $\theta$} = ($\tau$, EEP, [Fe/H], $D$, $A_V$) for each individual star. We then assume that the probability for a given set of hyperparameters $\alpha$ = ($\tau_\mathrm{pop.}$, $\sigma_\mathrm{pop.}$) comprised of hierarchical population age $\tau_\mathrm{pop.}$ and variance $\sigma_\mathrm{pop.}$ can be defined as

\begin{equation}
p(\alpha | x) = \frac{1}{\sigma_\mathrm{pop.}\sqrt{2\pi}}  \mathrm{exp}\Big(-{ \frac{(x-\tau_\mathrm{pop.})^2}{2\sigma_\mathrm{pop.}^2}}\Big)
\end{equation}

where $\tau_\mathrm{pop.}$ is the hierarchical age of the population, $\sigma_\mathrm{pop.}$ is the variance on that age, and $x$ is the age of a given star in our sample.

In addition, prior probability distributions must be specified for the model parameters {\bf $\theta$}. For this study, we use the default priors of \texttt{isochrones} \citep{morton2015} except for the prior distribution in metallicity, where we use a uniform prior with bounds of (-2, 0.5) implemented as in \citet{pricewhelan2019}. 

To compute the likelihood for our hierarchical model, we use the posterior samplings for model parameters for individual stars in our sample to marginalize over the per-source stellar parameters {\bf $\theta_n$}, where $n = (0,1,...,N)$, and $N$ is the number of sources in a population. We then compute the marginal likelihood p($m_n | \alpha$), where $m_n$ represents our data for the $n$th star in a population. This likelihood can be approximated using the `importance sampling trick', and here we follow the implementation of \citet{pricewhelan2019} with slight modification. We thus rewrite the marginal likelihood as 

\begin{equation}
    p( m_n | {\bf \alpha}) \approx \frac{\mathcal{Z}_n}{K}\sum_{k}^{K}\frac{p({\bf \theta}_{nk} | {\bf \alpha})}{p({\bf \theta}_{nk} | {\bf \alpha}_0)}
\end{equation}

where the index $k$ specifies the index of one of $K$ posterior samples generated from the independent samplings (described above), $\mathcal{Z}_n$ is a constant, and the denominator, p(${\bf \theta}_{nk}$ | ${\bf \alpha}_0$), are the values of the interim prior used to constrain the independent samplings. In this work, $N$ = 4 for the \GE\, 6 for the halo, and 3 for the disk population. In all cases, we adopt K = 2048, following convention in the literature \citep{pricewhelan2019}.

With the marginal likelihood (Equation 7), we then specify uniform prior probability distributions for the hyperparameters {\bf $\alpha$}. We place a uniform prior on both hyperparameters with boundaries of (0.1 Myr, 14 Gyr). Since we are sampling the logarithm of our hyperparamters, this is equivalent to placing a Jeffreys prior on ($\tau_\mathrm{pop.}, \sigma^2_\mathrm{pop.}$) \citep{jeffreys1946}. We then use \texttt{emcee} \citep{foremanmackey2013, goodman2010} to sample from the posterior probability distribution for the hyperparameters given all of our measured observables, 

\begin{equation}
    p({\bf \alpha} | m_n) \propto p({\bf \alpha}) \prod_{n}^{N} p(m_n | {\bf \alpha}) .
\end{equation}

Here we use 100 walkers and run for an initial 1000 steps to burn-in the sampler before running for a final 9000 steps. We again compute the Gelman-Rubin \citep{gelman1992} convergence diagnostic and find that all chains have R $<$ 1.01 and are thus likely converged.

We use our posterior distribution of $\alpha = (\tau_\mathrm{pop.}, \sigma_\mathrm{pop.})$ to determine an age for the \GE\ of \hbmage. We then perform similar hierarchical age determinations for the remaining halo stars (TIC453888381, 1008989, 91556382, 279510617, 159509702, 300938910) and disk stars (TIC25079002, 177242602, 9113677) as individual populations using the same likelihood equations and constants as described above. We display the posterior distribution of individual stellar ages as well as the age and age variance determined by this method for each population in Figures \ref{fig:hbm_age} and \ref{fig:hbm_age2}. We find an age $\tau_\mathrm{pop.}$ of 9.5 Gyr and uncertainty $\sigma^2_\mathrm{pop.}$ of 0.9 Gyr for the halo star population and 4$^{+7}_{-3}$ Gyr for the disk star population at 1-$\sigma$ confidence limits. We note that for the {\it in situ} halo stars, the distribution in age variance $\sigma_\mathrm{pop.}$ appears smallest, and is effectively consistent with 0 (i.e., all stars are the exact same age) within uncertainties despite the wider range in median ages of individual stars for this population. Comparable distributions of the ages of halo stars have been found by other surveys \citep{montalban2020}. 

We note that mass loss on the red giant branch is notoriously ill-constrained, and prevents accurate age estimation for red clump stars \citep[e.g.,][]{casagrande+2016}. We therefore recalculate our age estimate for \GE\ after removing TIC341816936 from the sample, which we classify as a red clump star. We note that excluding TIC341816936 does not significantly impact our age posterior distribution. This is likely due to the relatively large statistical age uncertainty for this star.

Additionally, we note that we have not accounted for all systematic uncertainties in age introduced by the use of different stellar models and stellar parameter constraints. For example, uncertainties in mixing length theory, convective overshoot and metallicity scale have been shown to cause issues in the determination of stellar ages, which have not yet been accurately determined for red giant stars \citep{silvaaguirre2020b}, and can introduce systematic age uncertainties that have been shown to be $\sim$30\% for evolved stars \citep{tayar2021}. We aim to account for these uncertainties by determining an average age using different age determination packages. This does not account for systematic uncertainties which affect all age determination packages in the same way; however, given that all of the stars in this sample are metal-poor, it should affect all stars in the sample in a similar fashion, and thus have only insignificant impacts on the relative ages and age rankings of the populations studied here. We suggest the systematic uncertainty of 1 Gyr found in the previous informal analysis should be added to the statistical uncertainties found by the hierarchical Bayesian age analysis done here.

Given the wide distribution of average age and variance for each individual population of stars, we conclude that our current sample of stars is not large enough and does not have the age accuracy or precision necessary to clearly distinguish between the \GE, galactic halo and disk structures. However, we can comment on some of the relative differences in the distributions of age that we measure in this study, and how future surveys may be able to further constrain these differences to better understand the assembly of our Galaxy. We expand on this in the following Section.









\section{Discussion}

\subsection{Age Distribution of Stars in the Nearby Halo}

As the kinematic properties of these stars allow clear distinction of different kinematic subpopulations, we can use our stellar age estimates to produce an age distribution of stars in our sample, and then use this distribution to determine the age of early satellite merger events such as that which created the \GE.

Following the reasoning of \citet{grand2020}, we can use the 20th percentile of the age distribution of our selected stars as a proxy for the true \GE\ merger time, as all stars from \GE\ should be as old as the merger or older. Following this reasoning, we determine an average age from the two youngest \GE\ ages using each age determination method, and find that the estimated age ranges for \GE\ merger event fall between 3.3 and 7.7 Gyr. Furthermore, evidence for a starburst induced by the \GE\ merger event should provide further constraints on the merger time. However, given our relatively small sample and wide distribution of stellar ages we have measured, evidence for a starburst cannot be clearly identified in our dataset.

Looking at our population of stars in more detail, we analyze the stars studied here in three distinct subgroups--the disk stars, halo stars likely formed {\it in situ} within our own Galaxy, and the \GE\ population of halo stars accreted from elsewhere. We note that we divide the halo population using both chemical and kinematic cuts, which agree for those stars where both data sets can be explored \citep{nissen2010, helmi2018}. We note that both the disk stars and halo stars show a wide spread in ages, but the age distribution of the disk stars skews younger while the halo star sample skews older. The stars we designate as \GE\ members all have consistent ages which are not the youngest or oldest in our sample. We estimate a population age from which the ages of these stars stars are drawn as \hbmage, in agreement with previous theoretical estimates which predict ages of 10 Gyr or greater for the \GE. We encourage future studies of stellar ages in the halo with larger sample sizes to either support or refute this claim.



We note we are not able to constrain stellar ages in our sample as well as that of $\nu$ Indi, a naked-eye halo star for which modeling individual asteroseismic frequencies was possible with short-cadence (2-minute) data. Our \GE\ ages as determined using a majority of age determination methods were in agreement with the estimate for $\nu$ Indi by \citet{chaplin2020} of 11.0 $\pm$ 1.5 Gyr. However, unlike $\nu$ Indi, the median age of the stars in our \GE\ sample is less than 10 Gyr. Our age estimate for the \GE\ and halo populations are also statistically in agreement with the age of $\nu$ Indi determined by \citet{chaplin2020}. We do not determine the age of $\nu$ Indi in this study because its oscillations are above the Nyquist frequency of the 30-minute cadence data. We also note that within uncertainties, all the \GE\ stars have ages consistent with 4 Gyr, and thus any estimate that place \GE\ ages above 4 Gyr seems possible for this population, in agreement with our result following the reasoning of \citet{grand2020}. 



We find that for those stars with spectra available, the metallicities of the population agree with previous measurements of the bulk metallicity of the Gaia-Enceladus population \citep{feuillet2020}. Furthermore, we find a divide in high-$\alpha$ and low-$\alpha$ element populations between our \GE\ stars and `halo' stars, additional evidence for an {\it in situ} halo population of stars excited by a major merger event along with an accreted population of stars from another galaxy. The stars in our sample with spectroscopic data occupy both sides of the divide identified in \citet{nissen2010} in $\alpha$ element abundance and metallicity, suggesting both {\it in situ} and accreted stars in our sample. This kinematic and spectroscopic divide is less clearly seen photometrically, as not all \GE\ stars appear bluer at the same temperature as stars formed in our Galaxy, as seen in Figure~\ref{fig:gaiahrd}. However, kinematic differences seen in Figure~\ref{fig:kinem} clearly reflect the spectroscopic divides. This provides further evidence that we have identified stars excited as part of the \GE\ merger, and that at least two distinct populations of stars do exist in the Galactic halo.

\citet{matsuno2020} performed a similar study combining spectral information from APOGEE and LAMOST with asteroseismic information from \emph{Kepler} to determine masses for 26 halo stars. They find that the average mass of the star in their sample is 0.97 M$_\odot$, with a scatter of 0.04 M$_\odot$. They speculate that these masses correspond to an age of $\sim$8 Gyr. This is in very good agreement with our individual stellar ages for \GE\ stars as well as the hierarchical Bayesian age estimate for the \GE\ population determined here. Furthermore, we note that the mass estimates of all the \GE\ members in this sample have masses consistent with the \citet{matsuno2020} mass range, with all stars with spectral information agreeing at a 95\% confidence level. This study gives further credence to the argument that the \GE\ was accreted less than 10 Gyr ago.

\citet{montalban2020} performs more detailed asteroseismic modeling of 95 \emph{Kepler} halo stars with \emph{APOGEE} spectra, fitting individual frequencies to calculate stellar ages using \texttt{AIMS} \citep{reese2016,lund2018,rendle2019}. They find slightly older average age estimates for the stars in their sample as compared to \citet{matsuno2020}. \citet{montalban2020} also distinguish between the $\alpha$-rich and $\alpha$-poor halo populations, and then further distinguish between the $\alpha$-poor halo population based on kinematic properties, to distinguish \GE\ stars in a complementary but independent approach to our own. \citet{montalban2020} find that the population of stars which they refer to as the \GE\ have a mean population age of 9.7 $\pm$ 0.6 Gyr, while the {\it in situ} halo population of stars have a mean population age of 10.4 $\pm$ 0.3 Gyr. This is in strong agreement with our population age estimates for both \GE\ stars (\hbmage) and {\it in situ} halo stars (9.5 $\pm$ 0.9 Gyr). Furthermore, the overall age distributions found by \citet{montalban2020} reproduce similar qualitative results to those found here in that {\it in situ} halo stars have a slightly older average population age and smaller mean age variance, yet wider tails in age distribution, when compared to the \GE\ age distribution. In addition, \citet{montalban2020} check their effective temperatures, chemical compositions, stellar surface effects, evolutionary state effects, and \emph{Gaia} constraints on stellar properties and find that these errors cannot account for the systematic differences in ages of stellar populations, further suggesting that the similar properties in age distributions in the populations that are observed is astrophysical.


\citet{donlon2020} perform an analysis of the shell substructure of the Milky Way in order to identify evidence of excess kinematic energy due to merger events in the Galaxy's history. Using $N$-body simulations of radial merger events, they find that shell structure similar to that seen in the Milky Way dissipates within 5 Gyr after collision with the Galactic center. This suggests that the infall of the \GE\ progenitor galaxy may have taken longer than previously thought, or that the \GE\ structure is younger than previous estimates have suggested. Our age estimates of the \GE\ agree with this shell substructure age analysis within uncertainties.

\subsection{Potential Mass Bias}

Each star in our sample had to be visually confirmed to show genuine oscillations despite independent confirmation of asteroseismic parameters due to the poorly understood nature of \tess\ systematics. Thus the mass distribution of stars in our sample may not be necessarily representative of the entire \GE\ population as a whole. This is entirely possible, but to address this we attempted to use the most reliable detections to establish the mass distribution of all stars in our sample. Using the \texttt{SLOSH} package \cite{hon2018}, we vet the probability of asteroseismic detection in each light curves, and then if the numax detected by this pipeline agrees with the SYD pipeline \citep{huber2009} within 40\%, we calculate a mass for this star, and visually inspect the power spectrum to ensure the asteroseismology is reliable. We find that only a small fraction ($<$~20\%) seem to have reliable detections, and note that the mass distribution of these reliable detections fall between 0.8 and 1.4 M$_\odot$ in all cases. We note that this agrees with the mass ranges for \GE\ stars found by other studies \citep{matsuno2020,montalban2020}. Thus, we believe that we are not biasing our inference of the mass distribution of \GE\ stars, as all stars we designate as \GE\ members have masses securely within this range, but highlight that this depends on our visual vetting of asteroseismic detection, which may have implicit mass bias we have not accounted for.




\section{Summary}

By combining light curves from the \tess\ Mission, \emph{Gaia} kinematic information and spectroscopic data from the APOGEE and GALAH surveys, we are able to robustly characterize 3 disk stars, and 10 halo stars. Of these halo stars, 4 appear to have been accreted from other galaxies in the distant past. We separate these stars into three groups: the disk population, the {\it in situ} halo population, and the accreted \GE\ population. Our findings can be summarized as follows:
\begin{itemize}
  \item When considering age rankings among a population of 13 galactic disk and halo stars, there exists a wider age range in the galactic disk population than the galactic halo population of stars. The stars kinematically and/or spectroscopically associated with the \GE\ appear to be more tightly clustered in age ranking than other halo or disk stars.
  \item The \GE\ substructure, recently identified our galactic halo in kinematic and spectroscopic space, do not appear to constitute the oldest nor youngest stars in the galactic halo.
  \item The four stars in this study we find are most likely to have been accreted from other galaxies as part of the \GE\ merger based on kinematic and spectroscopic properties (where available) have a median age of 8 $\pm$ 3 (stat.) $\pm$ 1 (sys.) Gyr, as determined from a weighted average resulting from passing the same sets of input parameters to four different age determination methods.
  \item Determining a hierarchical Bayesian age for the population of stars the \GE\ target are drawn from, we find an age of \hbmage, which is indistinguishable from the hierarchical Bayesian population ages determined for the {\it in situ} halo stars (9.5 $\pm$ 0.9 Gyr) or galactic disk stars (4$^{+7}_{-3}$ Gyr).
  \item Despite difficulties in determining absolute as well as relative ages for stars dissimilar to the Sun, age rankings for these stars appear relatively robust across multiple age determination methods.
\end{itemize} 

We also note that the age range and variance of \GE\ and halo stars is smaller the disk star population, and the halo population appears to robustly include stars both older and younger than the \GE\ stars (although our halo sample appears to have a more tightly constrained mean population age than our \GE\ sample). The disk population includes robustly younger stars than either of the other two halo populations. Furthermore, we caution that deriving absolute ages for these stars relies on proper modeling of the effects of composition, which are particularly difficult given the lack of nearby analogues for these stars. However, much more detailed asteroseismic and spectroscopic analysis combined with a detailed, careful description of the stellar interior structure can address some of this systematic uncertainty.


Additional spectra and longer, more precise light curves of our initial stellar sample will be crucial to confirming these results. Longer time baselines and higher cadence observations make the stochastically excited stellar oscillations more easily observable, and more precisely constrained. The extended missions of \tess\ will be crucial in obtaining these measurements \citep[][Stello, {\it in prep.}]{mackereth2021}. In addition, further spectroscopic coverage of these stars will be extremely valuable for reducing the uncertainties of the stellar ages estimated for these stars. Future releases of astrometric information from \emph{Gaia} will reduce parallax uncertainties for these targets as well, improving stellar model fits.

\acknowledgements{The authors thank David Hogg, Sarah Pearson and the attendees of the tess.ninja 3 Workshop for helpful discussions. S.G. acknowledges support by the National Aeronautics and Space Administration under Grant 80NSSC19K0110 issued through the K2 Guest Observer Program. JCZ is supported by an NSF Astronomy and Astrophysics Postdoctoral Fellowship under award AST-2001869.
Funding for the Stellar Astrophysics Centre is provided by The Danish National Research Foundation (Grant agreement no.: DNRF106). JLR acknowledges support from the Carlsberg Foundation (grant agreement CF19-0649).}


\software{
Astropy \citep{astropy13, astropy18}, 
gala \citep{gala, gala:version}, 
IPython \citep{ipython}, 
numpy \citep{numpy}, 
pyia \citep{pyia}, 
scipy \citep{scipy}, 
lightkurve \citep{lightkurve}
}

\bibliography{main.bib}

\appendix{

\begin{figure*}[h!]
    \centering
    \includegraphics[width=\linewidth]{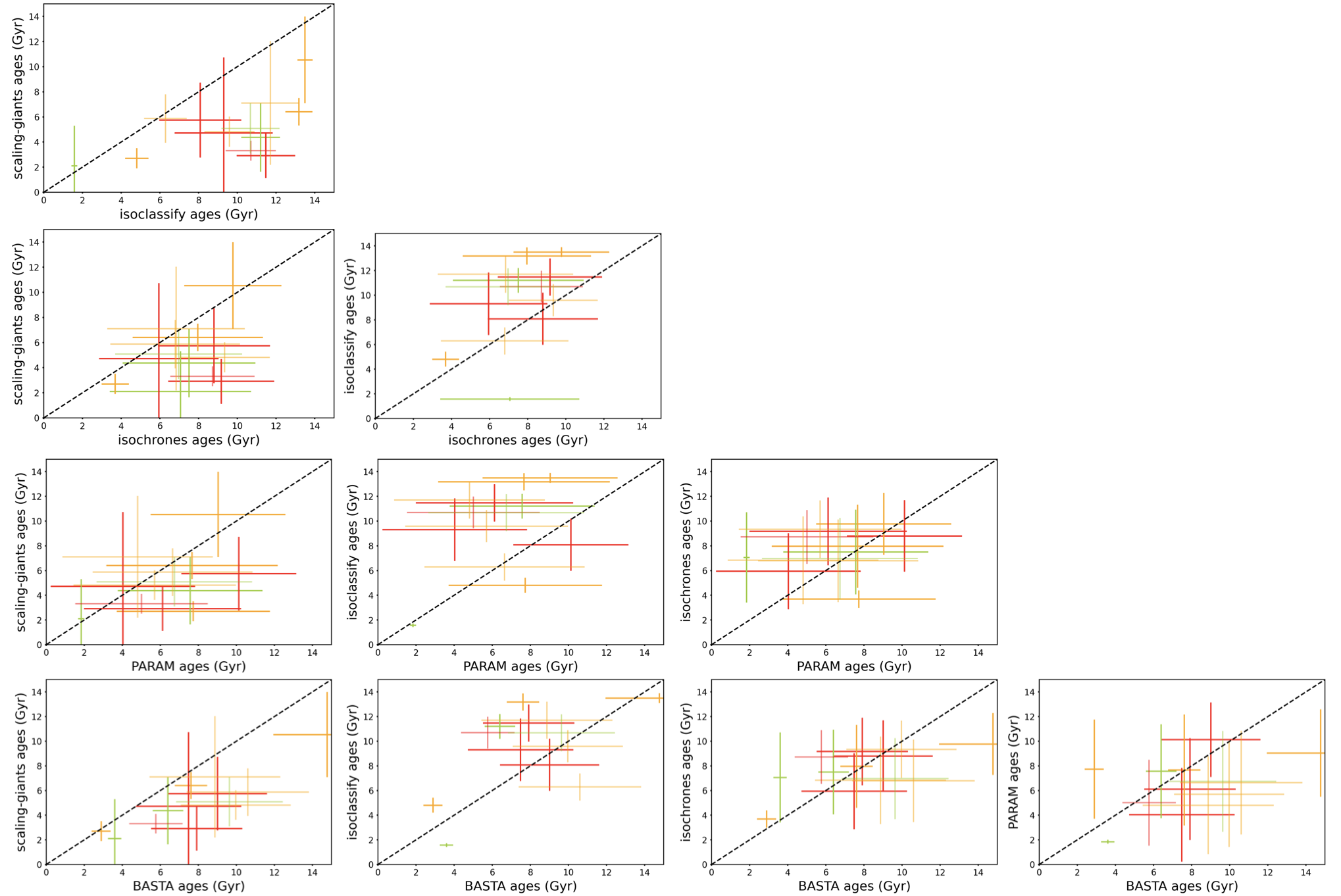}
    \caption{Median ages determined for our stellar sample, comparing 5 different age estimation methods against one another.  Errors are determined by each age determination package separately. Greater age agreement and smaller systematic offsets are seen between ages determined by the \texttt{PARAM} and \texttt{BASTA} packages, whose age determinations for red giant stars have been previously tested in the literature. In addition, \GE\ stars appear to occupy a smaller range of ages than the {\it in situ} halo and disk populations. However, given the large uncertainties on individual age measurements, stellar population distributions are largely in agreement within uncertainties. No clear difference in age determination can be seen between stars with (opaque) and without (translucent) spectral parameters, or between stars of different origins.}
    \label{fig:agecomp}
\end{figure*}

\begin{figure}
    \centering
    \includegraphics[width=.5\linewidth]{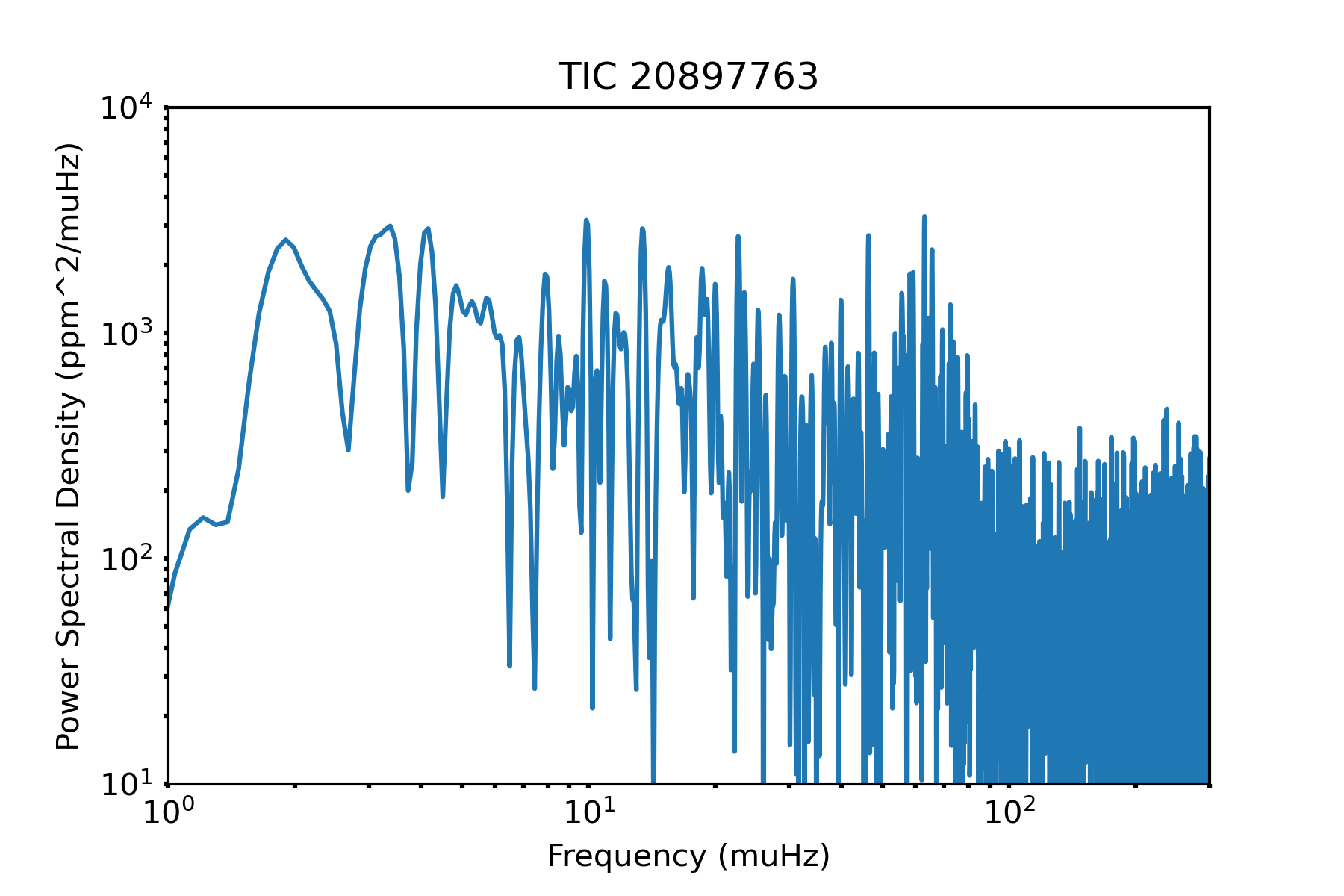}
    \includegraphics[width=.45\linewidth]{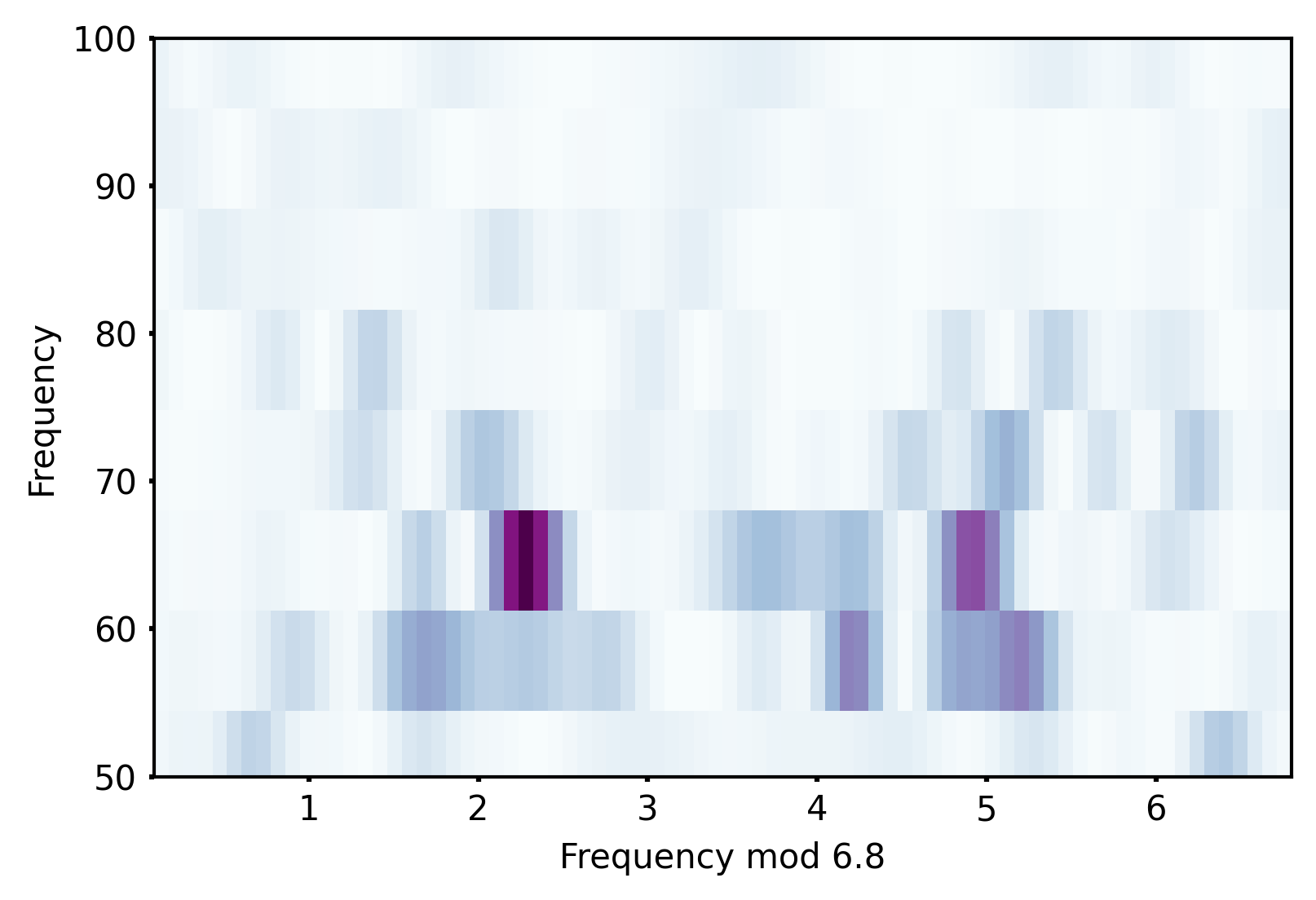}
    \caption{Power spectral density versus frequency (left) and echelle diagram (right) for the \emph{} light curve of TIC20897763.}
\end{figure}

\begin{figure}
    \centering
    \includegraphics[width=.5\linewidth]{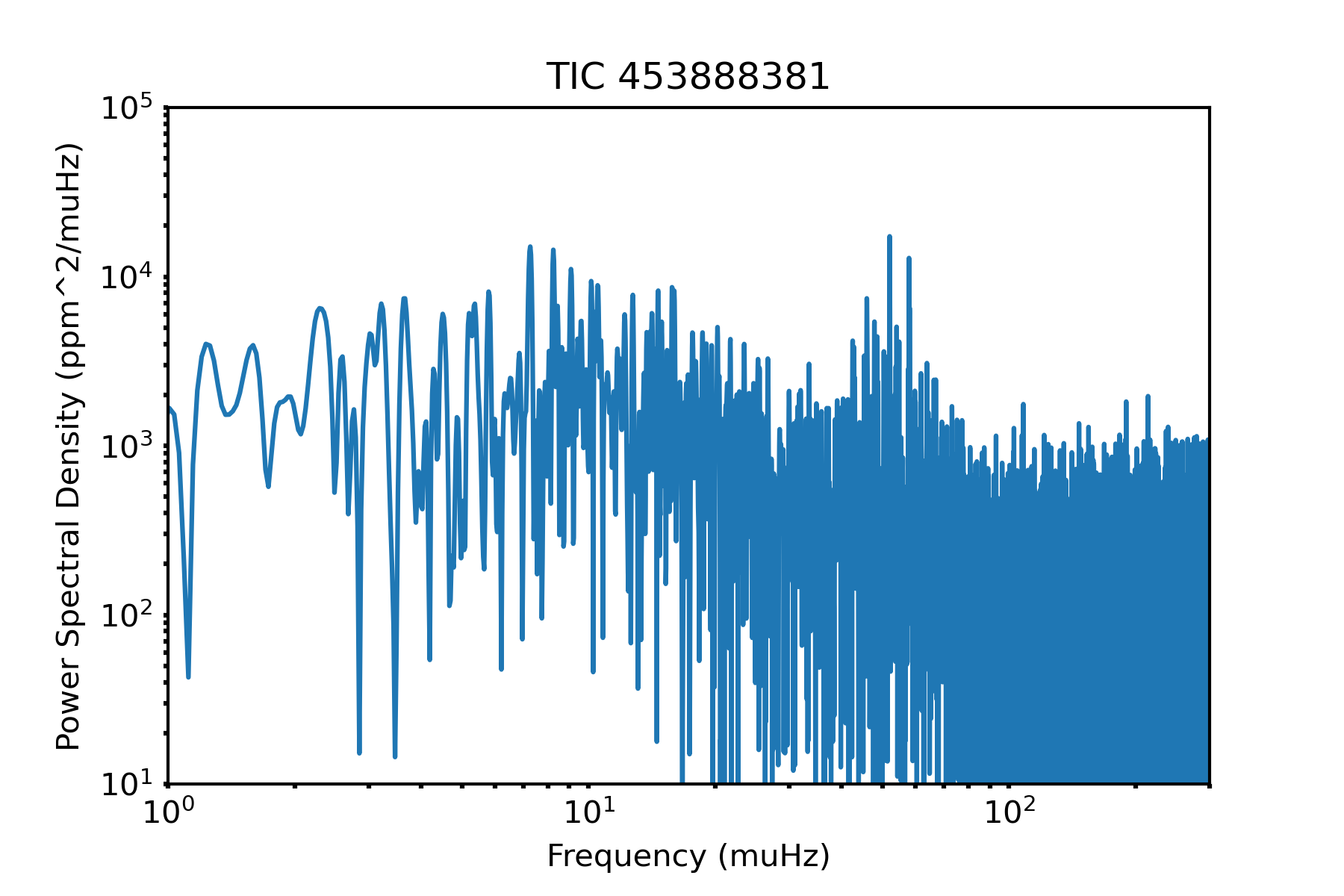}
    \includegraphics[width=.45\linewidth]{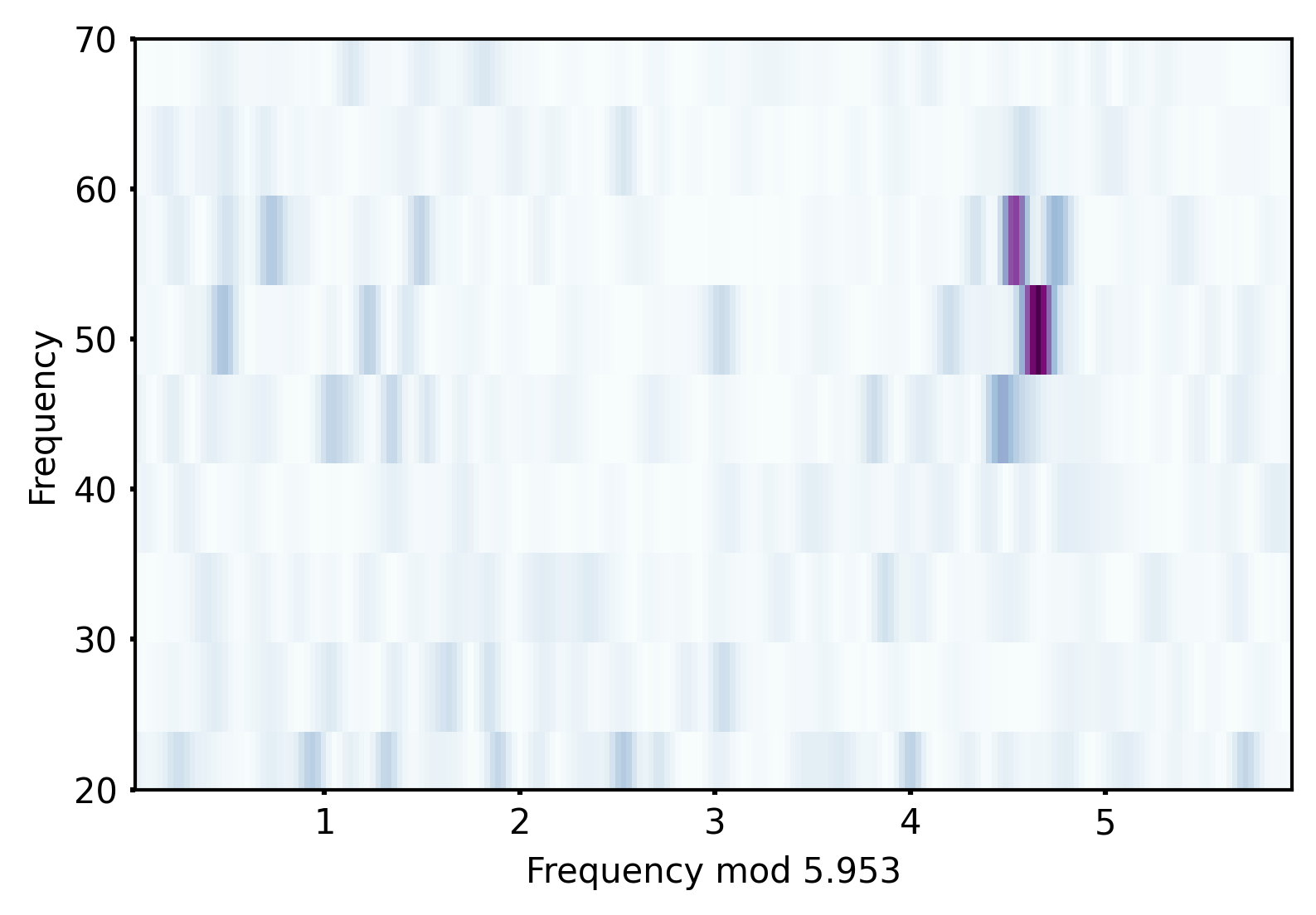}
    \caption{Power spectral density versus frequency (left) and echelle diagram (right) for the \emph{} light curve of TIC453888381.}
\end{figure}

\begin{figure}
    \centering
    \includegraphics[width=.5\linewidth]{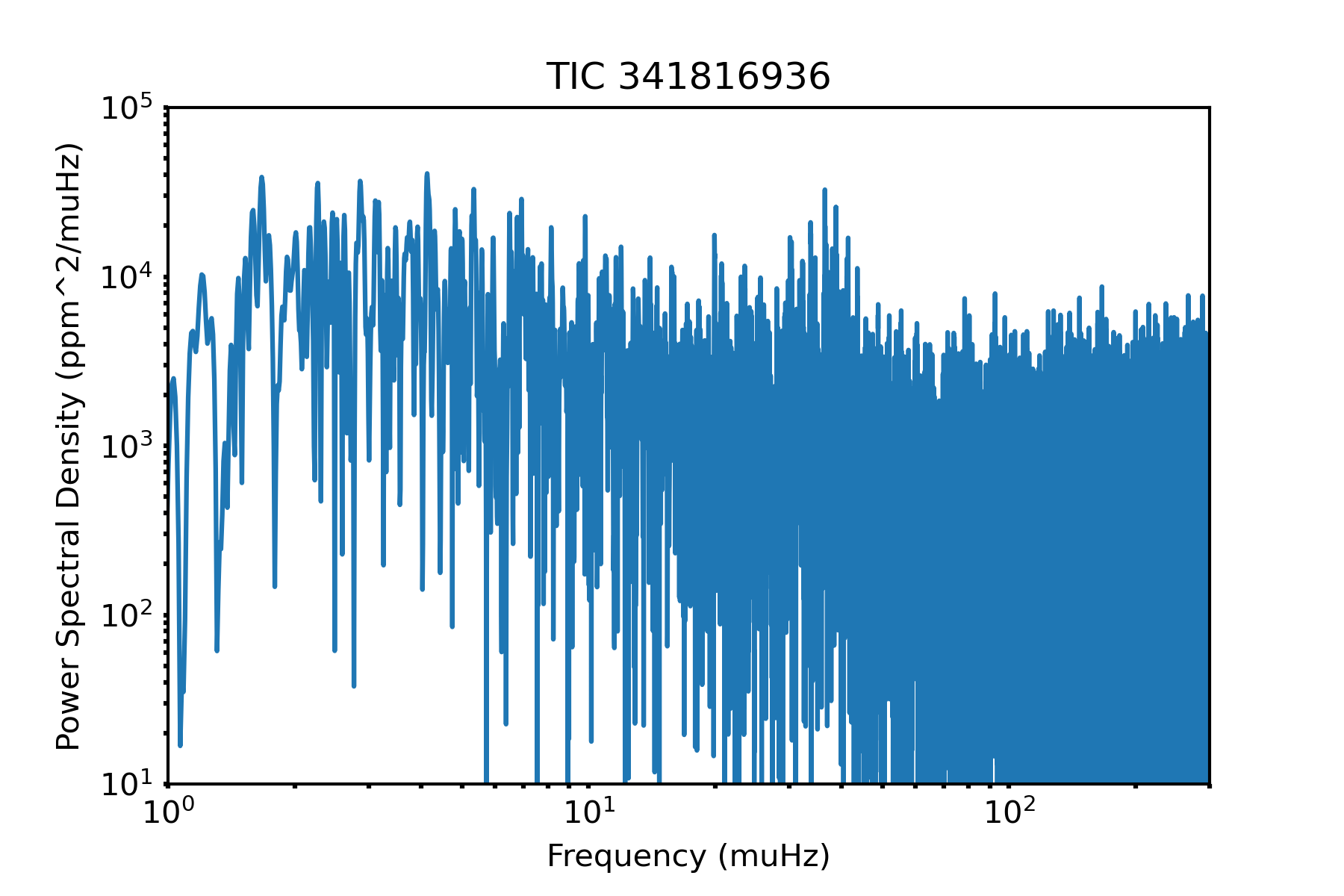}
    \includegraphics[width=.45\linewidth]{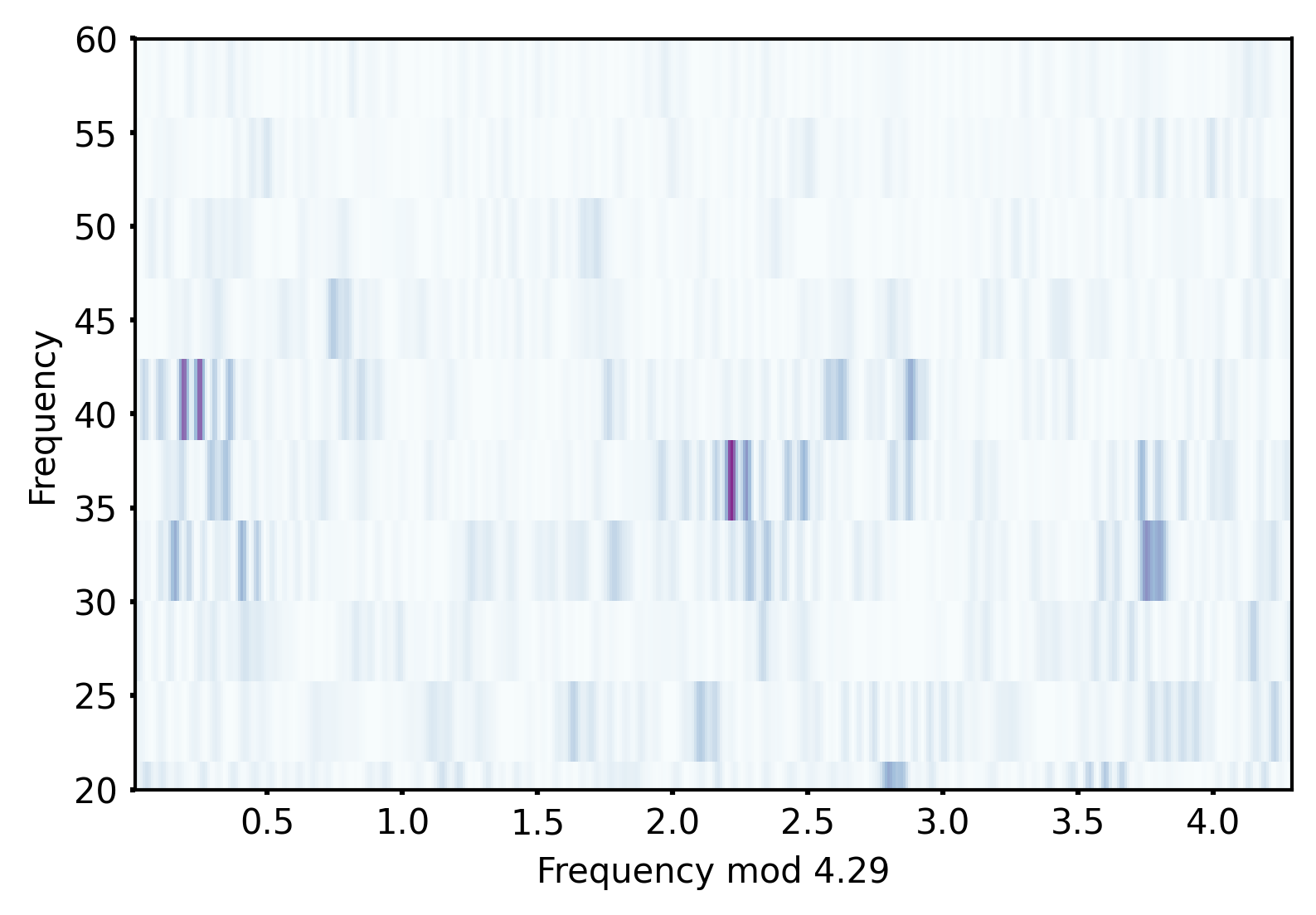}
    \caption{Power spectral density versus frequency (left) and echelle diagram (right) for the \emph{} light curve of TIC341816936.}
\end{figure}

\begin{figure}
    \centering
    \includegraphics[width=.5\linewidth]{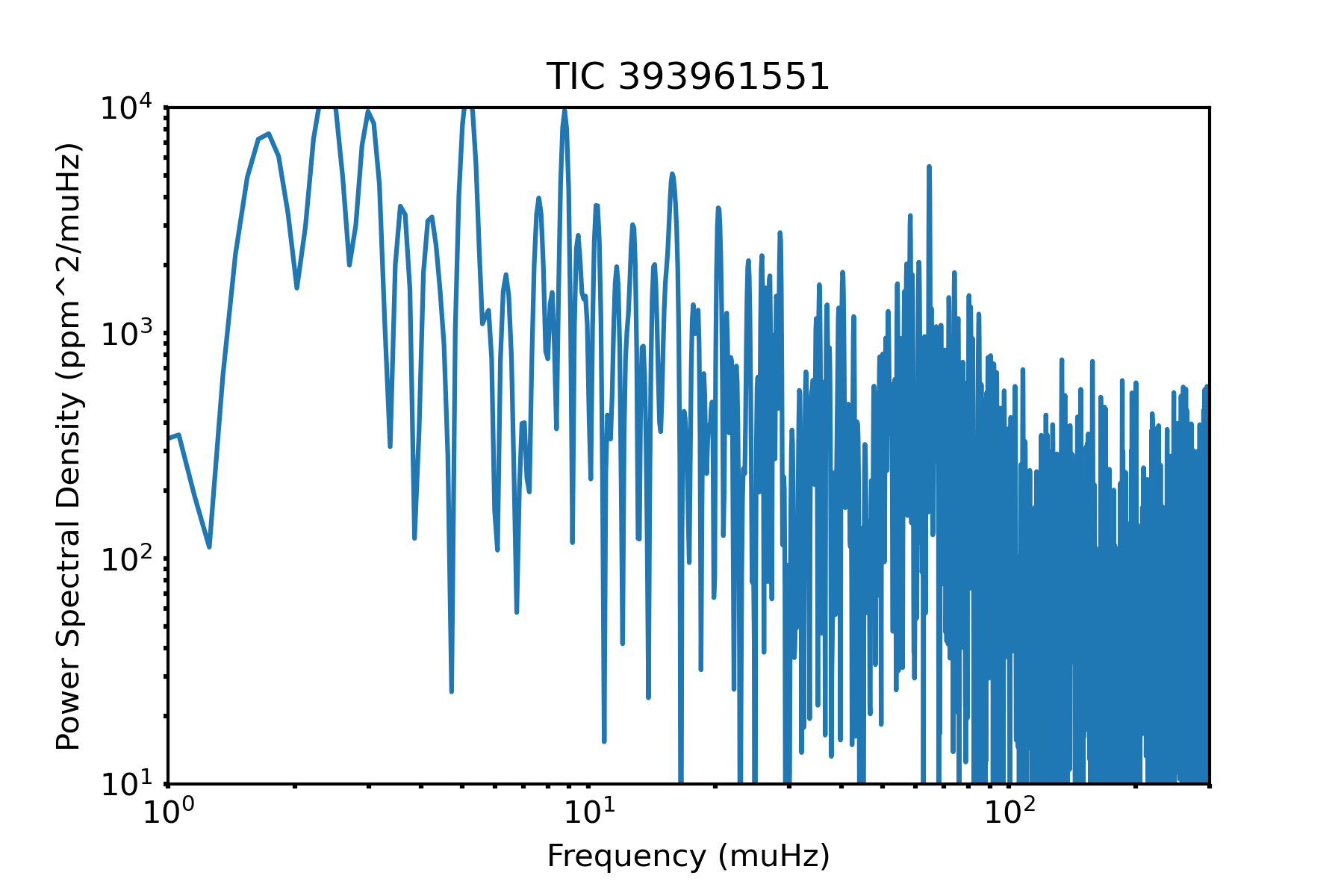}
    \includegraphics[width=.45\linewidth]{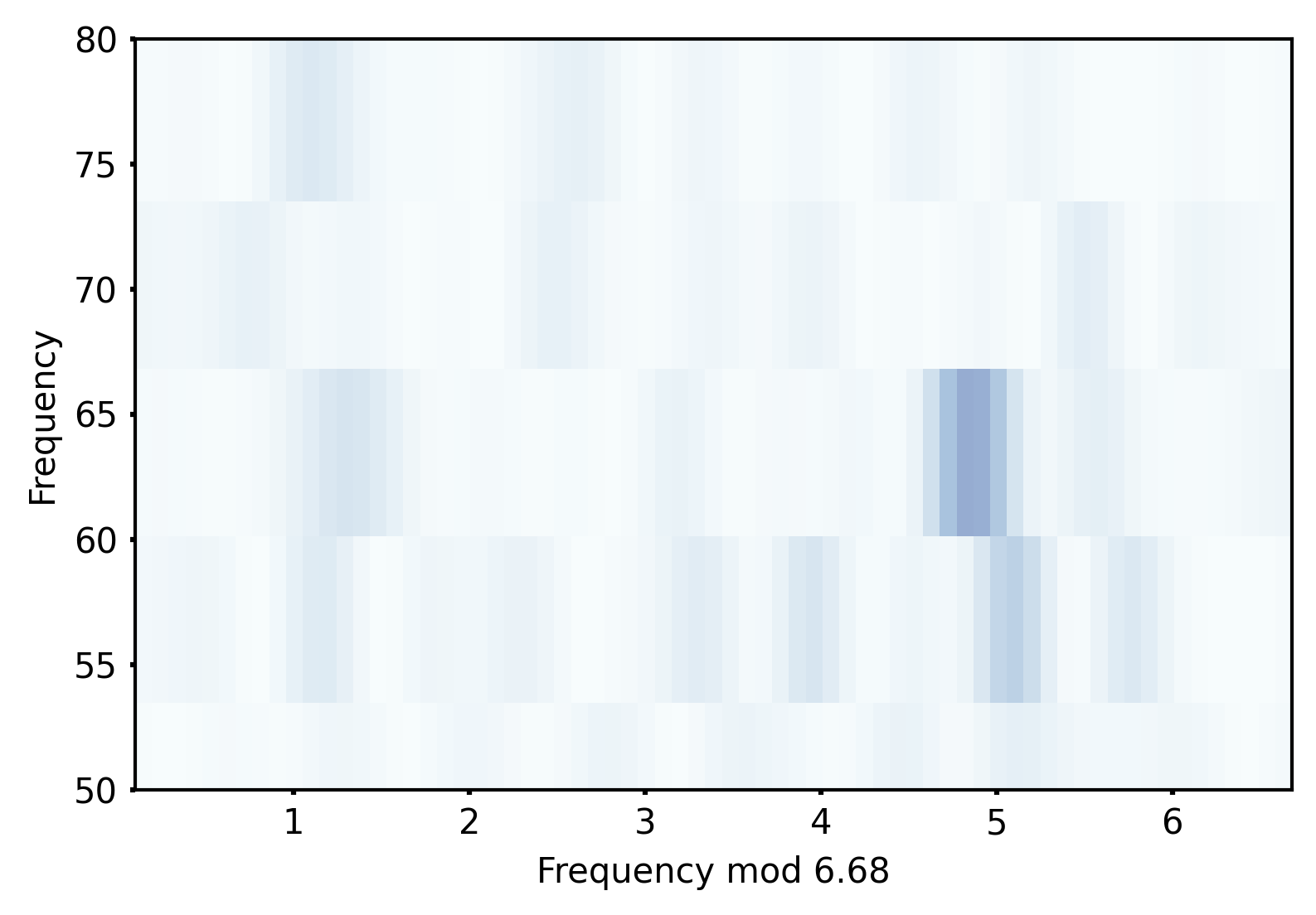}
    \caption{Power spectral density versus frequency (left) and echelle diagram (right) for the \emph{} light curve of TIC393961551.}
\end{figure}

\begin{figure}
    \centering
    \includegraphics[width=.5\linewidth]{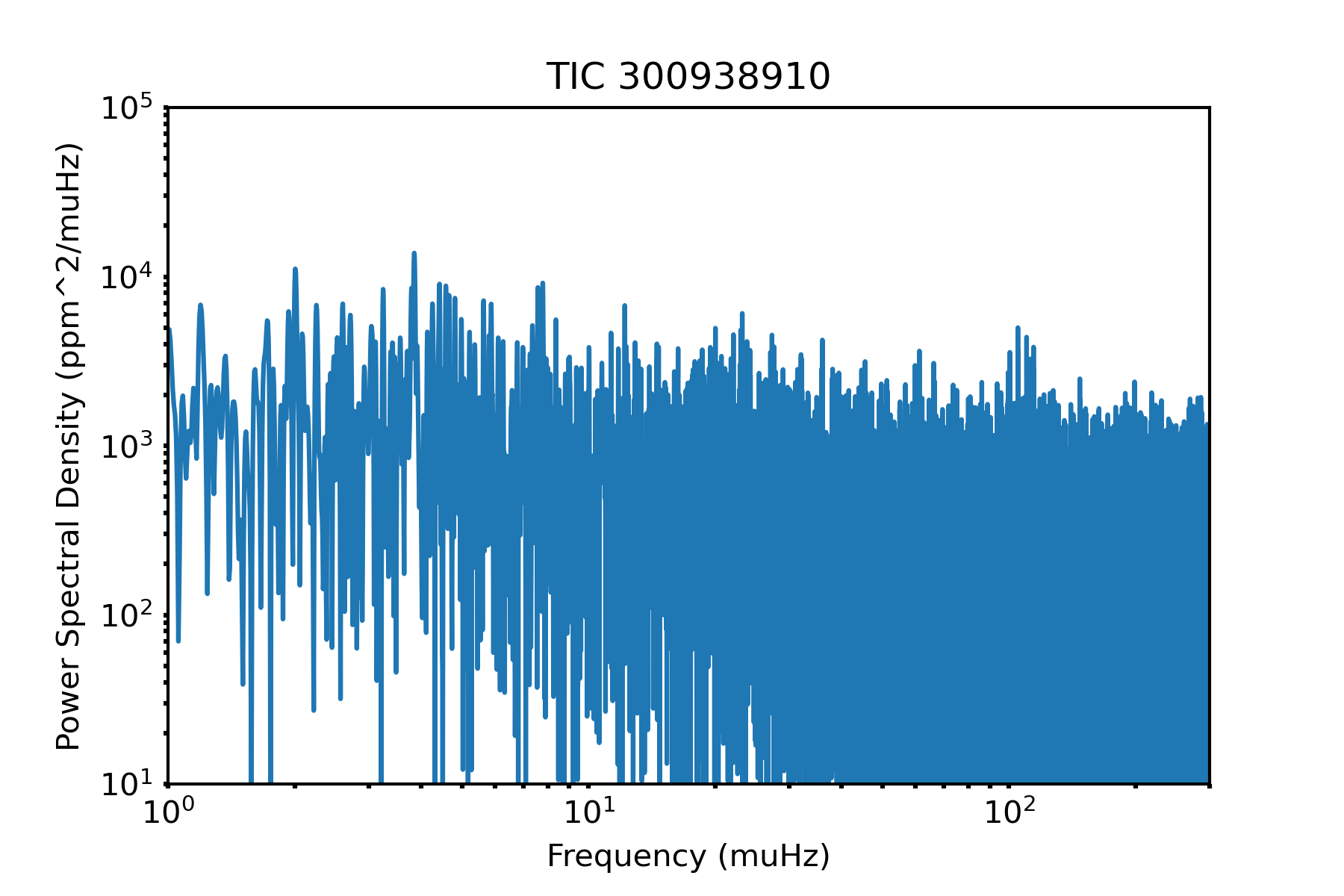}
    \includegraphics[width=.45\linewidth]{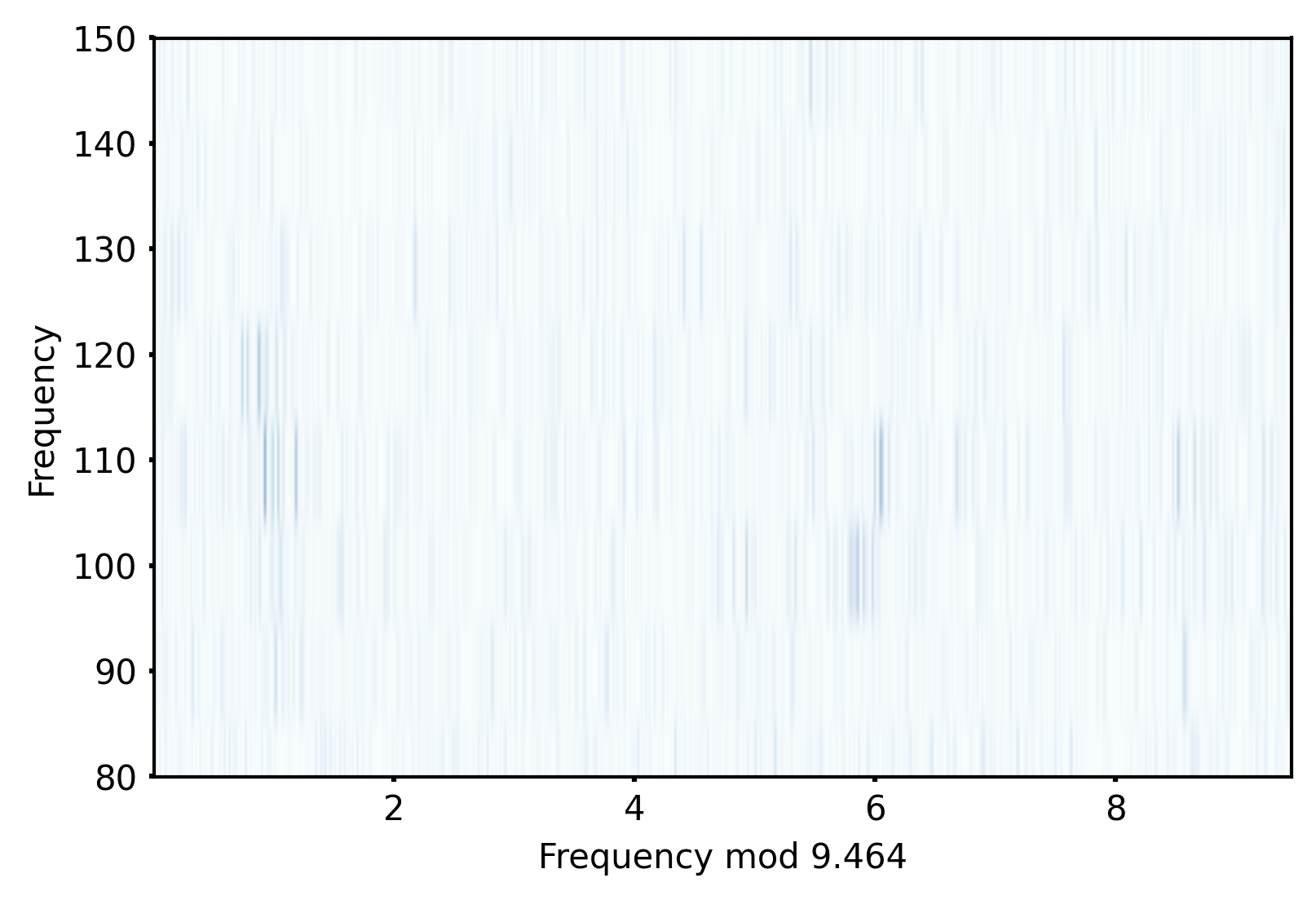}
    \caption{Power spectral density versus frequency (left) and echelle diagram (right) for the \emph{} light curve of TIC300938910.}
\end{figure}

\begin{figure}
    \centering
    \includegraphics[width=.5\linewidth]{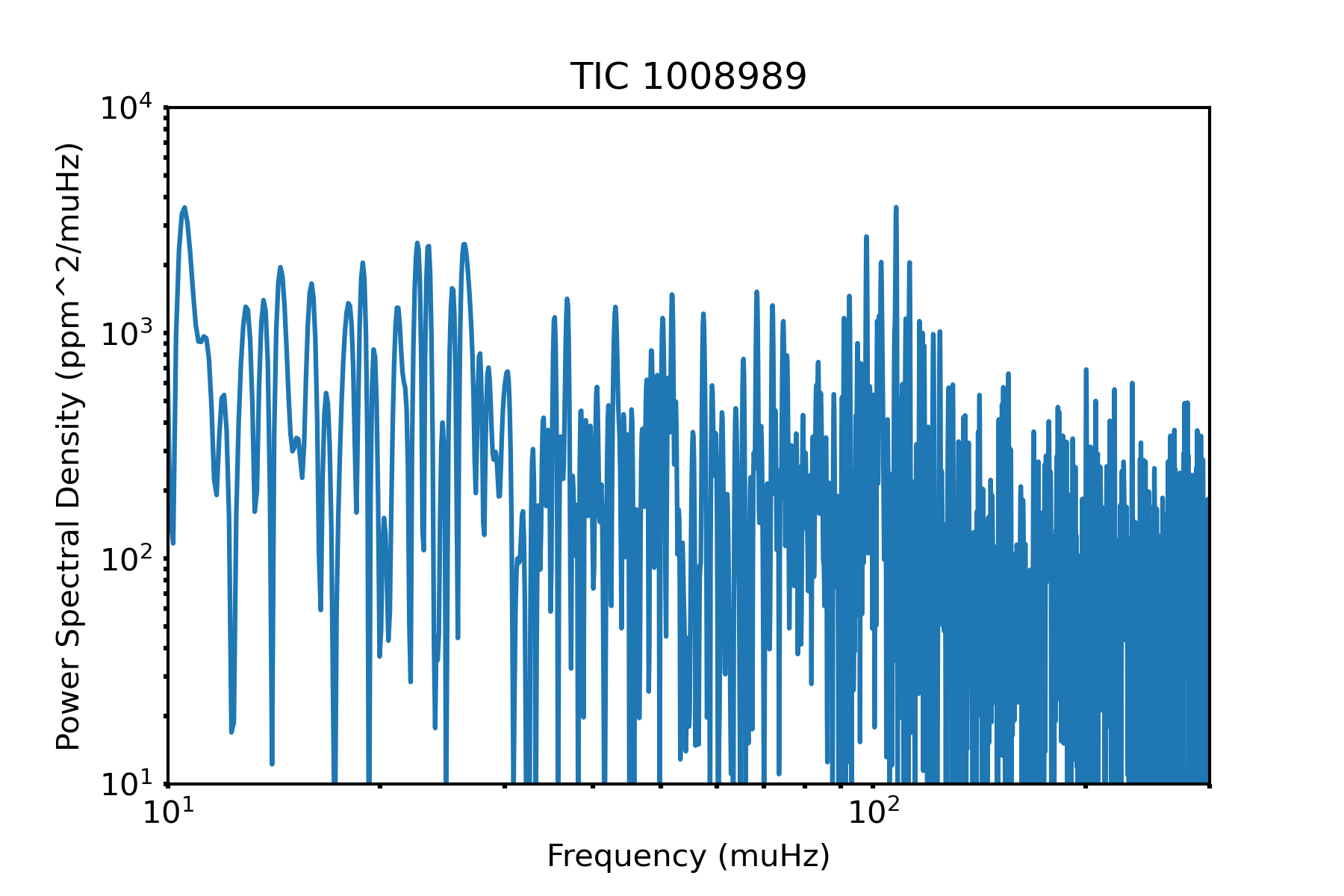}
    \includegraphics[width=.45\linewidth]{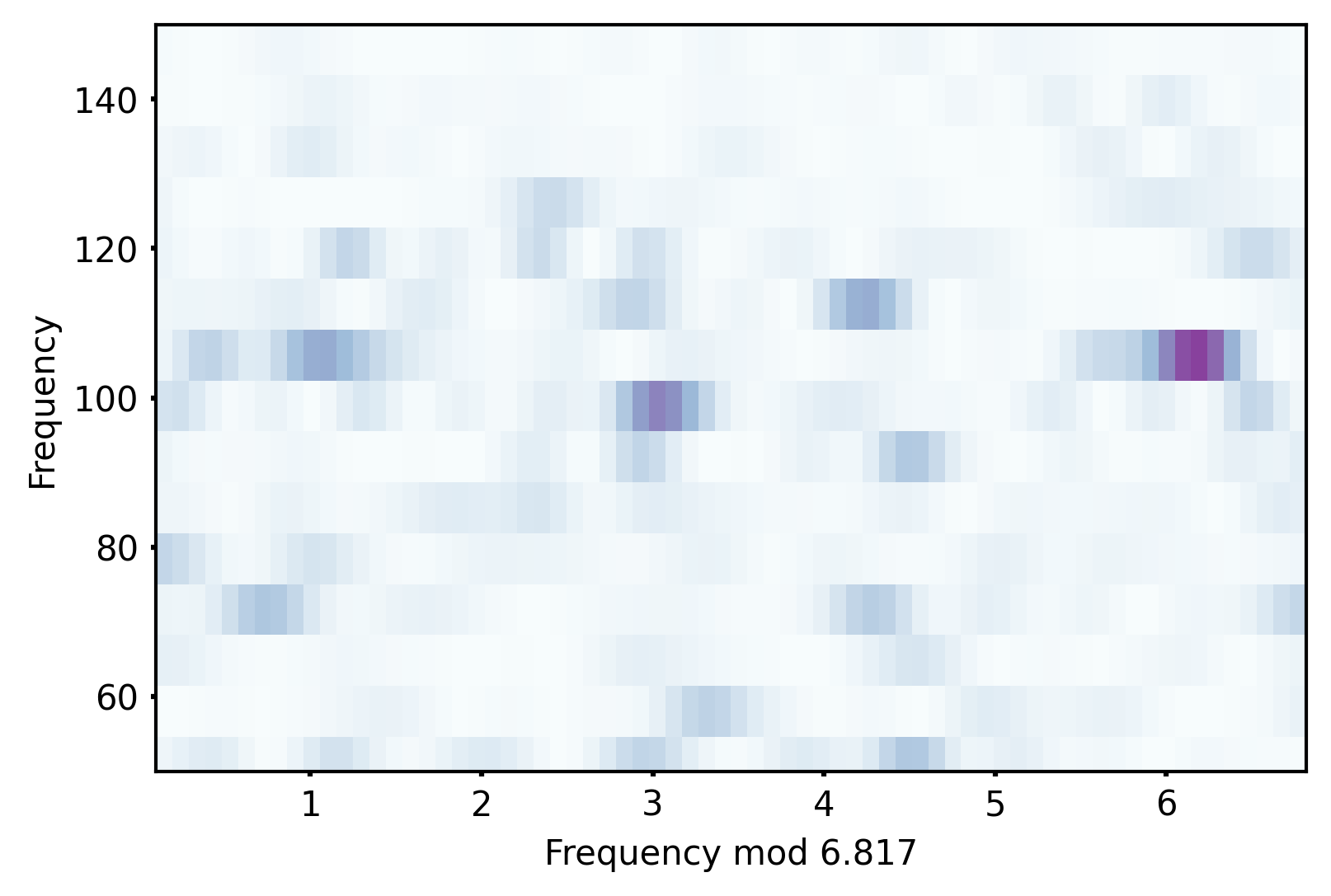}
    \caption{Power spectral density versus frequency (left) and echelle diagram (right) for the \emph{} light curve of TIC1008989.}
\end{figure}

\begin{figure}
    \centering
    \includegraphics[width=.5\linewidth]{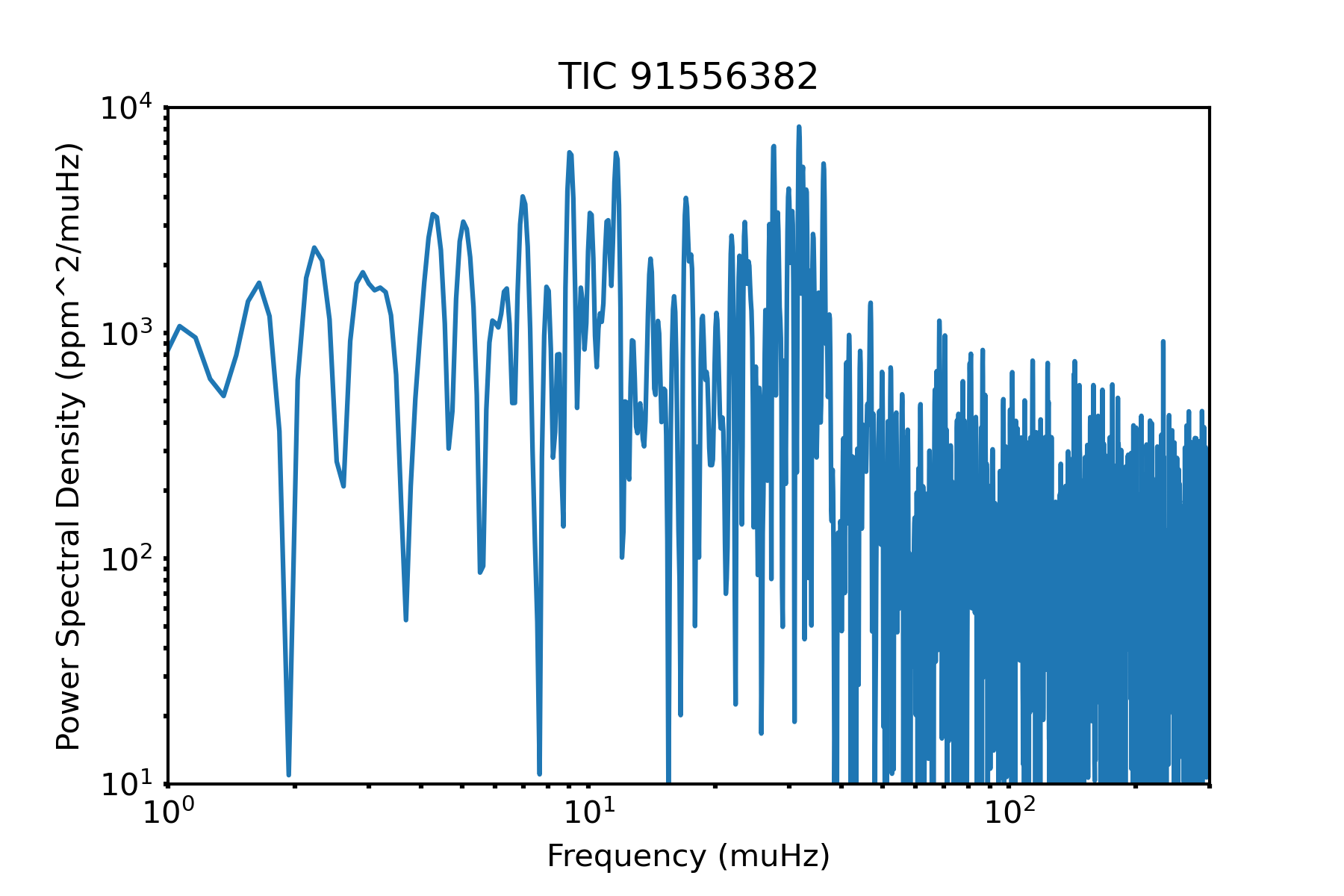}
    \includegraphics[width=.45\linewidth]{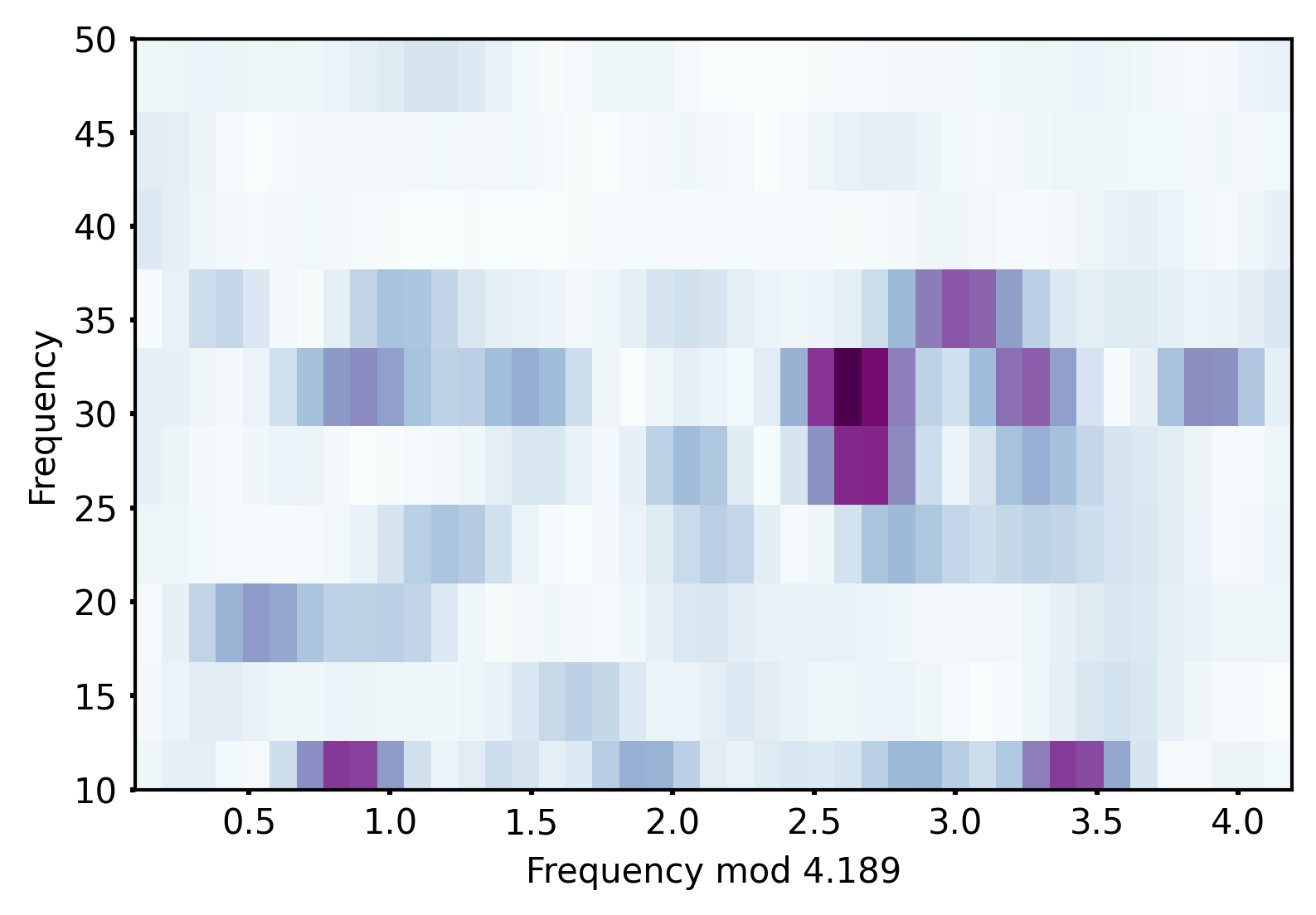}
    \caption{Power spectral density versus frequency (left) and echelle diagram (right) for the \emph{} light curve of TIC91556382.}
\end{figure}

\begin{figure}
    \centering
    \includegraphics[width=.5\linewidth]{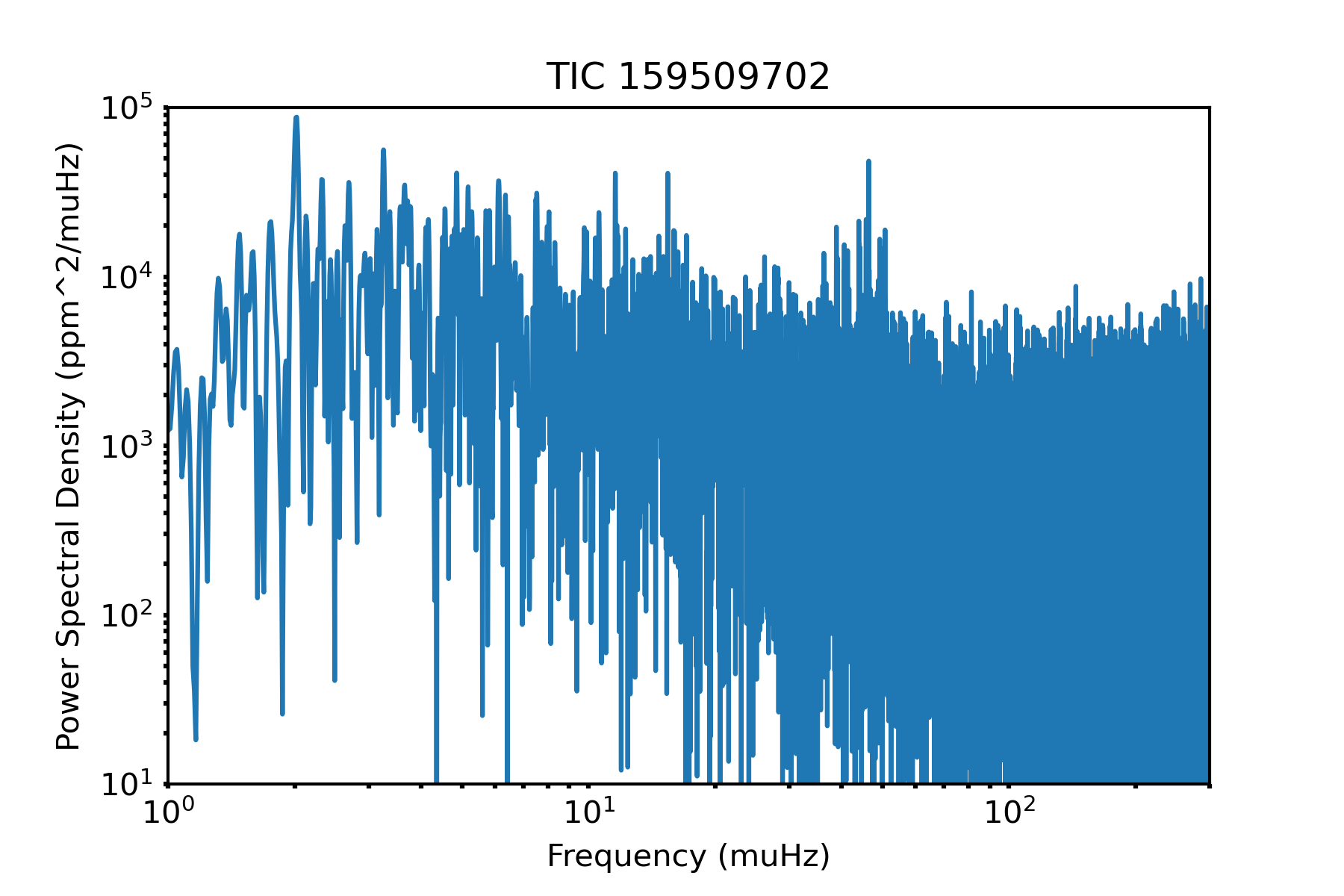}
    \includegraphics[width=.45\linewidth]{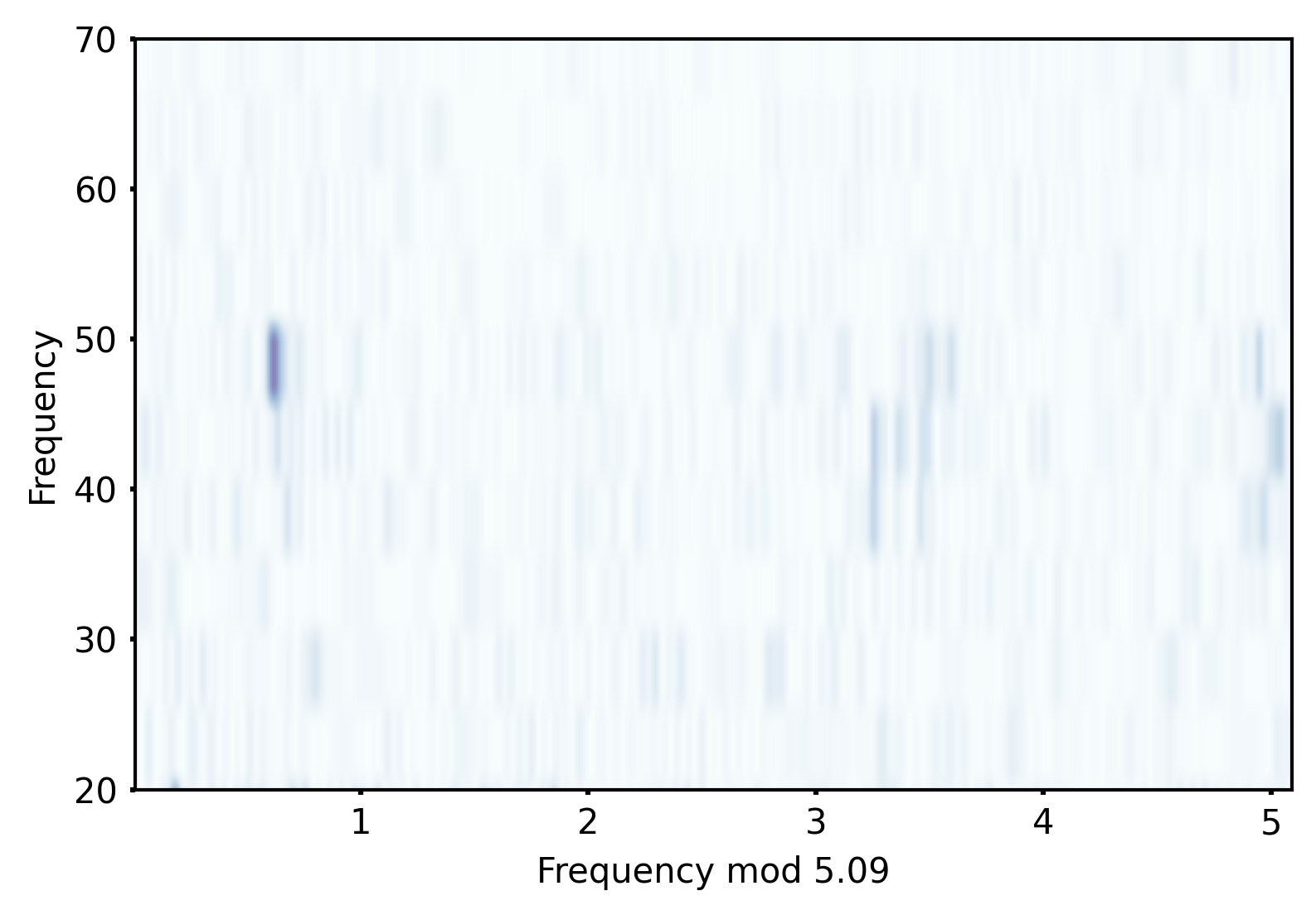}
    \caption{Power spectral density versus frequency (left) and echelle diagram (right) for the \emph{} light curve of TIC159509702.}
\end{figure}

\begin{figure}
    \centering
    \includegraphics[width=.5\linewidth]{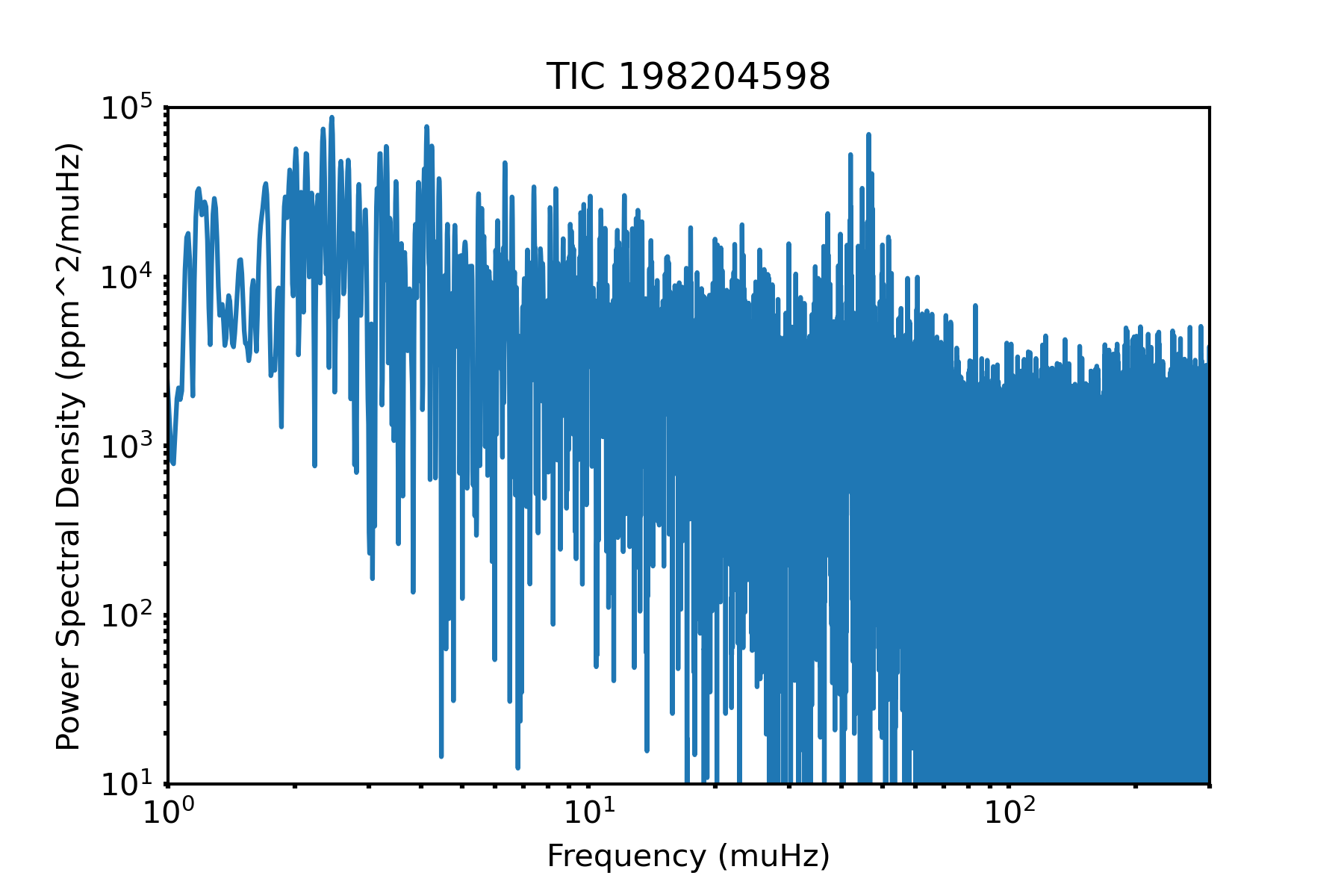}
    \includegraphics[width=.45\linewidth]{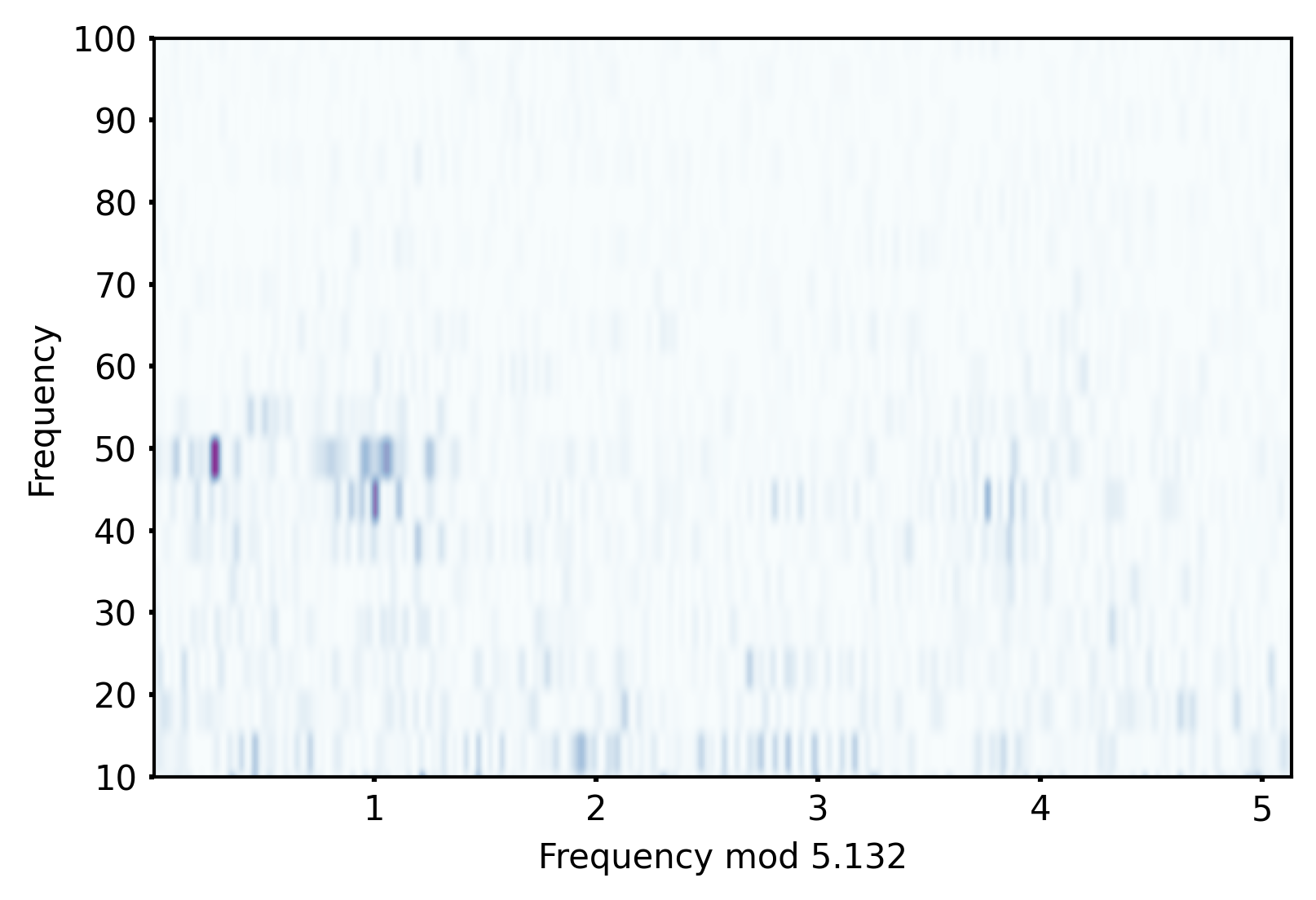}
    \caption{Power spectral density versus frequency (left) and echelle diagram (right) for the \emph{} light curve of TIC198204598.}
\end{figure}

\begin{figure}
    \centering
    \includegraphics[width=.5\linewidth]{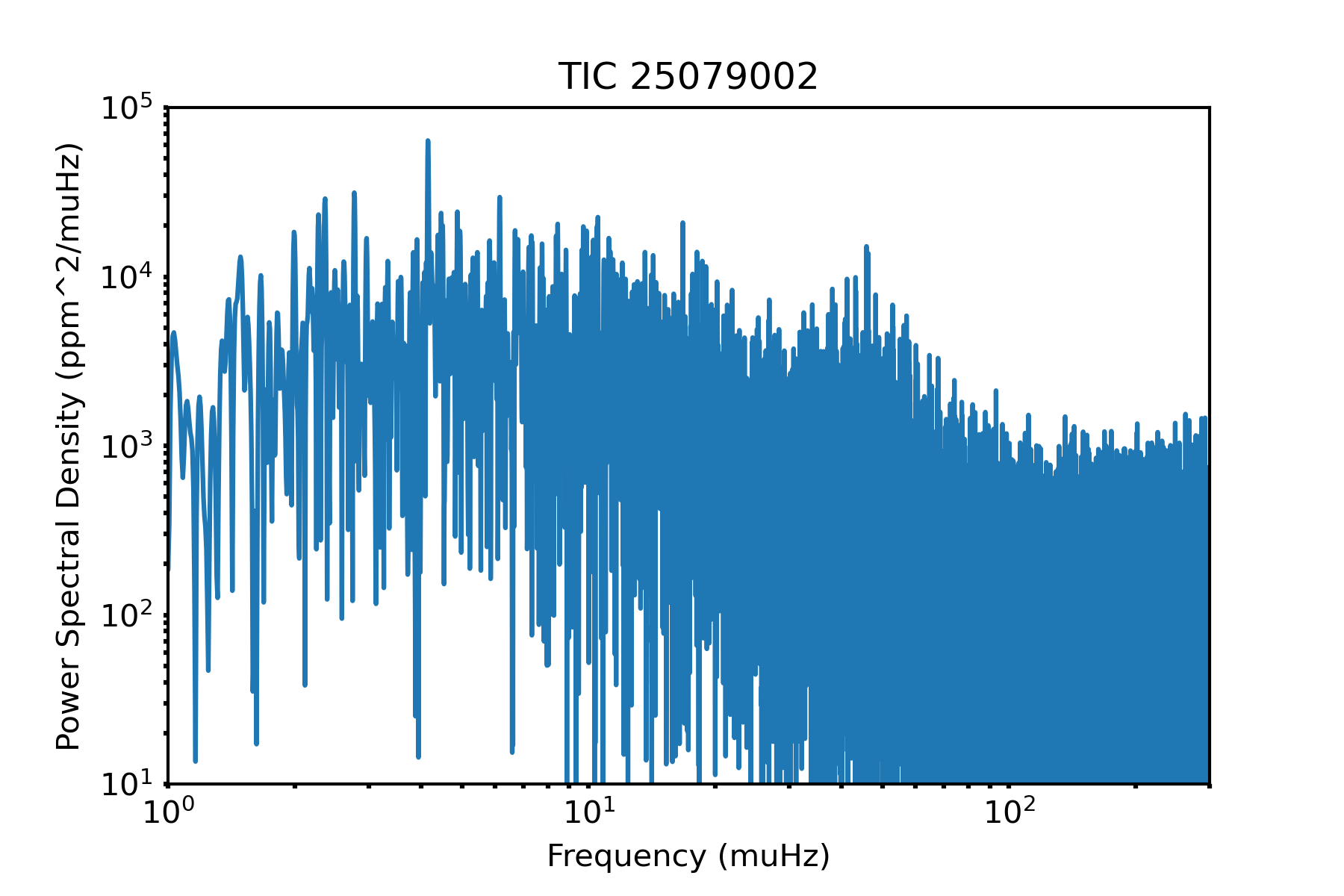}
    \includegraphics[width=.45\linewidth]{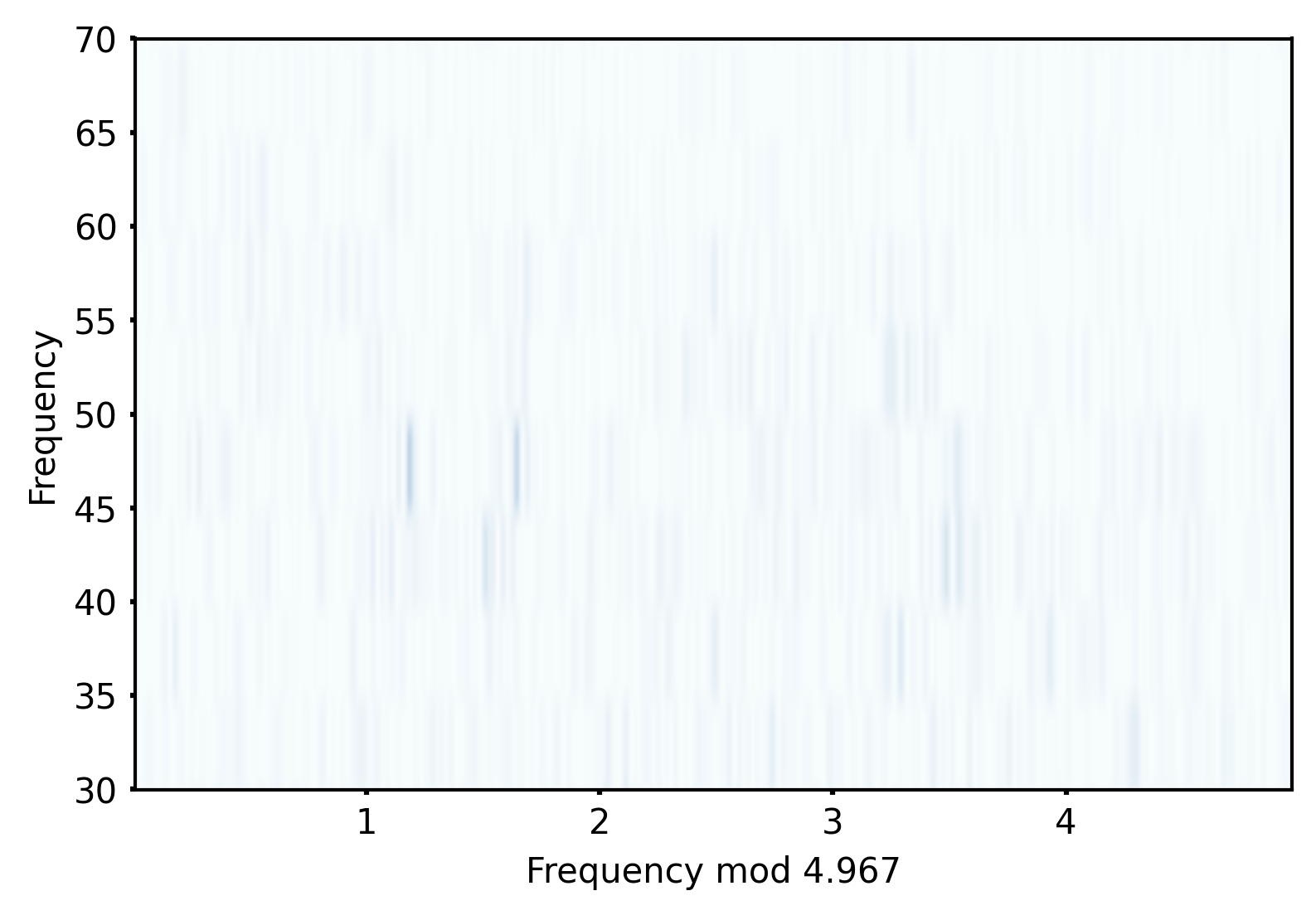}
    \caption{Power spectral density versus frequency (left) and echelle diagram (right) for the \emph{} light curve of TIC25079002.}
\end{figure}

\begin{figure}
    \centering
    \includegraphics[width=.5\linewidth]{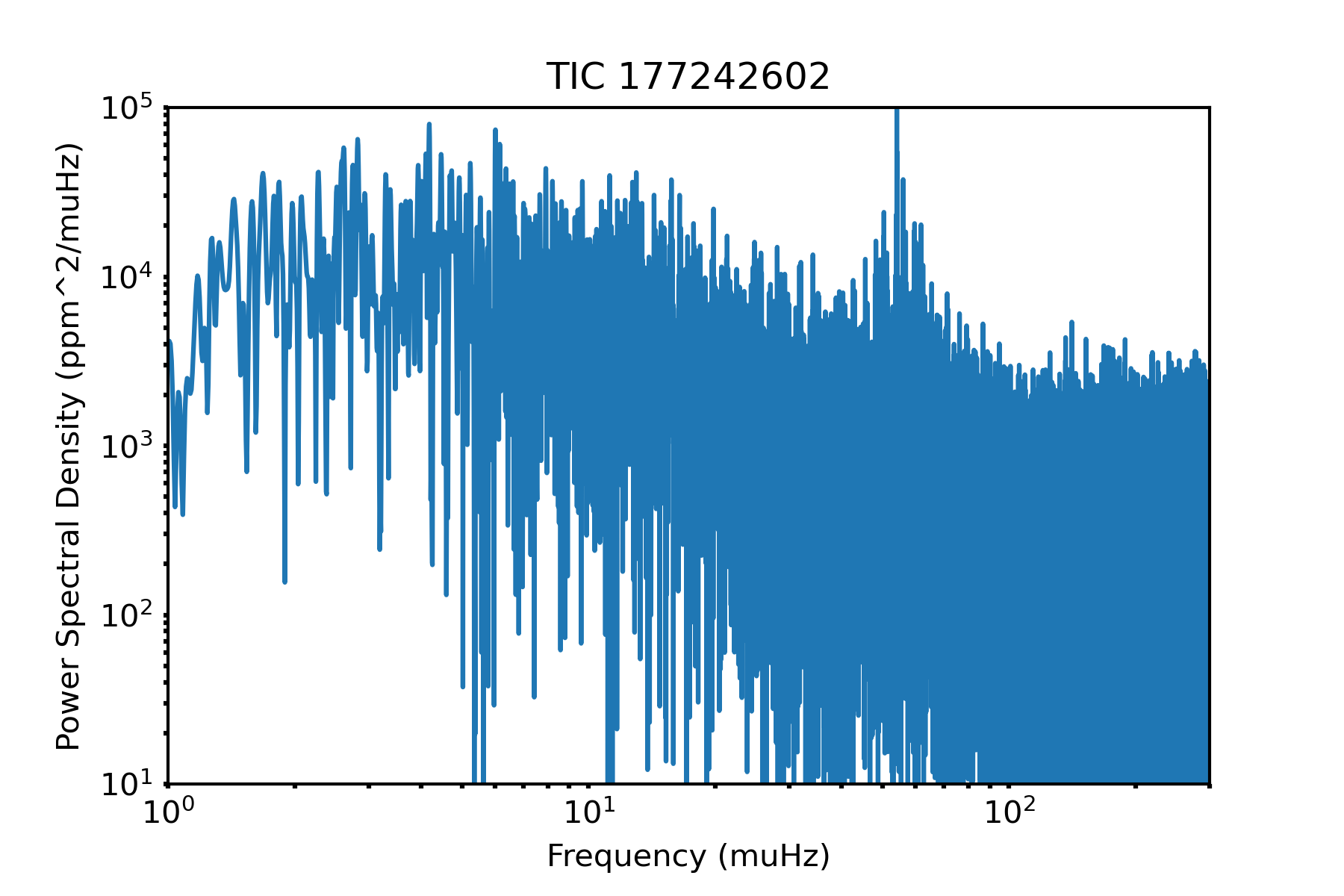}
    \includegraphics[width=.45\linewidth]{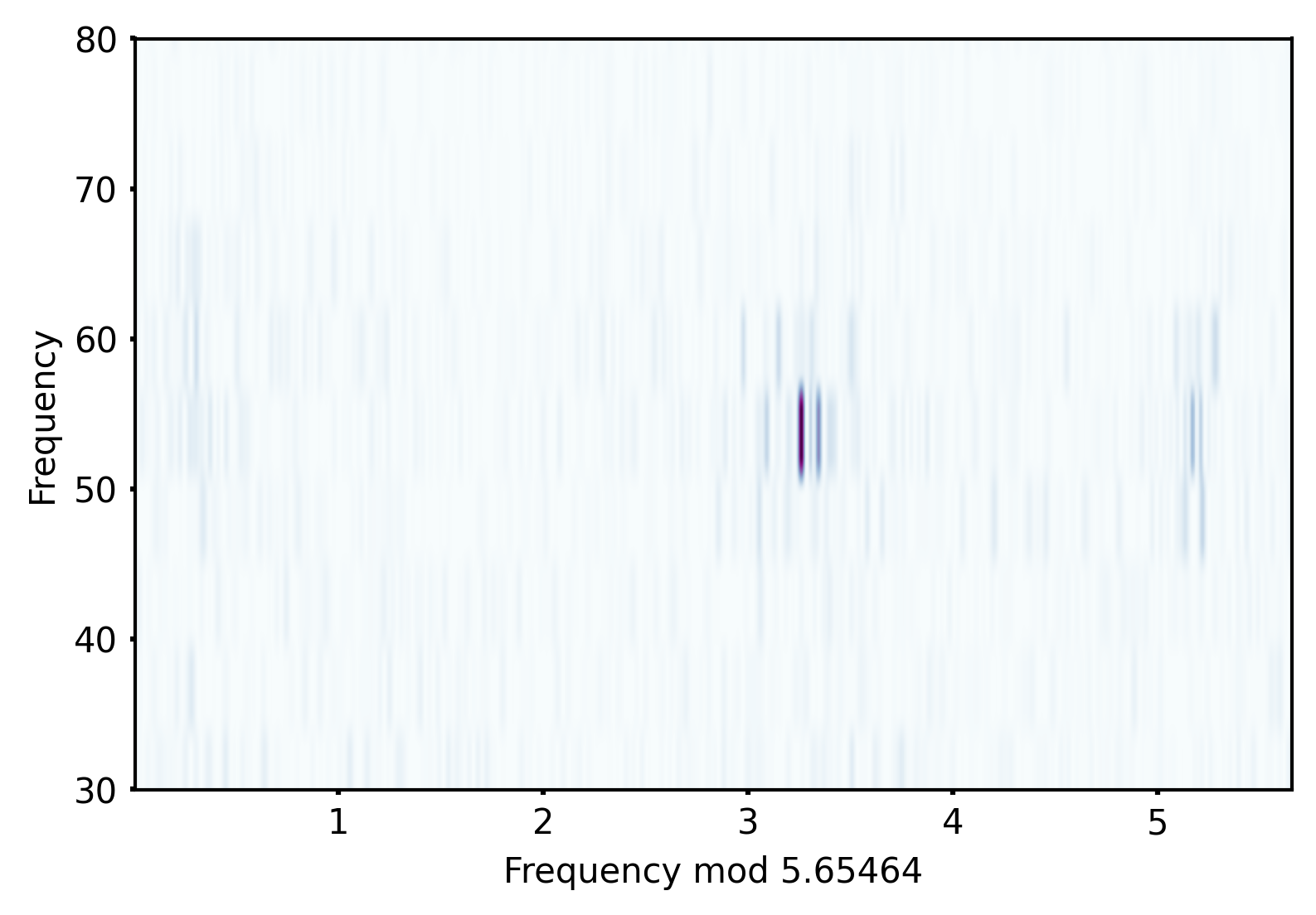}
    \caption{Power spectral density versus frequency (left) and echelle diagram (right) for the \emph{} light curve of TIC177242602.}
\end{figure}

\begin{figure}
    \centering
    \includegraphics[width=.5\linewidth]{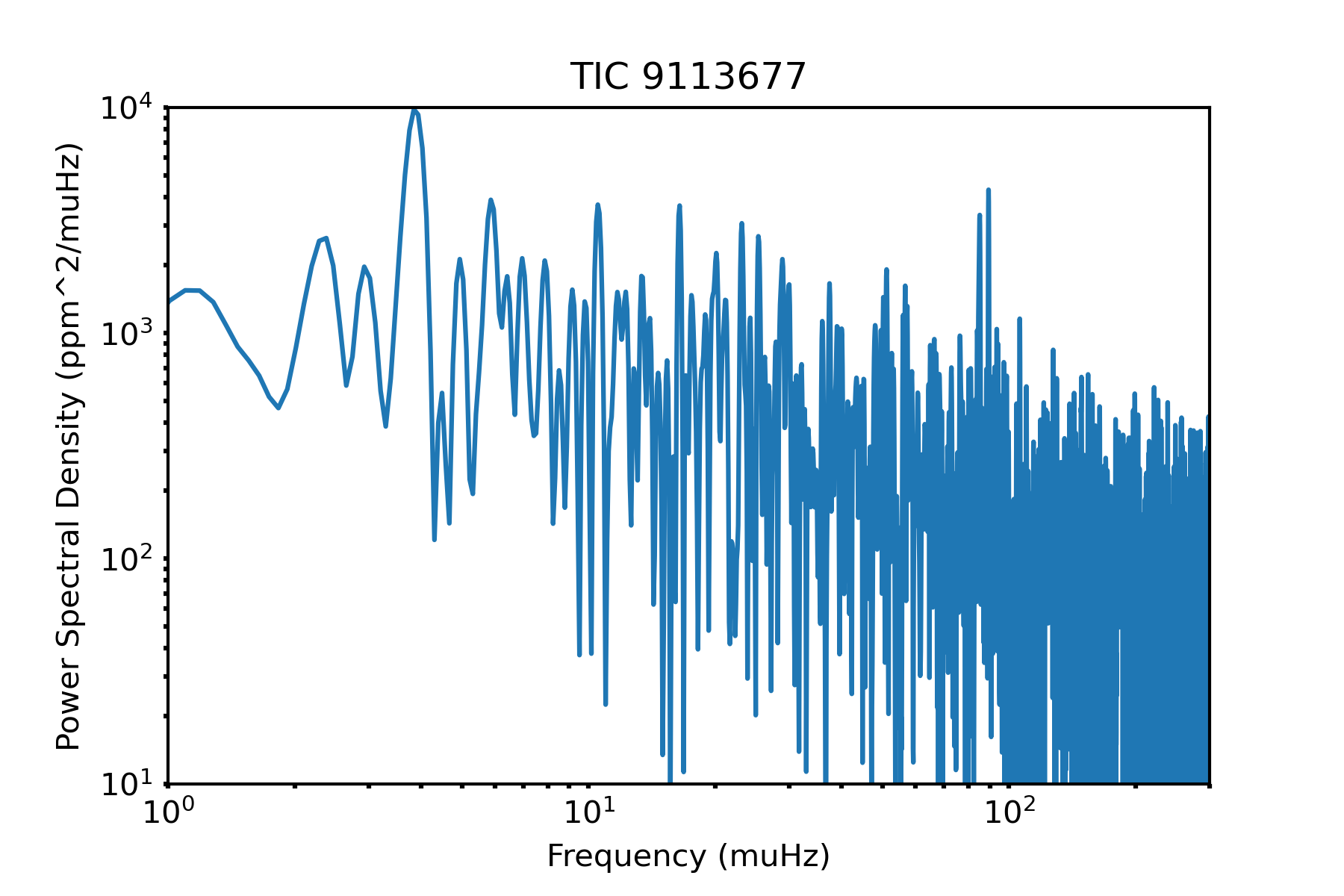}
    \includegraphics[width=.45\linewidth]{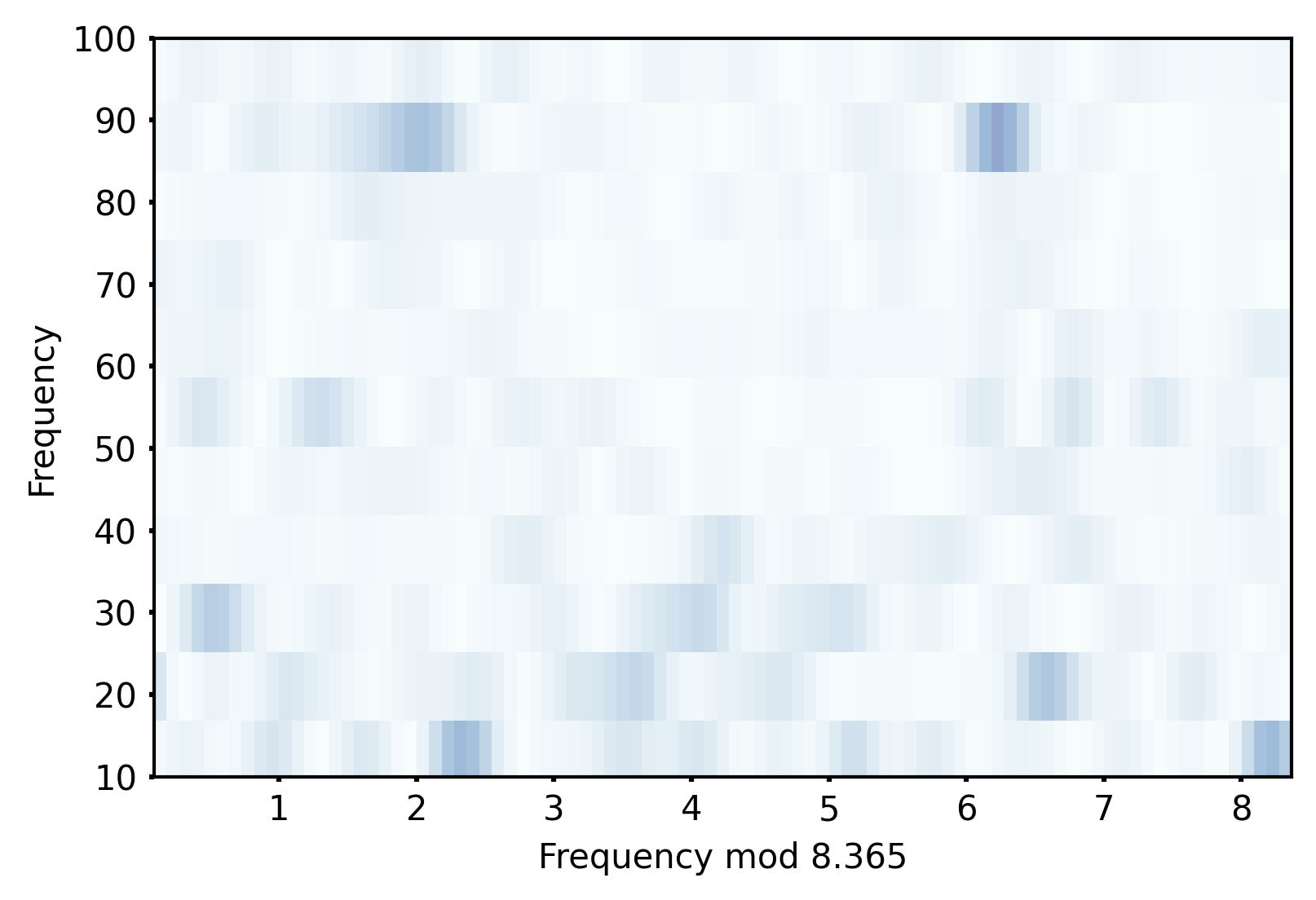}
    \caption{Power spectral density versus frequency (left) and echelle diagram (right) for the \emph{} light curve of TIC9113677.}
\end{figure}




}



\end{document}